\documentclass[twocolumn]{aastex631}

\def\pluto{{\sc pluto}}
\def\mesa{{\sc mesa}}
\def\snec{{\sc snec}}
\def\sn14c{SN~2014C}
\def\mlrate{$M_{\odot}$~yr$^{-1}$}
\def\mdot{\dot{M}}
\def\kms{km~s$^{-1}$}

\begin{document}


\title{Constraining the CSM structure and progenitor mass-loss history of interacting supernovae \\through 3D hydrodynamic modeling: The case of SN 2014C}

\author[0000-0003-2836-540X]{Salvatore Orlando}
\affiliation{INAF -- Osservatorio Astronomico di Palermo, Piazza del Parlamento 1, I-90134 Palermo, Italy}
\email{salvatore.orlando@inaf.it}

\author[0000-0001-5792-0690]{Emanuele Greco}
\affiliation{INAF -- Osservatorio Astronomico di Palermo, Piazza del Parlamento 1, I-90134 Palermo, Italy}

\author[0000-0002-8032-8174]{Ryosuke Hirai}
\affiliation{Astrophysical Big Bang Laboratory (ABBL), RIKEN Cluster for Pioneering Research, 2-1 Hirosawa, Wako, Saitama 351-0198, Japan}
\affiliation{School of Physics and Astronomy Monash University, Clayton, VIC 3800, Australia}
\affiliation{OzGrav: The Australian Research Council Centre of Excellence for Gravitational Wave Discovery, Clayton, VIC 3800, Australia}

\author[0000-0002-6916-3559]{Tomoki Matsuoka}
\affiliation{Institute of Astronomy and Astrophysics, Academia Sinica, No.1, Sec.4, Roosevelt Road, Taipei 106216, Taiwan}

\author[0000-0003-0876-8391]{Marco Miceli}
\affiliation{Dip. di Fisica e Chimica, Universit\`a degli Studi di Palermo, Piazza del Parlamento 1, 90134 Palermo, Italy}
\affiliation{INAF -- Osservatorio Astronomico di Palermo, Piazza del Parlamento 1, I-90134 Palermo, Italy}

\author[0000-0002-7025-284X]{Shigheiro Nagataki}
\affiliation{Astrophysical Big Bang Laboratory (ABBL), RIKEN Cluster for Pioneering Research, 2-1 Hirosawa, Wako, Saitama 351-0198, Japan}
\affiliation{RIKEN Interdisciplinary Theoretical \& Mathematical Science Program (iTHEMS), 2-1 Hirosawa, Wako, Saitama 351-0198, Japan}
\affiliation{Astrophysical Big Bang Group (ABBG), Okinawa Institute of Science and Technology (OIST), 1919-1 Tancha, Onna-son, Kunigami-gun, Okinawa 904-0495, Japan.}

\author[0000-0002-0603-918X]{Masaomi Ono}
\affiliation{Institute of Astronomy and Astrophysics, Academia Sinica, No.1, Sec.4, Roosevelt Road, Taipei 106216, Taiwan}
\affiliation{Astrophysical Big Bang Laboratory (ABBL), RIKEN Cluster for Pioneering Research, 2-1 Hirosawa, Wako, Saitama 351-0198, Japan}
\affiliation{RIKEN Interdisciplinary Theoretical \& Mathematical Science Program (iTHEMS), 2-1 Hirosawa, Wako, Saitama 351-0198, Japan}

\author[0000-0002-4848-5508]{Ke-Jung Chen}
\affiliation{Institute of Astronomy and Astrophysics, Academia Sinica, No.1, Sec.4, Roosevelt Road, Taipei 106216, Taiwan}

\author[0000-0002-0763-3885]{Dan Milisavljevic}
\affiliation{Department of Physics and Astronomy, Purdue University, 525 Northwestern Avenue, West Lafayette, IN 47907, USA}

\author[0000-0002-7507-8115]{Daniel Patnaude}
\affiliation{Smithsonian Astrophysical Observatory, 60 Garden Street, Cambridge, MA 02138, USA}

\author[0000-0002-2321-5616]{Fabrizio Bocchino}
\affiliation{INAF -- Osservatorio Astronomico di Palermo, Piazza del Parlamento 1, I-90134 Palermo, Italy}


\author[0000-0002-1381-9125]{Nancy Elias-Rosa}
\affiliation{INAF -- Osservatorio Astronomico di Padova, Vicolo dell’Osservatorio 5, 35122 Padova, Italy}
\affiliation{Institute of Space Sciences (ICE, CSIC), Campus UAB, Carrer de Can Magrans s/n, 08193 Barcelona, Spain}



\begin{abstract}

We investigate \sn14c using three-dimensional hydrodynamic modeling, focusing on its early interaction with dense circumstellar medium (CSM). Our objective is to uncover the pre-supernova (SN) CSM structure and constrain the progenitor star's mass-loss history prior to core collapse. Our comprehensive model traces the evolution from the progenitor star through the SN event and into the SN remnant (SNR) phase. We simulate the remnant's expansion over approximately 15 years, incorporating a CSM derived from the progenitor star's outflows through dedicated hydrodynamic simulations. Analysis reveals that the remnant interacted with a dense toroidal nebula extending from $4.3\times 10^{16}$~cm to $1.5\times 10^{17}$~cm in the equatorial plane, with a thickness of approximately $1.2\times 10^{17}$~cm. The nebula's density peaks at approximately $3\times 10^6$~cm$^{-3}$ at the inner boundary, gradually decreasing as $\approx r^{-2}$ at greater distances. This nebula formed due to intense mass-loss from the progenitor star between approximately 5000 and 1000 years before collapse. During this period, the maximum mass-loss rate reached about $8\times 10^{-4}$ \mlrate, ejecting $\approx 2.5\,M_{\odot}$ of stellar material into the CSM. Our model accurately reproduces Chandra and NuSTAR spectra, including the Fe K line, throughout the remnant's evolution. Notably, the Fe line is self-consistently reproduced, originating from shocked ejecta, with $\approx 0.05\,M_{\odot}$ of pure-Fe ejecta shocked during the remnant-nebula interaction. These findings suggest that the 3D geometry and density distribution of the CSM, as well as the progenitor star's mass-loss history, align with a scenario where the star was stripped through binary interaction, specifically common envelope evolution.

\end{abstract}



\section{Introduction} 
\label{sec:intro}


Interacting supernovae (SNe) are a fascinating subclass of stellar explosions, where clear signs of interaction between the ejected stellar material and a dense, inhomogeneous circumstellar medium (CSM) are evident during the early stages of blast wave expansion, typically within the first year (e.g., \citealt{2017hsn..book..403S} and references therein). These transients suggest that the progenitor star's evolution was marked by significant instability as it approached the end of its life cycle. This instability likely manifested as violent and sporadic mass-loss events occurring over decades, centuries, or even millennia prior to the core collapse.

When the interaction between the blast wave and the CSM occurs shortly after the SN event, within a few days, it indicates substantial mass ejection from the progenitor in the years to centuries preceding the explosion. These phenomena are classified as interacting SNe, specifically types IIn and Ibn (see \citealt{2017hsn..book..403S} for a recent review). In some cases, observations have revealed precursor outbursts from the progenitor star occurring months to years before the SN explosion (e.g., \citealt{2007Natur.447..829P, 2013ApJ...767....1P, 2013ApJ...779L...8F, 2023ApJ...955L...8H, 2024ApJ...964..181H, 2024MNRAS.527.5366N, 2024arXiv240202924E}). Conversely, when the interaction between the blast wave and the CSM is delayed by hundreds of days after the SN, it suggests that the dense CSM may have resulted from massive eruptions that occurred millennia before the SN event.

Observations have also shown that some of these SNe originated from progenitor stars that had lost a significant portion of their hydrogen (H) envelope. Notable examples include SNe\,2014C (\citealt{2015ApJ...815..120M, 2017ApJ...835..140M}), 2017ens (\citealt{2018ApJ...867L..31C}), 2017dio (\citealt{2018ApJ...854L..14K}), 2019oys (\citealt{2020A&A...643A..79S}), and 2019yvr (\citealt{2021MNRAS.504.2073K}). The study of this class of SNe provides, therefore, a unique opportunity to investigate the mechanisms behind envelope stripping by probing the structure of the CSM within approximately 0.1~pc of the progenitor star. This can yield valuable insights into the mass-loss history, including the processes driving eruptive mass-loss episodes, and the final stages of the progenitor star's evolution in the decades to millennia preceding core collapse.

{\sn14c} is a particularly interesting example of this class. This event occurred in the galaxy NGC~7331, at a distance of $14.7\pm 0.6$~Mpc (\citealt{2001ApJ...553...47F}). Its evolution was closely monitored across various wavelengths, from radio and infrared to X-rays, for several years (\citealt{2019ApJ...887...75T, 2021MNRAS.502.1694B, twd22, bmm22}). Initially classified as a typical H-stripped core-collapse SN of type Ib based on its spectroscopic features, {\sn14c} exhibited unusual characteristics approximately 200 days after the explosion, including a rapid increase in radio and X-ray emission. This led to its reclassification as an H-rich SN of type IIn (\citealt{2015ApJ...815..120M, 2017ApJ...835..140M, 2021MNRAS.502.1694B}). Observations revealed that the SN initially exploded within a cavity before encountering dense circumstellar material, indicating a complex mass-loss history for the progenitor star (\citealt{2017MNRAS.466.3648A, twd22, bmm22}).

This scenario is further supported by one-dimensional (1D) simulations that describe the SN evolution and its subsequent interaction with a massive circumstellar shell (\citealt{2022ApJ...930..150V}). Additionally, X-ray, optical, and radio VLBI observations suggest that the CSM around {\sn14c} is highly asymmetric, resembling an equatorial disk or torus (\citealt{2021MNRAS.502.1694B, twd22, bmm22}). Other SNe of this class have also shown evidence of interaction with an equatorial disk-like CSM (e.g., \citealt{2015MNRAS.449.1876S}), which is often considered indicative of binary interaction in the progenitor system (e.g., \citealt{2017hsn..book..403S}).

A binary origin for the progenitor of \sn14c was also proposed by \cite{2020MNRAS.497.5118S}, who estimated the cluster age to be around 20 million years based on Hubble Space Telescope observations of \sn14c\ host star cluster. Their analysis, using binary stellar evolution models, suggests that the progenitor was an $\approx 11\,M_{\odot}$ star in a relatively wide binary system that had lost its H-rich envelope centuries to millennia before the collapse (see also \citealt{2015ApJ...815..120M}), likely due to mass transfer via binary interaction. The unique properties of {\sn14c} have generated significant interest and ongoing research, offering valuable insights into the processes responsible for intense mass loss from progenitor stars as they approach core collapse.

In this study, we present the complete three-dimensional (3D) evolution of {\sn14c}, describing the transition from the SN event to the interaction of the SN shock and ejecta with the inhomogeneous CSM resulting from the progenitor star's mass-loss history. To achieve this, we combined stellar evolution and core-collapse SN models with state-of-the-art 3D hydrodynamic simulations of its remnant's evolution, following the methodology outlined in previous studies (e.g., \citealt{2020A&A...636A..22O, 2021A&A...645A..66O, 2022A&A...666A...2O}). Our modeling includes both 2D and 3D hydrodynamic simulations, capturing the CSM's evolution over millennia leading up to the SN event, thereby establishing the ambient medium through which the SN blast propagates. By comparing our model results with observations of \sn14c, we are able to link the observed properties of the SN remnant (SNR) with the 3D structure of the CSM, providing insights into the mass-loss history of the progenitor star.

The paper is organized as follows. In Sect.~\ref{sec:model}, we present the model setup to describe the evolution from the progenitor star, to the SN and the SNR, along with the modeling of the CSM; in Sect.~\ref{sec:mod_par} we constrain the large-scale structure of the CSM through a parametrized model; in Sect.~\ref{sec:phys_wind} we present the results of the 3D evolution of the SN blast and ejecta through the CSM resulting from the mass-loss history of the progenitor star; in Sect.~\ref{sec:model_obs} we revisit the analysis of the X-ray data of {\sn14c} in light of the model results; in Sect.~\ref{sec:summary} we summarize the main results and draw our conclusions. In the Appendix, we discuss the convergence test for the SNR model (\ref{app:convergence}), we provide the details of the analysis of X-ray data (\ref{app:obs}), and we compare the model results with all the X-ray spectra for which almost simultaneous Chandra and NuSTAR observations are available (\ref{app:spectra}).


\section{Approach and numerical setup}
\label{sec:model}

Our approach involves 3D hydrodynamic modeling and synthesizing observables for comparison with actual observations of \sn14c. The modeling integrates cutting-edge models of stellar evolution, core-collapse SN, and SNR (e.g., \citealt{2023arXiv231105612O} and references therein) to follow coherently the evolution from the progenitor massive star to the SN and the full-fledged SNR. Below, we provide details of the numerical setups used to describe the pre-SN evolution of the progenitor star (Sect.~\ref{sec:progenitor}) and of its CSM (Sect.~\ref{sec:modcsm}), the evolution of the core-collapse SN (Sect.~\ref{sec:sn}), and the expansion of the SN blast wave through the inhomogeneous CSM (Sect.~\ref{sec:snr}). Finally, we describe the process of synthesizing X-ray emissions from the models (Sect.~\ref{sec:synth}), enabling a direct comparison of model results with actual observations of \sn14c.

\subsection{Pre-SN evolution of the progenitor star}
\label{sec:progenitor}

Given the initial Ib classification of the SN and a relatively low inferred ejecta mass ($\sim1.7~M_\odot$; \citealt{2017ApJ...835..140M}), the progenitor of \sn14c was most likely a helium star whose H-rich envelope was stripped through a binary interaction. Assuming a neutron star baryonic mass of $\sim 1.6~M_\odot$, the progenitor mass at collapse should have been around $\sim 3$--$3.5~M_\odot$, considering the ejecta mass estimated from observations (\citealt{2017ApJ...835..140M}). A star that has a helium core in this mass range should have a zero-age main-sequence mass of roughly $10$--$12~M_\odot$ (see Fig. 7 in \citealt{2018ApJ...860...93S}), which is consistent with the host star cluster age of about $20$~Myr for {\sn14c} (\citealt{2020MNRAS.497.5118S}). Furthermore, the presence of close-by H-rich CSM indicates that the last bit of the H-rich envelope was ejected very recently (in the last $\lesssim1000$~yr). This implies that the envelope stripping occurred late in its evolution, where the envelope thermal timescale and the remaining lifetime of the star are comparable so that the mass transfer phase lasts until the core collapse. 

Based on the above points, we assert that the progenitor was a $\sim 11~M_\odot$ star that engaged in Case-C mass transfer (i.e., the mass-donor star has already completed core helium burning by the time the mass transfer occurs). This is consistent with the inference made by \citet{2020MNRAS.497.5118S} based on host-cluster age analysis. The companion could have been a relatively high-mass star moderating stable mass transfer, or a relatively low-mass star triggering unstable mass transfer leading to a common-envelope phase. Stable mass transfer has a higher chance of continuing its mass-loss until core collapse, but recent studies argue that the common-envelope phases for massive stars should also last for about a thermal timescale \citep{2022ApJ...937L..42H}, enabling the creation of close-by CSM.

We modeled the evolution of the progenitor star from the main-sequence phase up to the pre-SN stage using the Modules for Experiments in Stellar Astrophysics ({\mesa}; e.g., \citealt{2011ApJS..192....3P}). We initiated the modeling with an $11\,M_\odot$ star and tracked its evolution until it reached a radius of $700\,R_\odot$, corresponding to its core carbon-burning stage. At this stage, the helium core mass was $\sim3.3~M_\odot$. Subsequently, we mimicked Case-C mass transfer by applying a high mass-loss rate until the star depleted its H-rich envelope. The {\mesa} inlists used for creating the progenitor model are provided in a Zenodo repository\footnote{\dataset[DOI: 10.5281/zenodo.13933512]{https://doi.org/10.5281/zenodo.13933512}.}.

We note that spectra from the early photospheric epochs revealed absorption features indicative of high-velocity H (\citealt{2015ApJ...815..120M}), suggesting the presence of a thin H-rich layer at the time of the explosion. However, this thin layer does not significantly influence the subsequent modeling of the SN and SNR. Additionally, while understanding the mass-loss history is crucial for discerning the CSM structure, the specific mass-loss profile we applied in this study does not affect the progenitor's structure significantly, as there is not enough time for the helium core mass to grow appreciably after core He depletion. The star was then allowed to continue to evolve until the core collapsed.

\subsection{Modeling the pre-SN inhomogeneous CSM}
\label{sec:modcsm}

The strategy employed to determine the structure and density distribution of the inhomogeneous CSM comprises two key steps. Initially, following the approach used in earlier studies (e.g., \citealt{2015ApJ...810..168O, 2020A&A...636A..22O}), we established an analytical description of the CSM (referred to as the ``TOY model'') defined by a set of parameters. Systematically exploring the parameter space, we aimed to identify a configuration capable of generating SN-SNR simulations that accurately reproduce the lightcurve and temperatures observed in X-ray observations of \sn14c. This initial phase facilitated the derivation of a preliminary description of the CSM density structure, serving as a guide for subsequent CSM modeling.

In the next step, we conducted 2D and 3D hydrodynamic simulations to model the outflow emanating from the progenitor, tracing the evolution of the CSM over the millennia preceding the core collapse. We investigated various mass-loss histories of the progenitor and diverse outflow asymmetries, seeking a combination that could faithfully replicate the CSM structure derived in the initial step (the TOY model). The subsequent simulations, describing the expansion of the SN blast through this CSM, allowed us to refine our understanding of the mass-loss history of the progenitor and the structure of the CSM. This refinement was achieved by comparing the model results with observations of \sn14c. The following sections provide a comprehensive overview of the CSM models employed in the current study.

\subsubsection{The parametrized (TOY) model}
\label{sec:toy}

In our simplified model, we assumed that the region near the progenitor star is influenced by the fast and diffuse outflow originating from the helium star, namely the star resulted after the removal of its H-rich envelope (e.g., \citealt{2020MNRAS.497.5118S}). Beyond this immediate vicinity, the CSM reflects the influence of denser and slower outflows from earlier evolutionary stages of the progenitor. This considered that the CSM likely exhibits asymmetry shaped by binary interactions.

The outflow emanating from the helium star is assumed to be spherically symmetric and extending up to a distance $R_0$. Its gas density distribution is given by,

\begin{equation}
\label{eq:blondin}
    \rho(r) = \frac{\mdot_{\rm hs}}{4\pi r^2 u_{\rm hs}}, ~~~~~~~~~~~~ (\mbox{at } r < R_0)
\end{equation}

\medskip
\noindent
where $r$ denotes the radial distance from the progenitor, {$\mdot_{\rm hs}$} is the mass-loss rate, and $u_{\rm hs}$ the outflow velocity of the helium star during the years immediately preceding the core collapse. We fixed the values of {$\mdot_{\rm hs}$} and $u_{\rm hs}$ to those inferred from the analysis of observations, namely $\dot{M}_{\rm hs} = 5\times 10^{-6}$~\mlrate, and $u_{\rm hs} = 1000$~km~s$^{-1}$, respectively (e.g., \citealt{2022ApJ...930..150V}).

At larger distances ($r > R_0$), the CSM is characterized by a background asymmetric outflow from the red supergiant (RSG) phase of the progenitor and by a dense circumstellar nebula. This dense nebula may be the result of a significant eruption from the progenitor star, likely the event responsible for stripping off the H-rich envelope. The description of the asymmetric outflow from the RSG phase is based on the model introduced by \cite{1996ApJ...472..257B}, as outlined below:

\begin{eqnarray}
    \rho(r, \theta) = 
    C \left(\frac{\mdot_{\rm rsg} }{4\pi r^2 u_{\rm rsg}}\right) \times ~~~~~~~~~~~~~~~~~~~~~~~~~~ \nonumber
    & \\ \displaystyle
    \left[1-A\frac{\exp(-2 B \cos^2 \theta)-1}{\exp(-2 B)-1} \right] + \rho_{\rm bkg}, \nonumber & \\ \displaystyle 
    (\mbox{at } r \geq R_0)~~~~~
    \label{wind:blondin}
\end{eqnarray}

\medskip
\noindent
where $\theta$ is the polar angle, and the parameters $A$ and $B$ govern the outflow asymmetry with respect to $\theta$; the parameter $A$ determines the ratio of density at the equator to density at the pole, expressed as $R_{\rm A} = 1/(1-A)$, while $B$ influences the steepness; $\rho_{\rm bkg}$ is a background density, defining the minimum density allowed in the system. Finally, $C$ is a normalization constant to ensure the same total mass-loss rate independent of the asymmetry. 

In our simulations, we fixed the mass-loss rate and the velocity of the RSG phase, assigning values of {$\mdot_{\rm rsg} = 10^{-4}$~\mlrate} and $u_{\rm rsg} = 10$~km~s$^{-1}$, respectively\footnote{We found that values of $\mdot_{\rm rsg}$ below $10^{-4}$~\mlrate do not change significantly the results.}. The outflow's asymmetry is characterized by $A=0.99$ and $B=1.0$; the background density in Eq.~\ref{wind:blondin} is assumed to be $\rho_{\rm bkg} = 2.17\times 10^{-25}$~g~cm$^{-3}$ (corresponding to a particle number density of $0.1$~cm$^{-3}$). It is worth noting that, due to the presence of the nebula, the SN blast propagates through the outflow from the RSG phase only in polar directions, where the outflow density is relatively low (thus not contributing significantly to X-ray emission). Consequently, the specific choice of parameters defining the density distribution of this outflow holds less significance, as long as the density in polar directions remains low enough to ensure there the rapid expansion of the blast wave, consistent with observations.

The dense circumstellar nebula is superimposed to the RSG outflow for $r \geq R_0$ and is assumed to be toroidal (axisymmetric). The height of the torus surface from the equatorial plane is defined as 

\begin{equation}
    h(r) = \sqrt{\frac{r-R_0}{L_{\rm H}}} \times L_{\rm pc}
    \label{eq:torus}
\end{equation}

\medskip
\noindent
where $L_{\rm H}$ is a scale height defining the thickness of the nebula and $L_{\rm pc} = 1$~pc. The nebula is thus defined in the volume bounded by $-h(r) < z < h(r)$. Within this region, we assume the nebula consists of three distinct zones: (a) an inner one representing the interface between the outflow from the helium star phase and those from earlier phases; (b) a middle zone corresponding to the nebula formed by a massive eruption; and (c) an outer zone that bridges the phase of the massive eruption with the preceding RSG phase. The density structure of the torus is specifically defined as:

\begin{equation}
    \rho(r) = \left\{
    {\begin{array}{cc}
        \displaystyle
        ~~~\rho_{\rm n}\left(\frac{R_0}{r}\right)^{a0} & ~~~~~~R_0 < r < R_1 \\
        \displaystyle
        C_1 \rho_{\rm n}\left(\frac{R_0}{r}\right)^{a1} & ~~~~~~R_1 < r < R_2 \\         
        \displaystyle
        C_2 \rho_{\rm n}\left(\frac{R_0}{r}\right)^{a2} & ~~~~~~r > R_2 \\
    \end{array}}
    \right.
    \label{eq:toywind}
\end{equation}

\begin{figure*}[!ht]
  \centering
  \includegraphics[width=18cm]{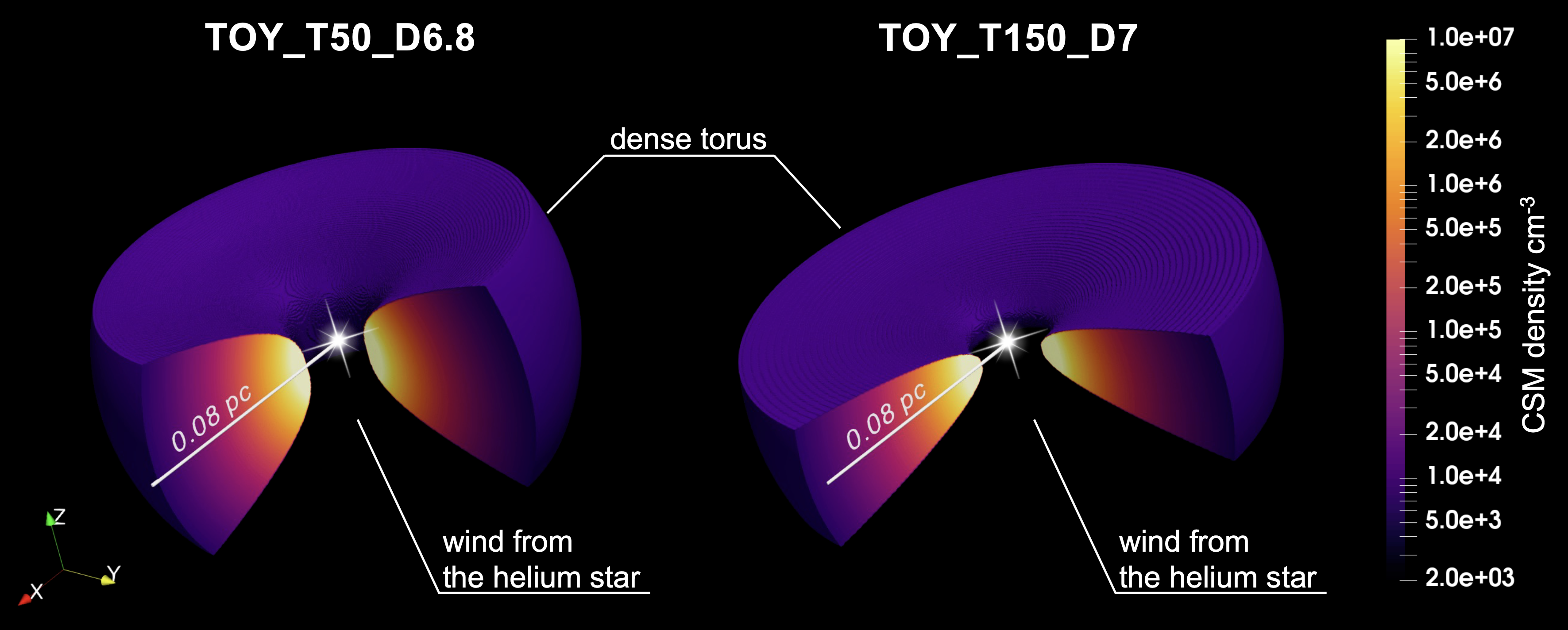}
  \caption{Parametrized models describing the possible density structure and geometry of the pre-SN nebula surrounding the progenitor star of \sn14c. The isosurfaces delineate material with a particle number density $n > 10^{4}$~cm$^{-3}$; color indicates the density value (color legend provided on the right). The star at the center indicates the position of the SN explosion. Both models presented successfully replicate the X-ray lightcurve of \sn14c. The model on the left (TOY\_T50\_D6.8) is tailored to match the temperature evolution inferred by \cite{bmm22}, while the one on the right (TOY\_T150\_D7) is needed to reproduce the temperature evolution observed by \cite{twd22}.}
  \label{csm_toy_model}
\end{figure*}

\medskip
\noindent
where $\rho_{\rm n}$ is the density of the nebula at $r = R_0$, $R_1$ and $R_2$ define the borders between the three zones in the nebula, and the parameters $a0$, $a1$, and $a2$ define the density gradient of the nebula in the radial direction; $C_1$ and $C_2$ are normalization constants defining the density jumps at $R_1$ and $R_2$. Figure~\ref{csm_toy_model} illustrates two examples of geometry and structure of the toroidal circumstellar nebula as described by our TOY model (see Sect.~\ref{sec:mod_par} for more details).

\subsubsection{Hydrodynamic modeling of the stellar outflow}
\label{sec:windmodel}

The TOY model outlined in the previous section enables us to derive the average structure of the pre-SN CSM, essential for explaining the variations in X-ray flux and temperature derived from the analysis of observations. However, this simplified model lacks details regarding the mass-loss history of the progenitor star and the final stages of its evolution before collapse. Addressing these issues necessitates the development of hydrodynamic models that track the CSM evolution in the millennia leading up to collapse. Such models aim to describe the CSM structure, reflecting the progenitor's mass-loss history.

In light of this, we derived a self-consistent description of the pre-SN CSM through dedicated 2D and 3D hydrodynamic simulations\footnote{The 2D simulations of the stellar outflow were used to explore the parameter space before running the 3D simulations for selected cases. Note that the 2D simulations were mapped onto the 3D domain (considering axisymmetry) before running the SNR simulations.}, using the results from the TOY model as a guide for explorig the parameter space. In this way, the structure of the CSM at the time of collapse reflects the mass-loss history and the asymmetries of the outflows emanated from the progenitor star in the final phases of its evolution. 

The stellar outflow is simulated by numerically solving the time-dependent fluid equations governing mass, momentum, and energy conservation in a spherical coordinate system $(r, \theta, \phi)$, including the effects of radiative losses from an optically thin plasma:

\begin{equation}
\begin{array}{l}\displaystyle
\frac{\partial \rho}{\partial t} + \nabla \cdot \rho \mbox{\bf u} = 0~,
\\ \\ \displaystyle
\frac{\partial \rho \mbox{\bf u}}{\partial t} + \nabla \cdot \rho
\mbox{\bf uu} + \nabla P = 0~,
\\ \\ \displaystyle
\frac{\partial \rho E}{\partial t} +\nabla\cdot (\rho E+P)\mbox{\bf u}
= -n_{\rm e} n_{\rm H} \Lambda(T)~,
\end{array}
\label{mod_eq}
\end{equation}

\medskip
\noindent
where $E = \epsilon + (1/2) |\mbox{\bf u}|^2$ is the total energy (internal energy, $\epsilon$, plus kinetic energy), $t$ is the time, $\rho = \mu m_{\rm H} n_{\rm H}$ is the mass density, $\mu = 1.3$ is the mean atomic mass (assuming abundances consistent with \citealt{2015ApJ...815..120M}), $m_{\rm H}$ is the mass of the H atom, $n_{\rm H}$ is the H number density, $n_{\rm e}$ is the electron number density, {\bf u} is the gas velocity, $T$ is the temperature, and $\Lambda(T)$ represents the radiative losses per unit emission measure (e.g. \citealt{rs77}; \citealt{mgv85}; \citealt{2000adnx.conf..161K}). We used the ideal gas law, $P=(\gamma-1) \rho \epsilon$, where $\gamma=5/3$ is the adiabatic index.

The simulations were conducted using \pluto\ v4.4-patch2, a parallel modular Godunov-type code designed primarily for astrophysical applications and high Mach number flows in multiple spatial dimensions (\citealt{2012ApJS..198....7M}). The code was configured to compute intercell fluxes using a two-shock Riemann solver, specifically the linearized Roe Riemann solver based on the characteristic decomposition of the Roe matrix. Fourth-order accuracy in time was achieved through a Runge–Kutta scheme (RK4) for advancing the solution to the next time level. The combination of the Roe solver with the RK4 algorithm in a spatially unsplit fashion results in the corner-transport upwind method (\citealt{1990JCoPh..87..171C, 2005ApJS..160..199M, 2005JCoPh.205..509G, chaplin15}), one of the most sophisticated and least diffusive algorithms available in \pluto. To prevent spurious oscillations in the presence of strong shocks, a monotonized central difference flux limiter (MC\_LIM, the least diffusive limiter in \pluto) is employed for the primitive variables.

The progenitor star is located at the center of the 3D spherical coordinate system $(r, \theta, \phi)$, and the initial density distribution of the stellar outflow is described by Eq.\ref{wind:blondin}, assuming $\mdot_{\rm rsg} = 10^{-4}$~{\mlrate}. Various initial outflow asymmetries were investigated by conducting simulations with different values for the parameters $A$ and $B$.

The computational domain spans from $R_{\rm min} = 0.002$~pc ($\sim 6\times 10^{15}$~cm) to $R_{\rm max} = 1.4$~pc ($\sim 4\times 10^{18}$~cm) in the radial direction. It covers an angular sector from $\theta_{\rm min} = 0^{\circ}$ to $\theta_{\rm max} = 90^{\circ}$ in the angular coordinate $\theta$ and, in 3D simulations, from $\phi_{\rm min} = 0^{\circ}$ to $\phi_{\rm max} = 90^{\circ}$ in the angular coordinate $\phi$. The radial coordinate $r$ is discretized on a logarithmic grid, with mesh size increasing as $r$ extends from the star, providing higher spatial resolution closer to the star. This is crucial for ensuring accuracy in describing the density structure of the CSM near the star and, consequently, for accurately representing the interaction of the SN blast during its expansion through the CSM. The radial grid consists of $N_{\rm r} = 1024$ points, with a maximum resolution of $\delta r \approx 4 \times 10^{13}$~cm near the star and a minimum resolution of $\delta r \approx 2.7 \times 10^{16}$~cm near the outer boundary. The angular coordinates $\theta$ and $\phi$ are uniformly discretized with $N_{\theta} = N_{\phi} = 256$ points, providing a resolution of $\delta \theta = \delta \phi = 0.35^{\circ}$.

Inflow and zero-gradient (outflow) boundary conditions are applied at the inner ($R_{\rm min}$) and outer ($R_{\rm max}$) boundaries, respectively, of the radial coordinate $r$. Axisymmetric and equatorial symmetric (reflective) boundary conditions are imposed at the inner ($\theta_{\rm min}$) and outer ($\theta_{\rm max}$) boundaries, respectively, of the angular coordinate $\theta$. Finally, periodic boundary conditions are assumed for the angular coordinate $\phi$ between $\phi_{\rm min}$ and $\phi_{\rm max}$.

The inflow at $R_{\rm min}$ is time-dependent and characterizes the outflow (mass density, pressure, and velocity) emitted from the progenitor star during various phases of its final evolution in the millennia leading up to the SN explosion. Two scenarios were considered. In the first case, the incoming flow exhibits the same dependence on $\theta$ as the outflow of a RSG progenitor (Eq.~\ref{wind:blondin}), but with time-dependent $\mdot(t)$ (in the following RSG models):

\begin{eqnarray}
    \rho(R_{\rm min},\theta,t) = C \left(\frac{\mdot(t) }{4\pi R_{\rm min}^2 u_{\rm rsg}}\right)\times \hspace{2cm}
    \nonumber \\  \displaystyle
    \left[1-A\frac{\exp(-2 B \cos^2 \theta)-1}{\exp(-2 B)-1} \right] + \rho_{\rm bkg}\,.
    \label{wind:rsg}
\end{eqnarray}

\medskip
\noindent
In the second case, the asymmetry is described by

\begin{equation}
    \rho(R_{\rm min},\theta,t) = C \left(\frac{\mdot(t) }{4\pi R_{\rm min}^2 u_{\rm mee}}\right)
    (1+ \alpha \cos^{\xi}\theta)^{-1}
    \label{wind:mee}
\end{equation}

\medskip\noindent
where $u_{\rm mee}$ is the velocity of the gas from the massive episodic eruption, $\alpha$ and $\xi$ are parameters defining the eruption asymmetry, and as before $\mdot(t)$ varies with time. Again, $C$ is a normalization constant to ensure the same total mass-loss rate independent of the asymmetry. This latter scenario aims to describe a massive eruption of the stellar envelope (in the following MEE models), primarily occurring along the equator. Such a structured, high-density circumstellar environment is expected, for instance, if a portion of the star's H-rich envelope is tidally ejected via interactions with a binary companion (e.g., \citealt{2012ApJ...744...52P, 2012ApJ...746...74R, 2015MNRAS.446.1716C, 2019MNRAS.484..631R, 2020A&A...644A..60S, 2023LRCA....9....2R}).

Regarding the mass-loss history, in both scenarios under investigation, we assumed a pattern in which $\mdot$ first increased a few millennia before the SN, representing the  ejection of the star's envelope, and subsequently diminished to align with the outflow emanating from the helium star in the centuries immediately preceding the SN event. During the envelope eruption, we allowed for two phases with different mass-loss rates. As discussed in Sect.~\ref{sec:phys_wind}, this change in mass-loss rate was necessary to better reproduce the observations. We fixed the initial mass-loss rate (after the RSG phase, before the eruption) as $\mdot_{\rm RSG} = 10^{-4}\,M_{\odot}$~yr$^{-1}$ and that during the final helium star phase, before the SN, as $\mdot_{\rm hs} = 5\times 10^{-6}\,M_{\odot}$~yr$^{-1}$. Thus, the parameters governing the exploration of mass-loss history that we considered are: the time $t_{\text{MEE1}}$ indicating the onset of the massive episodic eruption, the time $t_{\text{hs}}$ marking the conclusion of the eruption (and the start of the helium star phase), the time $t_{\text{MEE2}}$ delimiting the two phases with different $\mdot$ within the eruption, and the mass-loss rates $\mdot_{\text{MEE1}}$ and $\mdot_{\text{MEE2}}$ during the initial and subsequent phases of the eruption, respectively.

\subsection{Early evolution of the core-collapse SN}
\label{sec:sn}

The explosion of the mantle of the model  progenitor star was simulated using the SuperNova Explosion Code (\snec; \citealt{2015ApJ...814...63M}), employing the thermal bomb method. {\snec} adeptly tackles spherically symmetric radiation hydrodynamics within the Lagrangian coordinates, integrating the flux-limited diffusion method to accurately represent the intricate interplay between radiation and hydrodynamics.

The data generated from the progenitor model at core-collapse, as detailed in Sect.~\ref{sec:progenitor}, served as the foundation for establishing the initial conditions in the subsequent SN simulation. From the initial radial profile, we selectively removed a mass budget of $1.625\,M_\odot$ (representing the newly formed compact object) from the inner core. The remaining progenitor material, constituting the SN ejecta with a mass of $1.7\,M_\odot$, was discretized into 5000 mesh points employing uniform gridding with respect to mass coordinates. We initiated the explosion process by injecting $2.2\times 10^{51}\,{\rm erg}$ of internal energy into the inner $0.1\,M_\odot$ of the ejecta region. Following the energy redistribution during the explosion, the kinetic energy of the ejecta converged to $1.8\times10^{51}\,{\rm erg}$. Both the ejecta mass and kinetic energy were calibrated to align with values inferred from observational data \citep{2017ApJ...835..140M}.

We tracked the temporal evolution of the explosion up to the moment of shock breakout, defined as the point where the optical depth measured outward from the shock falls below $c/V_{\rm sh}$, where $c$ represents the speed of light and $V_{\rm sh}$ denotes the velocity of the SN shock. Upon reaching the phase of shock breakout (occurring approximately 105~s post-core-collapse), we captured the ultimate snapshot of the hydrodynamic profile. This final snapshot serves as the initial condition for our subsequent modeling of the SN-CSM interaction.

The SN model additionally tracks the intricate outcomes of explosive nucleosynthesis occurring within the initial seconds of the explosion. This comprehensive nuclear reaction network encompasses 19 distinct species, including protons ($^{1}$H), $^{3}$He, $^{4}$He, $^{12}$C, $^{14}$N, $^{16}$O, $^{20}$Ne, $^{24}$Mg, $^{28}$Si, $^{32}$S, $^{36}$Ar, $^{40}$Ca, $^{44}$Ti, $^{48}$Cr, $^{56}$Cr, $^{52}$Fe, $^{54}$Fe, $^{56}$Fe, and $^{56}$Ni. We note that, in the SN simulations, no mixing was initially assumed for the different species. However, we performed tests to evaluate the effects of mixing on the distribution of $^{56}$Ni. Our results indicate that the mixing broadens the radial distribution of the $^{56}$Ni mass fraction compared to a model without mixing. The most significant difference between the two scenarios is observed in the peak time of the bolometric lightcurve, which varies by up to 15\%. In any case, the mass of $^{56}$Ni synthesized within the ejecta amounts to $0.15\,M_\odot$, in agreement with \citet{2017ApJ...835..140M}.

\begin{figure}[!t]
  \centering
  \includegraphics[width=9cm]{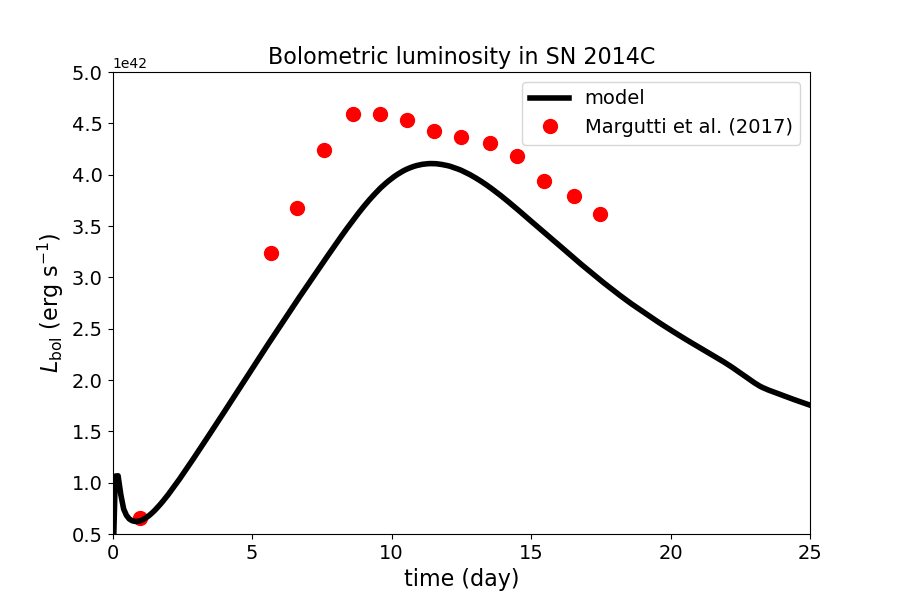}
  \caption{Bolometric lightcurve derived from our SN model compared with the actual one of \sn14c\ (\citealt{2017ApJ...835..140M}).}
  \label{bolometric_lc}
\end{figure}

One concern in this analysis is that the ejecta profile used in the simulations may not be entirely calibrated to match the observational characteristics during the early phase of \sn14c. To investigate further this point, we extended the 1D SN simulation to follow the evolution for about 30 days. Figure~\ref{bolometric_lc} compares the bolometric lightcurve derived from our SN model with the observed data. Notably, the figure highlights a discrepancy of a few tens of percent between our model and the observational data concerning the peak time and luminosity of the bolometric lightcurve, despite adopting the values for explosion energy and ejecta mass suggested by \citet{2017ApJ...835..140M}. A more precise alignment of the model with observational data would require incorporating the constraint of the time evolution of the photospheric radius, as determined from a Fe II line detection (\citealt{2015ApJ...815..120M}). By fitting three key observational features – peak time, peak bolometric luminosity, and photospheric radius – we could derive a more finely tuned solution for the primary parameters governing the calculation of the bolometric luminosity: the ejecta mass, kinetic energy, and the mass of $^{56}$Ni deposited. These parameters characterize the velocity of the forward shock and the amount of pure-Fe ejecta, both of which contribute to the X-ray emission properties such as bremsstrahlung and emission of the Fe K line. However, given the limited discrepancy between the model and the data, and considering that the X-ray emission in \sn14c\ is dominated by the shocked CSM, we expect that further refinement of the ejecta profile to match observational data would not substantially alter our conclusions, as discussed in Sect.~\ref{sec:compare_spec}. Furthermore, given our focus on a 1D SN model, we expect that neglecting prominent asymmetries emerging post-core-collapse (driven by stochastic processes like the standing accretion shock instability, convective overturning from neutrino heating, and Rayleigh-Taylor mixing) might have a more significant impact on the analysis of Fe K line (however, see discussion in Sect.~\ref{sec:compare_spec}).

\subsection{Expansion of the SNR through the CSM}
\label{sec:snr}

The modeling of the SNR expansion through the stellar outflow and its interaction with the nebula follows a strategy similar to that adopted in previous studies (e.g., \citealt{2020A&A...636A..22O, 2021A&A...645A..66O}). The output of the 1D SN simulation (Sect.~\ref{sec:sn}) at $\approx 105$~s after the core-collapse was mapped in 3D and used to initiate 3D hydrodynamic simulations which describe the long-term evolution ($\approx 15$~yr) of the blast wave and ejecta, from the shock breakout to the expansion of the remnant through the CSM. The CSM is described either by the TOY model (Sect.~\ref{sec:toy}) or by the output of hydrodynamic simulations of the CSM (RSG or MEE models; see Sect.~\ref{sec:windmodel}) mapped in the whole 3D domain.

We adopted the numerical setup of the SNR model
described in, e.g., \cite{2021A&A...645A..66O}. Specifically, the evolution of the blast wave and ejecta was simulated by solving the time-dependent hydrodynamic equations of mass, momentum, and energy conservation (Eqs.~\ref{mod_eq}) but now in a 3D Cartesian coordinate system $(x, y, z)$. Similar to the outflow model, we assumed an ideal gas law with $\gamma = 5/3$ (refer to Sect.~\ref{sec:windmodel}). Self-gravity effects were neglected during the remnant evolution, starting approximately 105~s after the core collapse, as the ejecta were in free expansion (see, e.g., \citealt{2021A&A...645A..66O}). Additionally, we omitted the effects of radiative losses from optically thin plasma, considering the remnant in its adiabatic expansion phase throughout the entire simulated evolution. This assumption remains valid since the duration of our simulations is shorter than the transition time from adiabatic to radiative phase for an expanding spherical blast (e.g., \citealt{1998ApJ...500..342B}; also see \citealt{2010ApJ...720L.195D}).

In addition, the SNR simulations include pertinent physical processes for describing the interaction between stellar debris and the CSM, crucial for synthesizing X-ray emission (see Sect.~\ref{sec:synth}). These include deviations from temperature equilibration between electrons and ions, involving almost instantaneous heating of electrons at shock fronts by lower hybrid waves and solving equations for Coulomb collisions \citep{2007ApJ...654L..69G}. Furthermore, the simulations consider deviations from the equilibrium ionization of the most abundant ions, accomplished through the computation of the maximum ionization age in each cell of the spatial domain. Detailed information on the implementation of these effects can be found in \cite{2015ApJ...810..168O, 2019A&A...622A..73O}.

Similar to the modeling of the stellar outflow, the simulations for the expanding SNR were conducted using \pluto\ v4.4-patch2, extended by additional computational modules designed to compute deviations from temperature equilibration between electrons and ions, as well as deviations from equilibrium ionization. The numerical setup adopts the same solvers and algorithms chosen for the stellar outflow model (Sect.~\ref{sec:windmodel}): the Roe solver with the RK4 algorithm in a spatially unsplit fashion, and the monotonized central difference flux limiter (MC\_LIM) for the primitive variables.

To monitor the evolution of distinct plasma components (ejecta and dense nebula), we introduced passive tracers, each associated with a specific component. Each tracer is initialized to one in cells belonging to the respective plasma component and to zero elsewhere. Alongside our set of hydrodynamic equations, the continuity equations of these tracers are also solved. Additional tracers were introduced to serve as repositories for information related to the shocked plasma, storing details such as time, shock velocity, and shock position (Lagrangian coordinates) when a cell of the mesh is shocked by either the forward or reverse shock. This information is essential for synthesizing emission in different wavelength bands (refer to Sect.~\ref{sec:synth}). These specific tracers are initialized to zero.

Furthermore, we monitored the chemical evolution of the ejecta using a multiple fluids approach as outlined by \cite{2016ApJ...822...22O}. In this approach, each fluid is linked to a specific species, modeled in the SN simulation (Sect.~\ref{sec:sn}), and it is initialized with the corresponding abundance obtained from the output of the SN simulations (at $\approx 105$~s after the core-collapse). This method enabled us to track the chemical composition of ejecta and to visualize the spatial distribution of heavy elements at various epochs throughout the evolution of \sn14c.

Our mesh strategy aligns with previous approaches (\citealt{2020A&A...636A..22O, 2021A&A...645A..66O}). It proves highly efficient in capturing the vast time and space scales inherent in the remnant's expansion. Initially, the computational domain is a Cartesian box ranging from $-3.2\times 10^{11}$~cm to $3.2\times 10^{11}$~cm in the $x$, $y$, and $z$ directions, encompassing the final domain of the SN simulation (see Sect.~\ref{sec:sn}). For the SNR evolution, a uniform grid with $512^3$ zones covers the box. The initial spatial resolution is, therefore, $\approx 1.2\times 10^{9}$~cm. Subsequently, we incrementally extended the computational domain by a factor of 1.2 as the forward shock propagates outward, remapping the physical values to the new domains. In the extended region, all physical quantities are set to the values of the pre-SN CSM, as derived in Sect.~\ref{sec:windmodel}. Approximately 90 remappings were required to monitor the interaction of the blast wave with the CSM over a 15-year evolution period. The final domain spans from $\approx -3.1\times 10^{18}$ to $3.1\times 10^{18}$~cm in the $x$, $y$, and $z$ directions, resulting in a spatial resolution of $\approx 1.1\times 10^{16}$~cm. Outflow boundary conditions were assumed at all boundaries.

We conducted supplementary simulations with grid resolutions of $256^3$ and $1024^3$ points to assess the impact of spatial resolution on our results. A high spatial resolution is required to describe appropriately the evolution of asymmetries characterizing the ejecta (e.g., \citealt{2020A&A...636A..22O, 2021A&A...645A..66O}). The findings indicated that the outcomes obtained with the highest resolution grid are consistent with those observed at the intermediate resolution as  that used in this paper (see Appendix~\ref{app:convergence}). This consistency is primarily because, in the present case, most of the X-ray emission originates from the shocked CSM, and our simulations do not account for the anisotropies that develop at smaller scales in the ejecta due to stochastic processes occurring soon after the core-collapse. Consequently, the majority of our simulations employed a grid with $512^3$ zones.

\subsection{Synthesis of X-ray emission}
\label{sec:synth}

We synthesized thermal X-ray emission from the SNR models and subsequently compared the results with observational data. This was realized by following the approach detailed, for instance, in \cite{2020A&A...636A..22O}. The key steps of this process are summarized below.

For each cell, $j$-th, within the spatial domain, the simulations provide the emission measure EM$_{\rm j} = n_{\rm e,j} n_{\rm H,j} V_{\rm j}$ (where $n_{\rm e,j}$ and $n_{\rm H,j}$ are the electron and H number densities in the $j$-th cell and $V_{\rm j}$ is the cell volume, assuming fully ionized plasma), the maximum ionization age $\tau_{\rm j} = n_{\rm e,j} \Delta t_{\rm j}$ (where $\Delta t_{\rm j}$ denotes the time since the plasma in the $j$-the cell of the domain was shocked), and the electron temperature $T_{\rm e,j}$ derived from ion temperature, plasma density, and $\Delta t_{\rm j}$, assuming Coulomb collisions (see \citealt{2015ApJ...810..168O} for the details of the implementation). The values of EM$_{\rm j}$, $\tau_{\rm j}$, and $T_{\rm e,j}$ within the $j$-th cell are then used to generate the X-ray emission in the $[0.1, 10]$~keV and $[0.1, 100]$~keV bands. This was done by using the non-equilibrium of ionization (NEI) emission model VPSHOCK from the XSPEC package (\citealt{1996ASPC..101...17A}), coupled with NEI atomic data sourced from ATOMDB (\citealt{2001ApJ...556L..91S}).

\begin{table*}
\caption{Explored and best-fit parameters of the CSM for the TOY models of \sn14c.}
\label{tab1}
\begin{center}
\begin{tabular}{lclccc}
\hline
\hline
CSM      & Parameters   & Units & Range    &  TOY\_T50\_D6.8  & TOY\_T150\_D7 \\
component&              &       & explored &    & \\
\hline
 HS outflow: & $\dot{M}_{\rm hs}$ & ($10^{-6} M_\odot$~yr$^{-1}$) &  - & $5$ & $5$   \\
          & $u_{\rm hs}$  & (km~s$^{-1}$) &  -          & 1000  & 1000  \\
\hline
RSG outflow:& $\dot{M}_{\rm rsg}$ & ($M_\odot$~yr$^{-1}$) &  - & $10^{-4}$ & $10^{-4}$   \\
         & $u_{\rm rsg}$  & (km~s$^{-1}$) &  - & 10  & 10  \\
         & $A$     &          &  -             & 0.99  & 0.99   \\
         & $B$ &          &  -             & 1.0   & 1.0    \\
         & $C$     &          &  $[1.0-5.0]$   & 1.8   & 1.8    \\
         & $\rho_{\rm bkg}$  & ($10^{-25}$~g cm$^{-3}$)  & - & $2.17$  &  $2.17$ \\
\hline
Nebula:  & $n_{\rm n} = \rho_{\rm n}/(\mu m_{\rm H})$ &  ($10^6$~cm$^{-3}$)  & $[1-10]$ & $6$ & $10$   \\
         & $R_0$  &  ($10^{16}$~cm)    &  $[1-5]$    & 4.3  & 4.3   \\
         & $R_1$  &  ($10^{16}$~cm)    &  $[4-6]$    & 4.8  & 4.8   \\
         & $R_2$  &  ($10^{16}$~cm)    &  $[10-20]$  & 14   & 14   \\
         & $L_{\rm H}$& (pc)    &  $[20-200]$    & 50   & 150    \\
         & $a_0$  &          &  -             & 10   & 10      \\
         & $a_1$  &          &  $[2-3]$       & 2    & 2       \\
         & $a_2$  &          &  $[5-10]$      & 5    & 5       \\
         & $C_1$  &          &  $[0.05-0.5]$  & 0.26 & 0.26       \\
         & $C_2$  &          &  $[1-30]$      & 7.8  & 7.8       \\
\hline
\end{tabular}
\end{center}
\end{table*}

The X-ray spectra were synthesized with the assumption that the source is located at a distance of $D = 14.7$~Mpc (namely the distance of the host galaxy NGC 7331; \citealt{2001ApJ...553...47F}). The synthesis also accounts for the effects of time-dependent photoelectric absorption caused by dense material along the line of sight, using absorption column density values derived by \cite{bmm22}. Additionally, it incorporates line broadening due to Doppler effects resulting from the bulk velocities of the approaching and receding parts of the emitting plasma (see \citealt{2009A&A...493.1049O, 2019NatAs...3..236M}). We omitted thermal broadening associated with the high temperatures of ions, as the actual spectra predominantly feature only one prominent line (Fe K$\alpha$ line), and we anticipate that thermal broadening is negligible compared to Doppler broadening (see \citealt{2019NatAs...3..236M}).

In the final step, we performed the integration of X-ray spectra derived from cells across the entire spatial domain. Subsequently, the resulting integrated spectrum was convolved with the instrumental response of either Chandra/ACIS or NuSTAR, yielding the corresponding focal-plane spectra. As the synthetic data are formatted to closely resemble actual X-ray observations, we analyzed these synthetic spectra using the standard data analysis system employed for actual spectra (XSPEC v.12.10; \citealt{1996ASPC..101...17A}).

\section{Constraints on the density structure and geometry of the CSM}
\label{sec:mod_par}

\begin{figure*}[!ht]
  \centering
  \includegraphics[width=16cm]{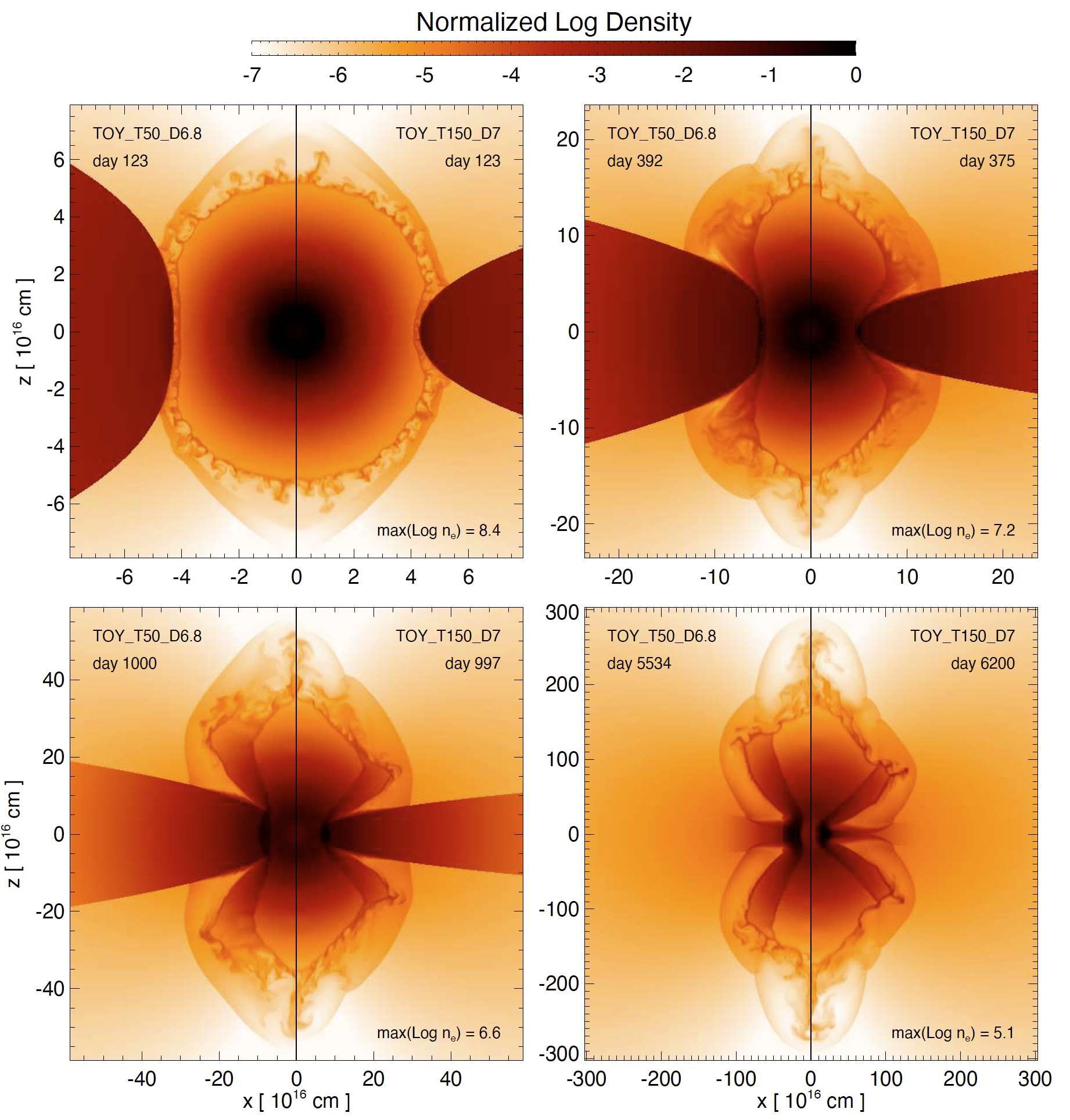}
  \caption{2D cross-sections in the $(x, z)$ plane show the spatial density distribution (in log scale) at the labeled times for models TOY\_T50\_D6.8 (left half panels) and TOY\_T150\_D7 (right half panels). Each image is normalized to its maximum, $\max (\log n_{\rm e})$, for enhanced visibility. The symmetry axis of the disk aligns with the $z$-axis.}
  \label{evol_snr_toy}
\end{figure*}

As discussed in Sect.~\ref{sec:modcsm}, we first adopted an approach similar with that discussed in \cite{2015ApJ...810..168O}. We used an analytical description of the CSM (see Sect.~\ref{sec:modcsm}) to constrain its geometry and density structure through simulations of the expanding SNR. This involved comparing the X-ray lightcurve and temperature evolution synthesized from our SNR model with those inferred from the analysis of X-ray observations \citep{twd22, bmm22}. 
The system evolution was followed from the SN phase to the SNR phase for $\approx 15$ years. Our exploration of the CSM parameter space involved key parameters such as the inner edge of the nebula ($R_0$) and its density at this edge ($\rho_{\rm n}$), along with the thickness of the nebula ($L_{\rm H}$) and the distances $R_1$ and $R_2$ defining three regions characterized by distinct density gradients (see Sect.~\ref{sec:modcsm} for the details); additionally, we explored different exponents in Eq.~\ref{eq:toywind}, which determine the density gradients in the three regions.

To streamline the exploration of the parameter space, especially given the computational costs of 3D simulations, we initially adopted fiducial values inferred from observations of {\sn14c} \citep{bmm22, twd22}: $R_0\approx 10^{16}$~cm, $\rho_{\rm n}\approx 10^5 - 10^6$~cm$^{-3}$, and a density gradient $\propto r^{-2.42\pm 0.17}$. Subsequently, we iteratively refined our approach through trial and error, adjusting model parameters to achieve a convergence that accurately reproduced the X-ray lightcurve and temperature evolution of \sn14c.

Due to a significant difference in temperature values reported by \cite{twd22} compared to those presented by \cite{bmm22} (although there is agreement in the derivation of X-ray fluxes), we formulated two distinct models: run TOY\_T50\_D6.8, 
designed to replicate the X-ray lightcurve and temperature evolution as inferred by \cite{bmm22}, and TOY\_T150\_D7, intended to match the results of \cite{twd22}. Table~\ref{tab1} summarizes the explored ranges of values and the model parameters for these two TOY models. Figure~\ref{csm_toy_model} illustrates the geometry and density structure of the dense nebula surrounding the explosion center at the time of core collapse for each model.

Table~\ref{tab1} and Figure~\ref{csm_toy_model} highlight the key distinction between the two TOY models: the thickness and density of the nebula. In fact, the observed X-ray flux, which depends on the square of the plasma density and the emitting volume, can be reproduced by varying combinations of nebula volumes and densities. This degeneracy is resolved by considering the average electron temperature of the X-ray emitting plasma, as higher densities correspond to lower temperatures. Consequently, the parameters for the two TOY models were fine-tuned to match the distinct electron temperature values reported by \cite{bmm22} and \cite{twd22}.

In all simulations performed, the evolution consistently follows a common general trend. Initially, the shockwave emanating from the SN propagates through the tenuous outflow of the progenitor helium star. Approximately 100 days after the explosion, the remnant encounters the dense nebula, triggering a transformative shift from an H-poor type Ib SN to an H-rich type IIn SN. The interaction of the remnant with the nebula leads to the generation of a forward shock advancing into the nebula, and a reverse shock rebounding into the ejecta (see Figure~\ref{evol_snr_toy}).

\begin{figure*}[!ht]
  \centering
  \includegraphics[width=15cm]{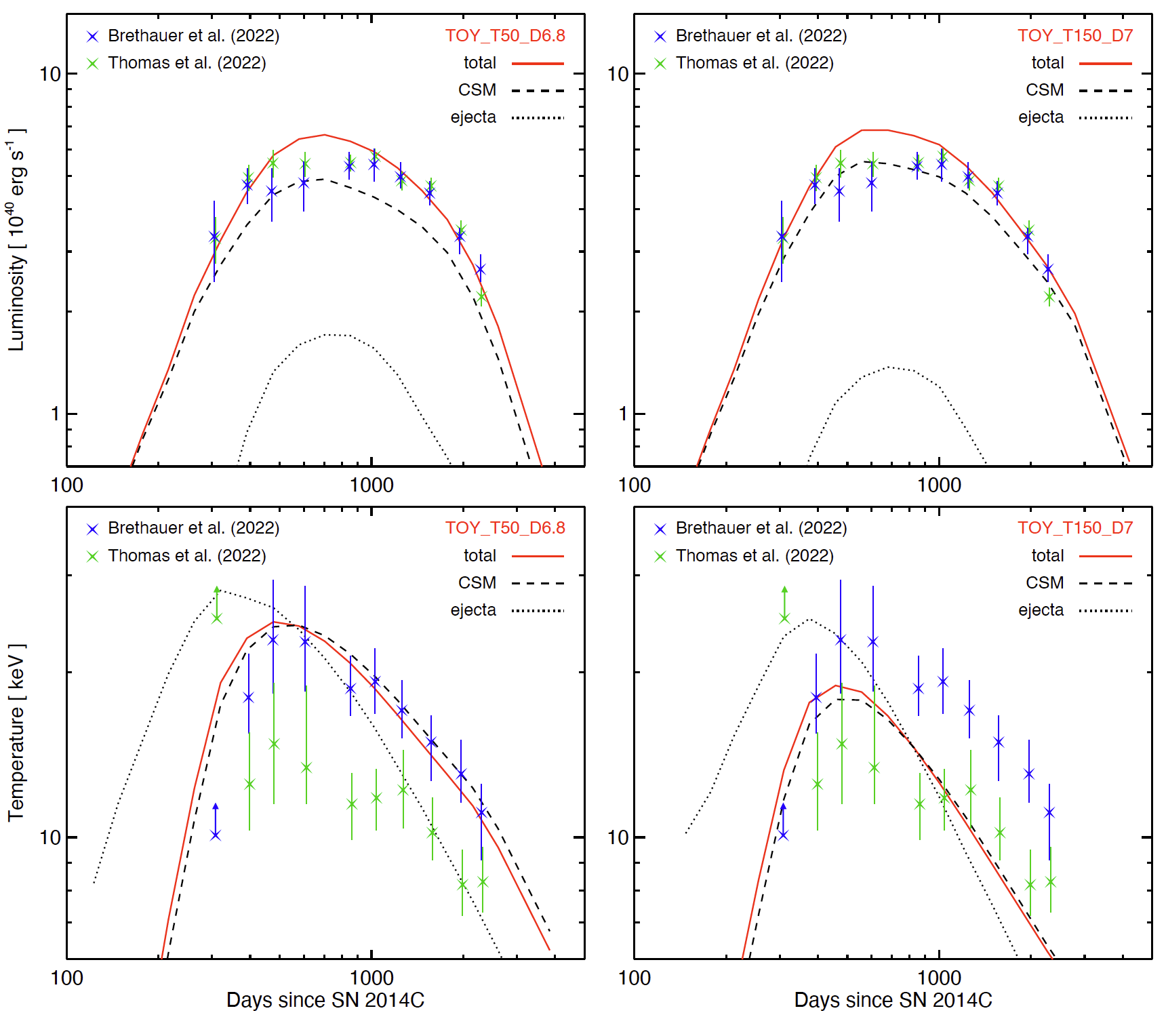}
  \caption{Upper panels: X-ray lightcurve in the $[0.5, 100]$~keV band (red line) synthesized from runs TOY\_T50\_D6.8 (on the left) and TOY\_T150\_D7 (on the right) compared with the observed lightcurve of \sn14c\ obtained with Chandra and NuSTAR (blue and green symbols; \citealt{bmm22} and \citealt{twd22}, respectively). Dashed and dotted lines mark the contributions to emission from the shocked plasma from the nebula and the shocked ejecta, respectively. Lower panels: Corresponding to the upper panels, the plots show the evolution of average X-ray emission-weighted electron temperature. Dashed and dotted lines indicate the average electron temperature of shocked plasma from the nebula and the shocked ejecta, respectively.}
  \label{lc_TOY}
\end{figure*}

During the initial phase of the interaction between the remnant and the nebula, the X-ray emission undergoes a rapid increase over time, primarily driven by the contribution of shocked plasma from the nebula and, to a lesser extent, by the shocked ejecta in the outer envelope of the SN. This is evident in the lightcurve where the dashed and dotted lines indicate the contributions of shocked plasma from the nebula and shocked ejecta, respectively, to the X-ray emission (Figure~\ref{lc_TOY}). During this phase, the average X-ray emission-weighted electron temperature rapidly increases, reaching values of tens of keV at day $\approx 500$. This rising phase persists for approximately 500 days, culminating in the peak of X-ray emission around day $\approx 700$ (see Figure~\ref{lc_TOY}).

Beyond day 700, the interaction between the remnant and the nebula enters into its second phase of evolution. The X-ray luminosity undergoes a gradual decline, indicating the blast's propagation through increasingly less dense regions of the nebula (see Figures~\ref{evol_snr_toy} and \ref{lc_TOY}). Simultaneously, the average electron temperature of the X-ray emitting plasma continues to decrease following its peak around day 500, eventually dropping below 10~keV after day 3000 in model TOY\_T50\_D6.8 and day 1000 in model TOY\_T150\_D7. 

During the evolution, the nebula plays a pivotal role in shaping the collimation of the blast wave along its symmetry axis (see Figure~\ref{evol_snr_toy}), resulting in a distinctive bipolar shock morphology reminiscent of that observed in SN~1987A (e.g., \citealt{2005ApJS..159...60S, 2015ApJ...810..168O}), in nova outbursts (e.g., \citealt{2010ApJ...720L.195D, 2012MNRAS.419.2329O}), or in the Great Eruption of Era Carinae (\citealt{1995ApJ...441L..77F, 1999ApJ...520L..49L, 2021MNRAS.503.4276H}), namely in cases in which the blastwave interacts with dense material placed in the equatorial plane. The forward shock advances through the nebula within the equatorial plane, exhibiting velocities on the order of $u_{\rm FS} \approx 5000$~km~s$^{-1}$ at the peak of emission, while poleward propagation through the tenuous outflow occurs at velocities reaching tens of thousands of km~s$^{-1}$. These velocities align well with those deduced from observational analyses: $u_{\rm FS}\approx 4000$~km~s$^{-1}$ as determined from X-ray observations (\citealt{bmm22}) and $u_{\rm FS} > 10000$~km~s$^{-1}$ according to very long baseline interferometry (VLBI) radio observations (\citealt{2021MNRAS.502.1694B}). 

The density structure of the nebula is characterized by three main regions (see Table~\ref{tab1}). In our preferred TOY models, the innermost region is characterized by maximum particle number densities of up to $10^7$~cm$^{-3}$ and is notably thin, confined between $R_0 = 4.3\times 10^{16}$~cm and $R_1 = 4.8\times 10^{16}$~cm. As discussed in Sect.~\ref{sec:phys_wind}, this region represents the interaction zone between the helium star's fast outflow and the dense nebula resulting from a previous episodic massive eruption. This region is responsible for the fast increase of X-ray luminosity\footnote{Without this region, the X-ray luminosity increases more gradually and smoothly at odds with the observations.} and electron temperature around day 200. The main body of the nebula is characterized by an intermediate region, spanning from $R_1$ to $R_2 = 1.4\times 10^{17}$~cm. At $r = R_1$, the maximum density reaches a few $10^6$~cm$^{-3}$, diminishing with a radial dependence of $r^{-2}$. This region is responsible for the main properties of the X-ray lightcurve and temperature evolution observed in \sn14c. Finally, beyond $R_2$, the density experiences a rapid decline following a $r^{-5}$ trend. This outer region is challenging to constrain with our model due to the faint X-ray emission from \sn14c after day 2000. It likely represents the phase of progenitor evolution just before the onset of the progenitor star's massive eruption. Nevertheless, we note that \cite{2019ApJ...887...75T} analyzed infrared observations of \sn14c collected 1–5 years post-explosion. Their findings revealed that the dense, H-rich region through which the SN blast propagated extends to approximately $1.4 \times 10^{17}$~cm (based on the shock's location at 1920 days). Interestingly, this result aligns quite well with our findings, even if our models were constrained with X-ray data.


\section{SNR expansion through the CSM sculpted by progenitor's outflows}
\label{sec:phys_wind}

In this section, we delve into the hydrodynamic models of stellar outflows outlined in Sect.~\ref{sec:modcsm} with the aim of constraining a progenitor mass-loss history capable of reproducing the X-ray lightcurve and temperature evolution inferred from the analysis of X-ray observations. The parametrized (TOY) models, as discussed in Sect.~\ref{sec:mod_par}, enabled us to constrain the large-scale structure and geometry of the nebula surrounding \sn14c. In particular, a valuable piece of information that serves as a guide in the search for a coherent physical model for the  CSM is the mass flux rate versus radial distance derived from the density distribution of the CSM in models TOY\_T50\_D6.8 and TOY\_T150\_D7 (see Figure~\ref{ml_rate_dist}) and considering a velocity of 1000~km~s$^{-1}$ for the outflow from the helium star and 10~km~s$^{-1}$ for outflows in previous phases of evolution.


\subsection{Evolution of the CSM in the millennia preceding the collapse of the progenitor}

\begin{figure}[!t]
  \centering
  \includegraphics[width=8.5cm]{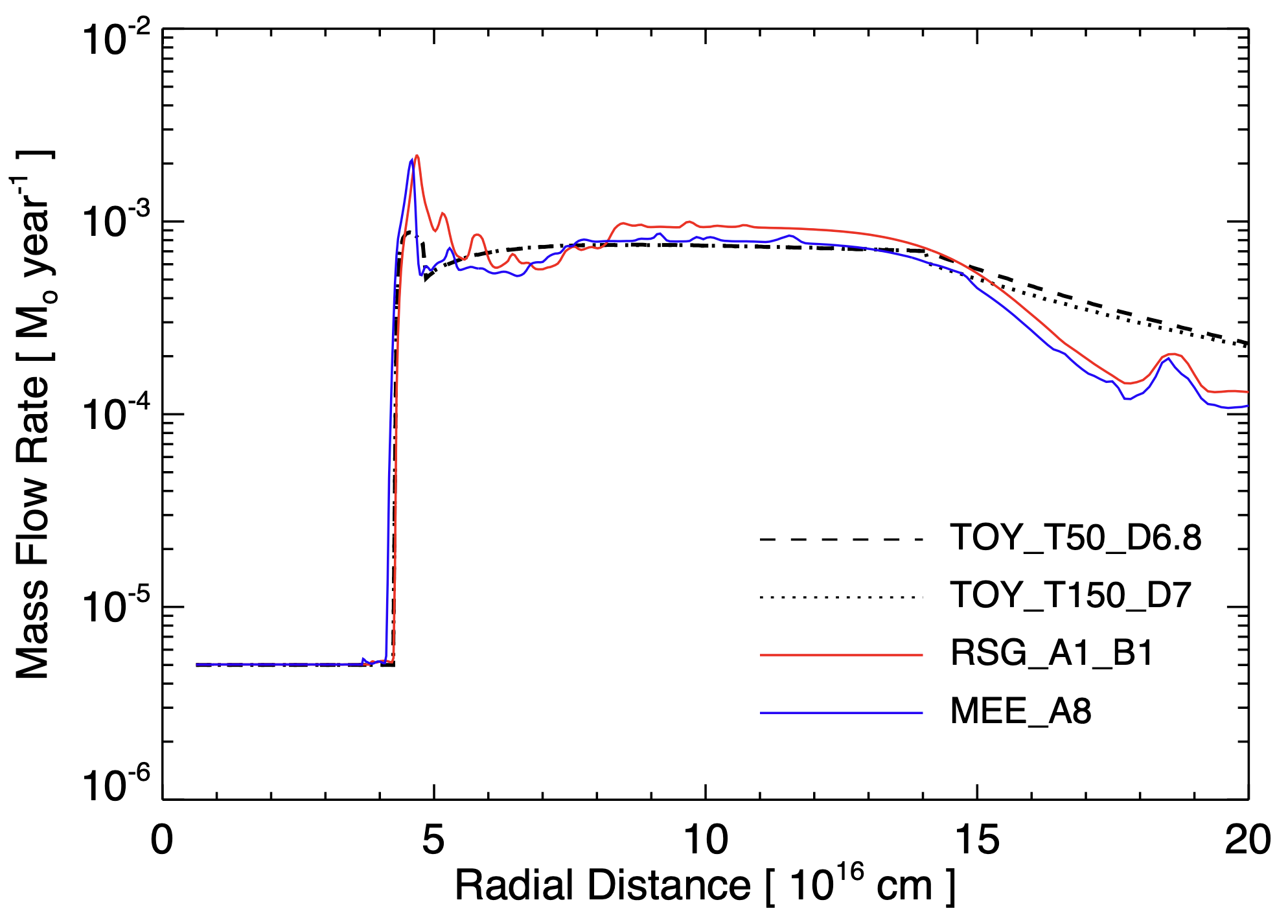}
  \caption{Mass flow rate at the progenitor star's collapse vs. radial distance, obtained from the TOY models outlined in Sect.~\ref{sec:mod_par} (dotted and dashed lines), along with the two reference models characterizing the remnant's expansion: one through the asymmetric outflow of a RSG progenitor with variable mass-loss rate (model RSG\_A1\_B1; red line), and the other through the nebula of a massive episodic eruption (model MEE\_A8; blue line).}
  \label{ml_rate_dist}
\end{figure}

\begin{figure}[!ht]
  \centering
  \includegraphics[width=8.5cm]{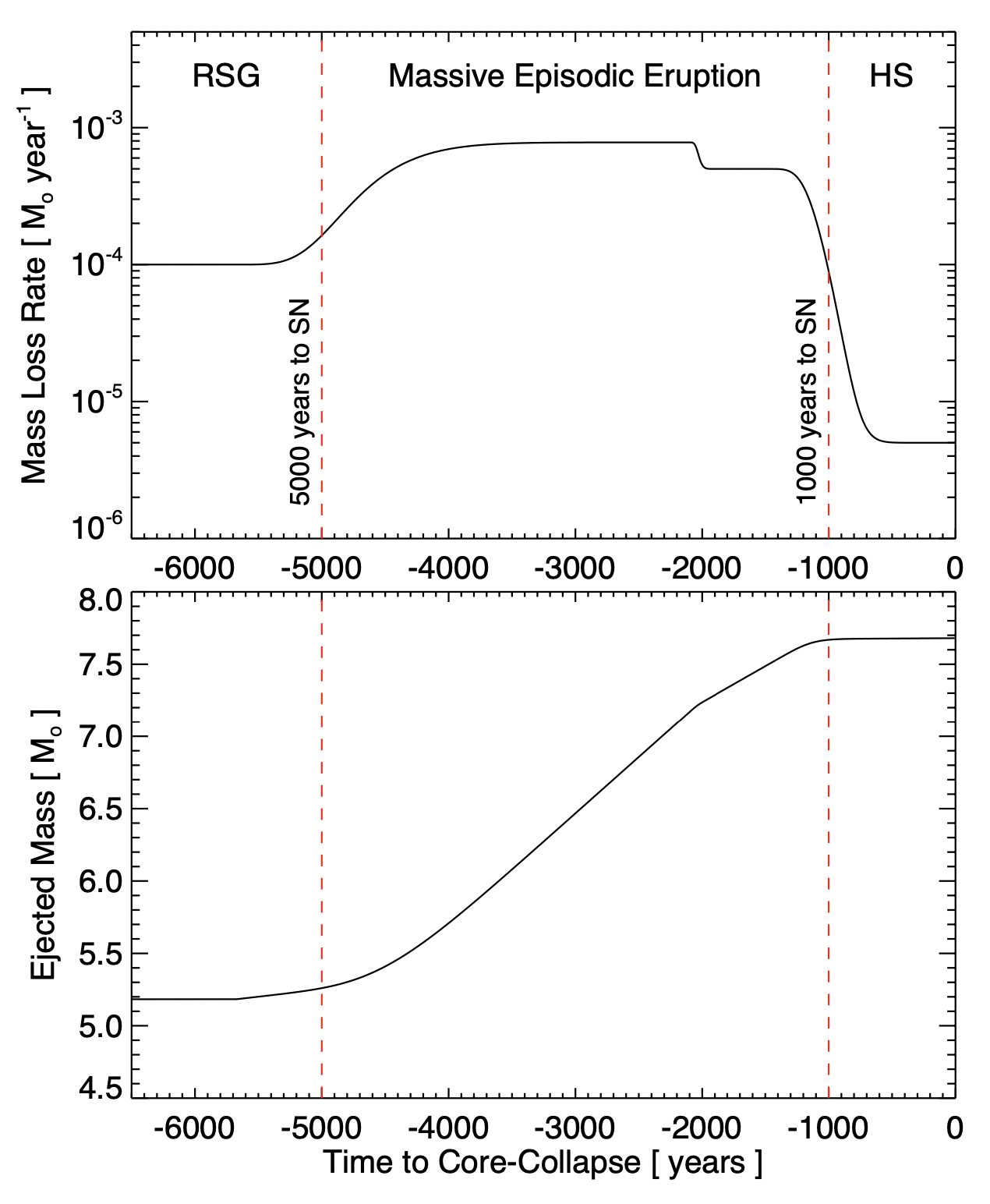}
  \caption{Upper panel: Mass-loss history used in model MEE\_A8 (see Table~\ref{tab:mod_wind}), resulting in X-ray lightcurve and temperature evolution of the SNR consistent with those inferred from the analysis of X-ray observations of {\sn14c} (e.g., \citealt{twd22, bmm22}). Lower panel: Corresponding amount of mass released by the progenitor star in the millennia leading up to its core-collapse.}
  \label{ml_histo}
\end{figure}

In both the outflow scenarios considered (either RSG or MEE models), we explored different levels of asymmetry in the outflow, varying the parameters regulating the dependence of outflow density on the polar angle $\theta$ (see Sect.~\ref{sec:windmodel}). In the case of a remnant expanding through the asymmetric outflow originating from a RSG progenitor (\citealt{1996ApJ...472..257B}) with a time-varying mass-loss rate (RSG models), we performed simulations with different parameters $A$ and $B$ (see Eq.~\ref{wind:rsg}). In the case of a remnant interacting with a dense nebula from a massive episodic eruption from the progenitor star (MEE models), we considered simulations with a different parameter $\alpha$, assuming $\xi = 2$ (see Eq.~\ref{wind:mee}). At the same time, for each of the models considered, we explored various mass-loss histories of the progenitor by varying the parameters $t_{\text{MEE1}}$, $t_{\text{MEE2}}$, $t_{\text{hs}}$, $\mdot_{\text{MEE1}}$, and $\mdot_{\text{MEE2}}$ (see Sect.~\ref{sec:windmodel}).

\begin{table*}
\caption{Parameters of the CSM and of the mass-loss history for the RSG and MEE models reproducing the observations of \sn14c. In all the models, the initial mass-loss rate at the end of the RSG phase is $\mdot_{\rm RSG} = 10^{-4}$~\mlrate, and that in the final helium star phase $\mdot_{\rm hs} = 5\times 10^{-6}$~\mlrate; the background density is assumed to be $\rho_{\rm bkg} = 2.17\times 10^{-25}$~g~cm$^{-3}$ in all the simulations. The four columns on the right report parameters obtained from the simulations (see text): the inner ($R_{\rm in}$) and outer ($R_{\rm out}$) radii of the nebula, the nebula thickness at $r= 6\times 10^{16}$~cm ($\sigma_{\rm sh}$), the peak density ($\max(n_{\rm sh})$) at the inner boundary of the nebula, and the total mass of the nebula ($M_{\rm sh}$).}
\label{tab:mod_wind}
\begin{center}
\begin{tabular}{lcccccccc|cccc}
\hline
\hline
RSG MODEL           & $A$  &$B$& $C$ & $t_{\rm MEE1}$ & $t_{\rm MEE2}$ & $t_{\rm hs}$ & $\mdot_{\rm MEE1}$ & $\mdot_{\rm MEE2}$ &  $(R_{\rm in}; R_{\rm out})$  &  $\sigma_{\rm sh}$ & $\max(n_{\rm sh})$ &  $M_{\rm sh}$ \\
  & & & & yr & yr & yr & $M_{\odot}$ yr$^{-1}$ & $M_{\odot}$ yr$^{-1}$  &  $10^{16}$~cm  &  $10^{16}$~cm  &  $10^6$~cm$^{-3}$  &  $M_{\odot}$ \\
\hline
RSG\_A1\_B5     & 1    & 5   & 4.1 & 5500 & 2200 & 1200 & $8.3\times 10^{-4}$ & $5.4\times 10^{-4}$ & $(4.3; 15.7)$ & $10.1$ & 4.6 & 2.2 \\
RSG\_A1\_B1     & 1    & 1   & 2.4 & 5500 & 2200 & 1200 & $9.3\times 10^{-4}$ & $5.6\times 10^{-4}$ & $(4.3; 14.9)$ & $15.1$ & 3.2 &  2.3 \\
RSG\_A1\_B0.5   & 1    & 0.5 & 2.5 & 5500 & 2200 & 1200 & $1.1\times 10^{-3}$ & $8.6\times 10^{-4}$ & $(4.4; 14.6)$ & $16.2$ & 2.9 &  2.2 \\
RSG\_A0.99\_B5  & 0.99 & 5   & 3.9 & 5500 & 2200 & 1200 & $8.1\times 10^{-4}$ & $5.3\times 10^{-4}$ & $(4.1; 15.5)$ & $18.9$ & 4.6 &  2.1\\
\hline
\hline
MEE MODEL           & $\alpha$  &$\xi$& $C$ & $t_{\rm MEE1}$ & $t_{\rm MEE2}$ & $t_{\rm hs}$ & $\mdot_{\rm MEE1}$ & $\mdot_{\rm MEE2}$ &  $(R_{\rm in}; R_{\rm out})$  &  $\sigma_{\rm sh}$ & $\max(n_{\rm sh})$ &  $M_{\rm sh}$  \\
  & & & & yr & yr & yr & $M_{\odot}$ yr$^{-1}$ & $M_{\odot}$ yr$^{-1}$ & $10^{16}$~cm  &  $10^{16}$~cm  &  $10^6$~cm$^{-3}$  &  $M_{\odot}$ \\
\hline
MEE\_A2         & 2   & 2   & 19.5 & 5500 & 2100 & 1200 & $6.7\times 10^{-4}$ & $5.8\times 10^{-4}$ & $(4.3; 16.1)$ & $6.8$ & 6.1 &  1.9\\
MEE\_A4         & 4   & 2   & 5.1  & 5500 & 1700 & 1200 & $7.3\times 10^{-4}$ & $5.1\times 10^{-4}$ & $(4.2; 15.7)$ & $8.6$ & 5.8 &  2.0\\
MEE\_A8         & 8   & 2   & 2.5  & 5500 & 1700 & 1200 & $7.8\times 10^{-4}$ & $5.0\times 10^{-4}$ & $(4.3; 15.0)$ & $11.6$ & 3.4 &  2.0\\
MEE\_A16        & 16  & 2   & 1.9  & 5500 & 2100 & 1200 & $9.0\times 10^{-4}$ & $6.3\times 10^{-4}$ & $(4.3; 14.5)$ & $16.2$ & 2.9 &  2.3\\
\hline
\hline
\end{tabular}
\end{center}
\end{table*}

Each outflow model was constrained using the mass flow rate profile shown in Figure~\ref{ml_rate_dist}. Subsequently, we simulated the expansion of the SNR through the CSM sculpted by the previously modeled outflow and compared the resulting synthetic X-ray lightcurves and temperature evolution with those derived from the analysis of Chandra and NuSTAR observations.

\begin{figure*}[!ht]
  \centering
  \includegraphics[width=0.97\textwidth]{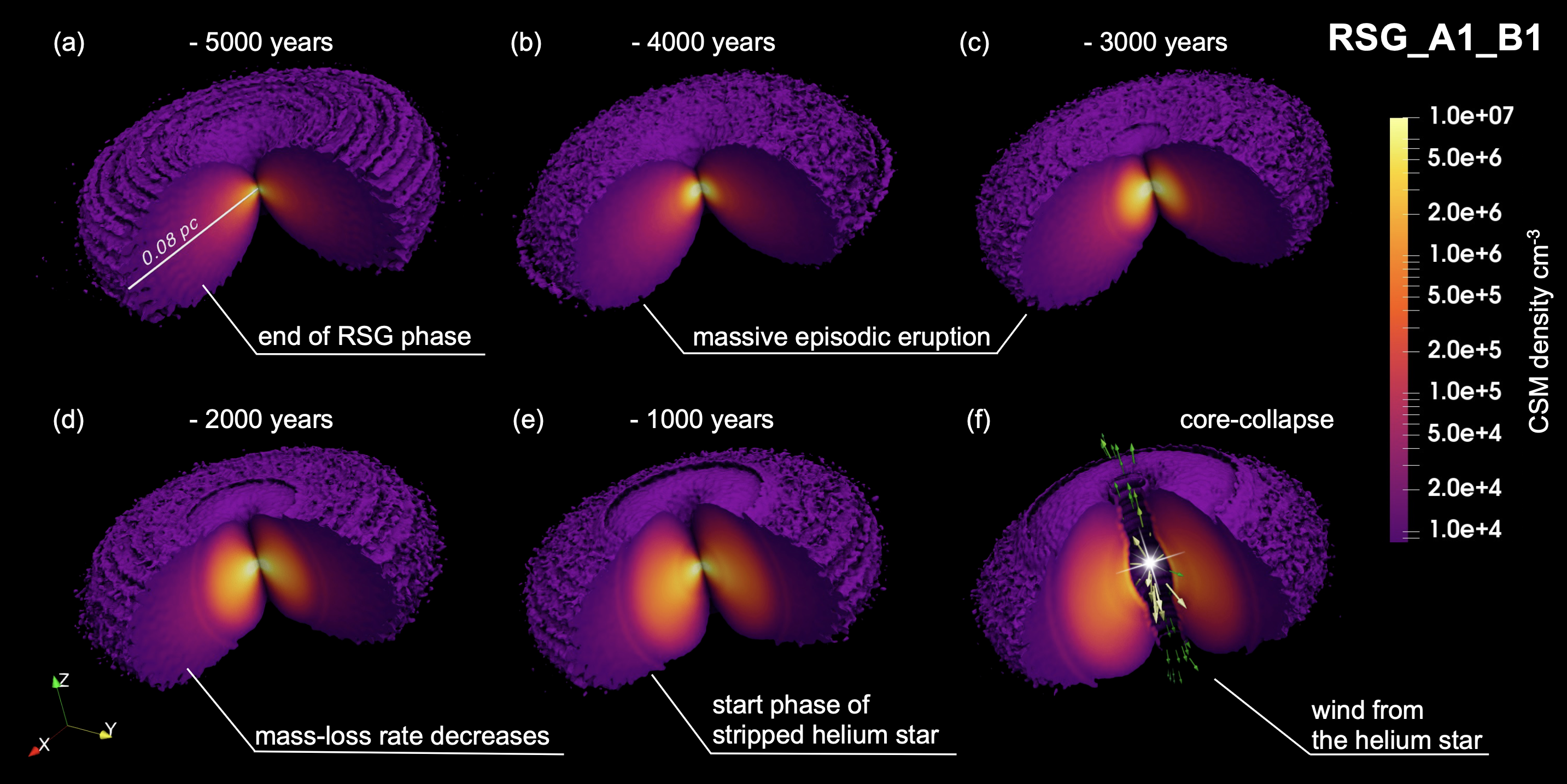}
  \centerline{ }
  \includegraphics[width=0.97\textwidth]{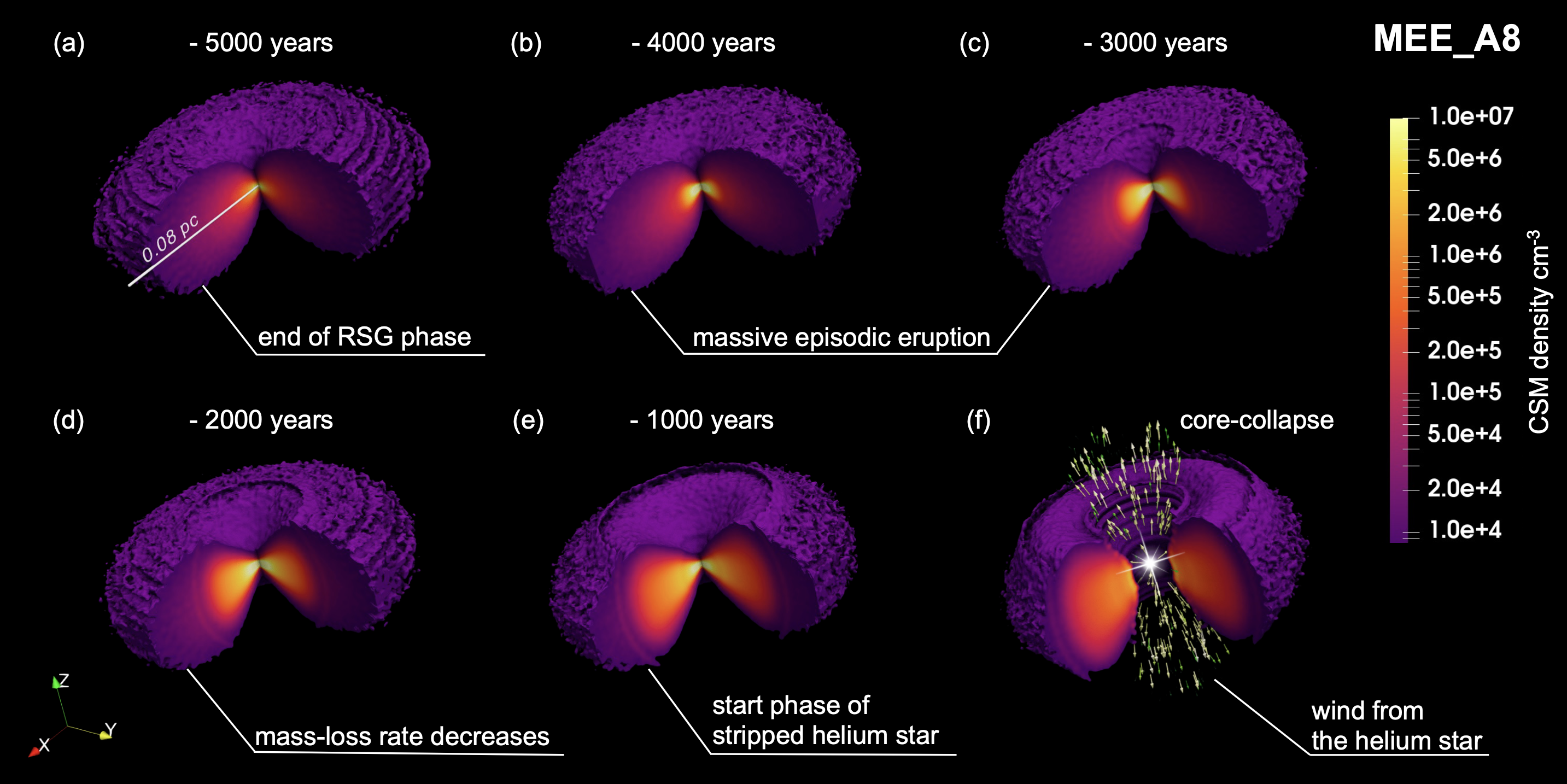}
  \caption{Evolution of the CSM in the reference models RSG\_A1\_B1 (frames (a)-(f) in the upper panel) and MEE\_A8 (lower panel); see Table~\ref{tab:mod_wind} for the details of the models. The frames in each panel illustrate the geometry and density distribution of the CSM at the labeled times. The isosurface delineates material with a particle number density $n > 10^{4}$~cm$^{-3}$; the color indicates the density value (color legend provided on the right). In frames (f), the star at the center indicates the position of the SN explosion and the arrows denote the outflow velocity during the helium star phase.}
  \label{csm_evol}
\end{figure*}

We note that models capable of accurately reproducing the X-ray observations exhibit very similar mass-loss histories for the progenitor star, regardless of the outflow scenario (RSG or MEE models) and level of outflow asymmetry. An example of these mass-loss histories is shown in the upper panel of Figure~\ref{ml_histo} for model MEE\_A8. The outflow parameters corresponding to the SNR models that best match the observations are summarized in Table~\ref{tab:mod_wind}. For each of these models, we defined the circumstellar nebula (corresponding to the torus in the TOY models) as the CSM material with a particle number density above $10^5$~cm$^{-3}$. Then we characterized the nebula at the time of core-collapse by determining its extension in the equatorial plane (from the inner, $R_{\rm in}$, to outer, $R_{\rm out}$, radius), its thickness perpendicular to the equatorial plane at $r= 6\times 10^{16}$~cm, $\sigma_{\rm sh}$, the maximum density in the inner boundary, $\max(n_{\rm sh})$, and the total mass of the nebula, $M_{\rm sh}$. Table~\ref{tab:mod_wind} also reports these parameters in the last four columns. Figure~\ref{csm_evol} illustrates the evolution of the CSM in the millennia leading up to the SN, as predicted by the reference models shown in Figure~\ref{ml_rate_dist} for the two scenarios considered (runs RSG\_A1\_B1 and MEE\_A8).

In these models, the CSM already exhibited denser regions in the equatorial plane over 5000 years before the SN due to the asymmetric outflow from the progenitor star (see frame (a) in each panel of Figure~\ref{csm_evol}). While the X-ray lightcurve does not allow us to precisely constrain this phase of evolution, our findings suggest that the onset of the eruption occurred not later than 5000 years before the SN. During this period, the mass-loss rate increased from $10^{-4}\,M_{\odot}$~yr$^{-1}$ to approximately $7\times 10^{-4}\,M_{\odot}$~yr$^{-1}$ within a few centuries (see upper panel of Figure~\ref{ml_histo}), though the exact timing and details of the rising phase remain unconstrained by the model. Consequently, the outflow density significantly increased, resulting in a substantial enhancement of density in the equatorial region, with values reaching about $10^7$~cm$^{-3}$ near the star (frames (b)-(e) in each panel of Figure~\ref{csm_evol}). This phase persisted for approximately 4000 years, substantially elevating the density of the CSM up to distances of approximately $0.05$~pc ($\approx 1.5\times 10^{17}$~cm). Interestingly, this phase generates the highest mass-loss rates in our models, which align closely with those inferred from the analysis of infrared data (i.e., $\sim 10^{-3}\,M_{\odot}$~yr$^{-1}$; \citealt{2019ApJ...887...75T}).

Notably, our SNR models required that, in the last phase of this period, the mass-loss rate slightly decreased by $10-40$\,\% (depending on the model). As mentioned in Sect.~\ref{sec:phys_wind}, we accounted for this by delineating two distinct phases during the eruption characterized by different $\mdot$ values. In fact, we found that, while our model requires the decrease of $\mdot$ from approximately 5000 years to 1000 years before the SN, it is not particularly sensitive to the specific way in which $\mdot$ decreases. Therefore, we adopted a step function to represent this decrease (see upper panel of Figure~\ref{ml_histo}), facilitating the exploration of the parameter space. 

The eruption phase concluded around 1000 years before the SN, marked by a sharp decrease in $\mdot$ to $5\times 10^{-6}\,M_{\odot}$~yr$^{-1}$ (upper panel of Figure~\ref{ml_histo}). In the final centuries leading up to the SN, the fast and tenuous outflow emitted by the progenitor helium star gradually displaced the dense outflow from the preceding mass eruption, producing a cavity in the surrounding environment of the star (frame (f) in each panel of Figure~\ref{csm_evol}). We found that the cavity extends in the equatorial plane up to a distance of approximately $4.5\times 10^{16}$~cm, which aligns with the inner radius of the torus obtained in the TOY models capable of replicating the X-ray observations of {\sn14c} (refer to Table~\ref{tab1}).

At the time of core-collapse, the CSM exhibits a nebula with peak densities in the range $10^6-10^7$~cm$^{-3}$ in the inner boundary (see Table~\ref{tab:mod_wind}), forming a dense torus encircling the star and extending from $\sim 4\times 10^{16}$~cm to $\sim 15\times 10^{16}$~cm in the equatorial plane. This configuration aligns with the idealized torus described in the TOY models (see Sect.~\ref{sec:mod_par} and Figure~\ref{csm_toy_model}) and with the average configuration inferred from the analysis of infrared and X-ray observations (\citealt{2017ApJ...835..140M, 2019ApJ...887...75T}). The density peaks significantly at the interaction interface between the outflow originating from the helium star and the surrounding nebula, resulting in a conspicuous spike in the mass flux rate, as evidenced in Figure~\ref{ml_rate_dist} (red and blue lines for the two reference models considered). This region of enhanced density plays a pivotal role in the rapid rise of X-ray emissions upon the impact of the SN blast on the nebula. Consequently, it justifies the necessity of incorporating a thin, high-density inner region within the TOY models to accurately replicate the X-ray lightcurve (see Sect.~\ref{sec:mod_par}). According to our simulations, the eruption led to the ejection of approximately $2.5\,M_{\odot}$ of stellar material into the CSM (lower panel of Figure~\ref{ml_histo}), thereby exposing the naked core of the progenitor star. 

\begin{figure*}[!ht]
  \centering
  \includegraphics[width=0.93\textwidth]{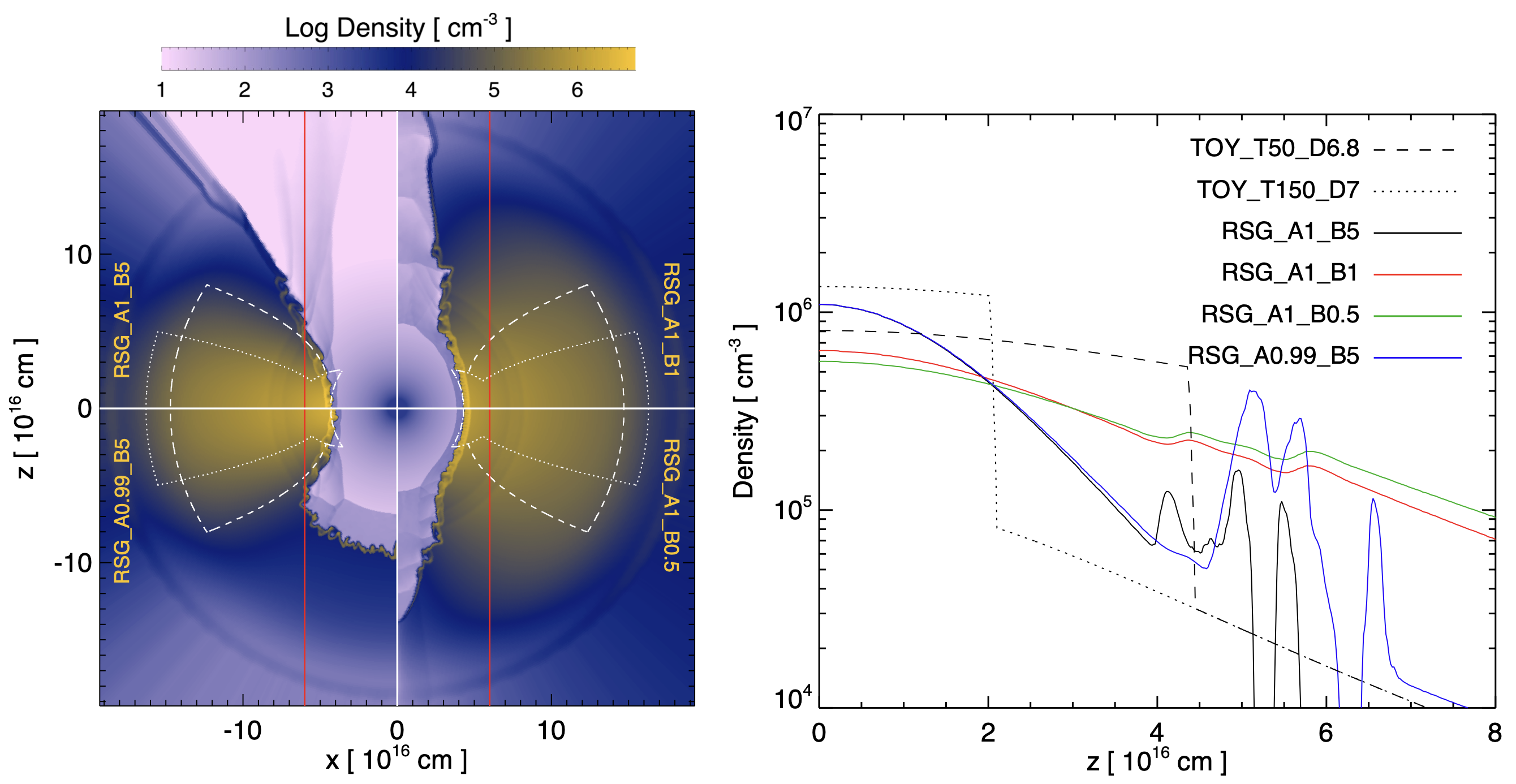}
  \includegraphics[width=0.93\textwidth]{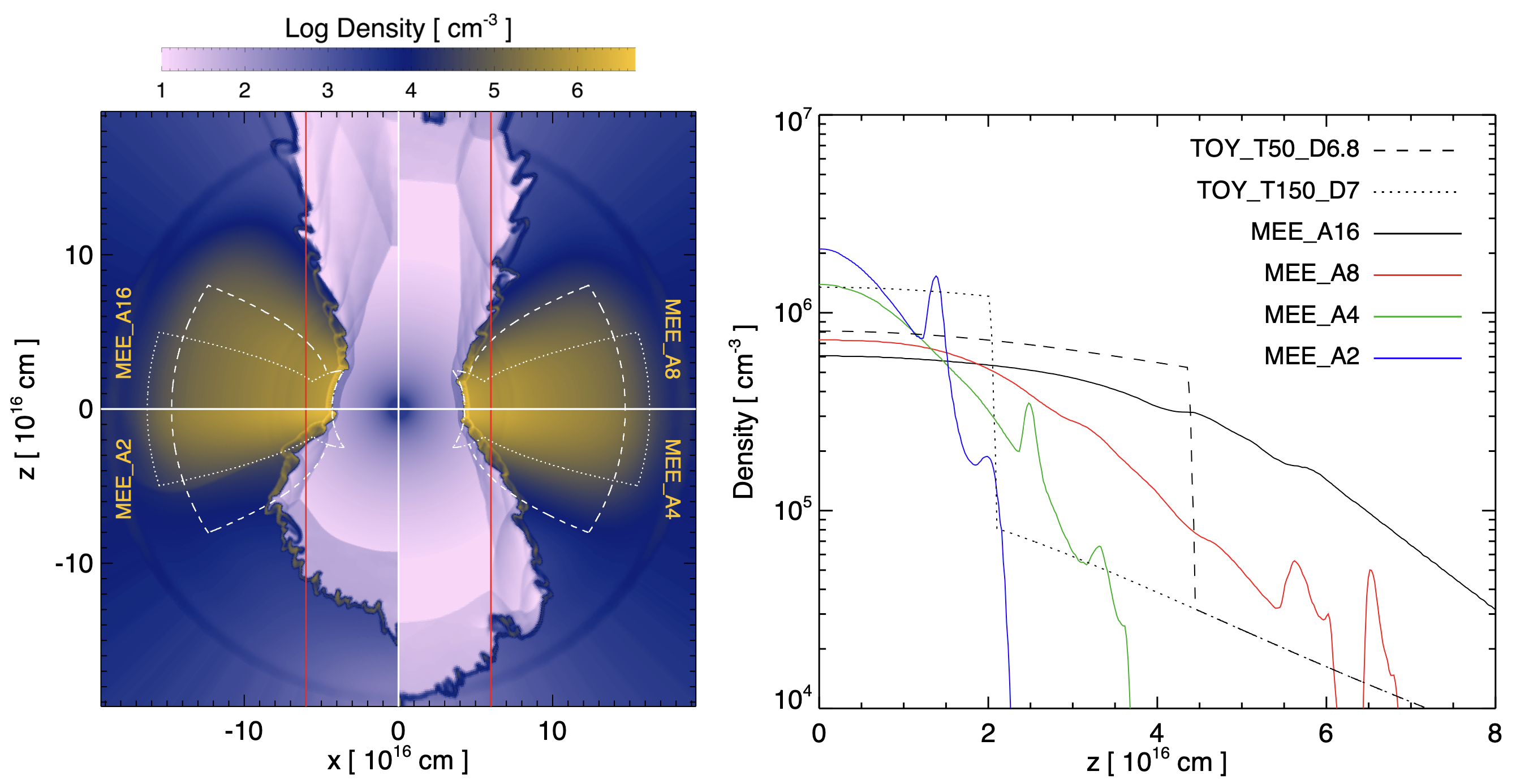}
  \caption{Left panels: The 2D cross-sections in the $(x, z)$ plane show the spatial density distribution (in a logarithmic scale) of the CSM at core-collapse from RSG (upper panel) and MEE (lower panel) models; each quadrant in the panels displays the result for each of the models listed in Table~\ref{tab:mod_wind}. The symmetry axis of the nebula aligns with the $z$-axis. The dashed and dotted contours show the cross-section of the torus in runs TOY\_T50\_D6.8 and TOY\_T150\_D7, respectively. Right panels: Density profiles along the red lines shown in the left panels are presented for the RSG (upper panel) and MEE (lower panel) models in comparison with the analogous profiles derived from the two TOY models (dashed and dotted lines).}
  \label{csm_map}
\end{figure*}

\begin{figure*}[!ht]
  \centering
  \includegraphics[width=1.\textwidth]{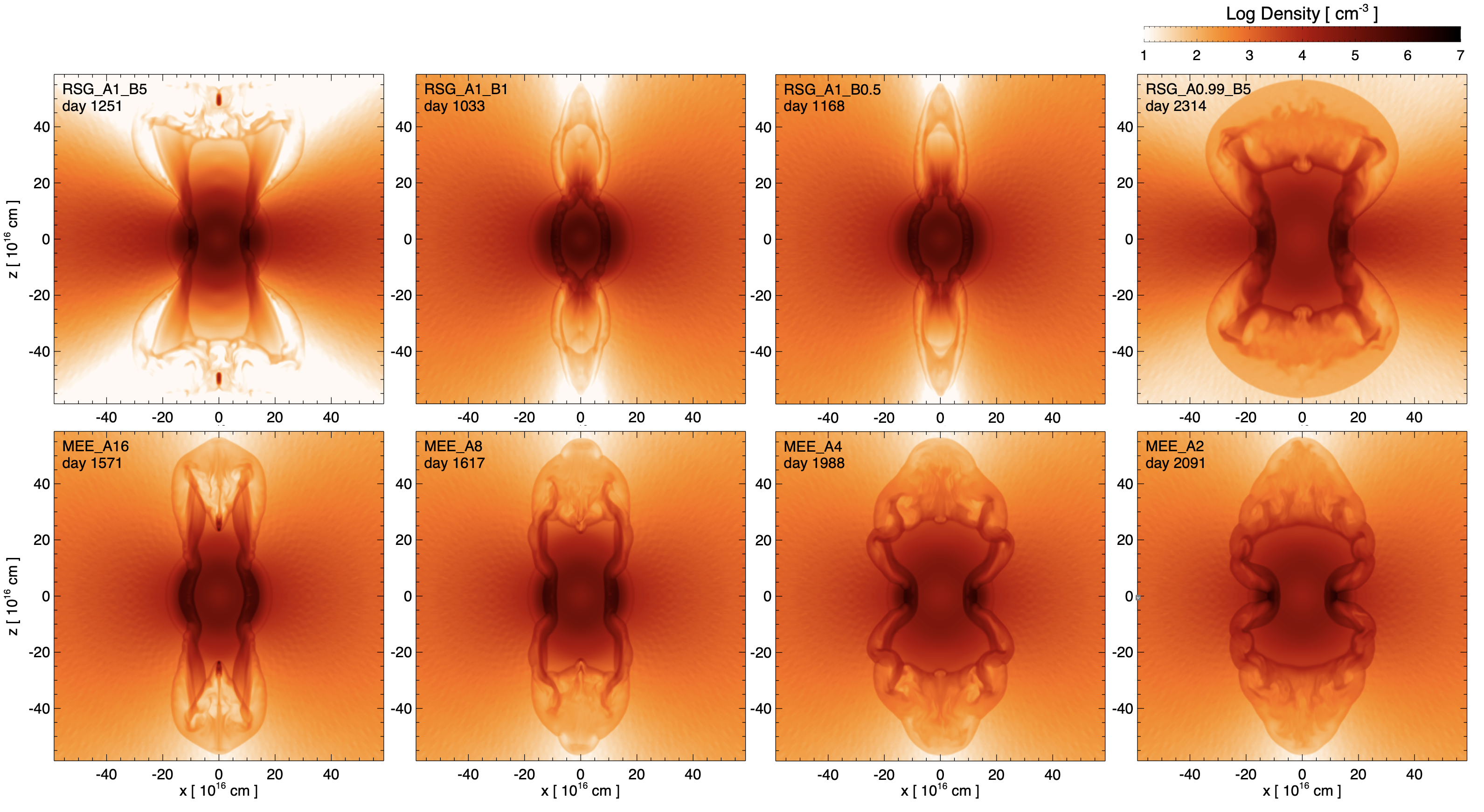}
  \caption{The 2D cross-sections in the $(x, z)$ plane show the spatial density distribution (in a logarithmic scale) of the SNR at the labeled times from RSG (upper panels) and MEE (lower panels) models. The symmetry axis of the bipolar blast wave aligns with the $z$-axis.}
  \label{snr_wind}
\end{figure*}

The outflow evolution outlined above for models RSG\_A1\_B1 and MEE\_A8 applies to the other RSG and MEE models listed in Table~\ref{tab:mod_wind}. Figure~\ref{csm_map} illustrates the cross-sections in the $(x,z)$ plane of the CSM at the time of core-collapse for all these models. They primarily differ from one another in terms of the thickness of the region of the CSM characterized by density values exceeding $10^5$~cm$^{-3}$ (marked as yellow areas in the left panels of the figure; see also the parameter $\sigma_{\rm sh}$ in Table~\ref{tab:mod_wind}) and the extent of the cavity carved out by the fast and tenuous outflow emanating from the helium star. The figure also shows how the interaction between this fast outflow and the CSM results in the formation of a shell which, due to radiative cooling, becomes extremely thin and prone to the Vishniac instability (\citealt{1983ApJ...274..152V, 1989ApJ...337..917V}). The densest portion of the CSM corresponds to the nebula resulting after the massive episodic eruption from the progenitor star.

Generally, the cavity appears much more developed in MEE models than in RSG models. The only exception is run RSG\_A1\_B5 (upper panel in Figure~\ref{csm_map}), which is the most asymmetric outflow model considered, exhibiting the highest extension of the cavity polewards. The nebula appears thinnest in models with large outflow asymmetries (e.g., RSG models with high $B$ or MEE models with low $A$; see Table~\ref{tab:mod_wind}). Conversely, in the less asymmetric outflow models considered, the nebula is the thickest. Our reference models, namely runs RSG\_A1\_B1 and MEE\_A8, fall between these two extremes, while runs RSG\_A0.99\_B5 and MEE\_A2 exhibit a thin nebula and a poorly extended cavity.

We compared the density profiles of the nebula perpendicular to the equatorial plane near the inner radius (at $r = 6\times 10^{16}$~cm) of the outflow (RSG and MEE) models with those of the TOY models (right panels in Figure~\ref{csm_map}). In RSG models, the density exhibits a smooth decline with distance from the equatorial plane, while MEE models display a slight density variation within the nebula and a sharper decline at its periphery. The latter profiles are more similar to those of the TOY models. 

In both scenarios, increasing the level of asymmetry of the outflow results in higher densities within the equatorial plane and a more rapid density decrease with distance from it. Conversely, models with lower levels of outflow asymmetry exhibit lower densities within the equatorial plane and a gentler density decrease with distance from it. Consequently, models with higher outflow asymmetry generate a CSM structure more resembling run TOY\_T150\_D7, while those with lower levels of outflow asymmetry lead to a CSM structure more resembling run TOY\_T50\_D6.8.

We also note that the density structure of the nebula shows larger asymmetries in MEE models than RSG models. In fact, different MEE models have significantly different densities in the equatorial plane, whereas RSG models show small variations (compare the values of $\max(n_{\rm sh})$ in Table~\ref{tab:mod_wind}). As a result, we anticipate that the temperatures of the shocked nebula after interaction with the SNR are more sensitive to the level of asymmetry of MEE models than RSG models (see Sect.~\ref{snr_evol}).

\begin{figure*}[!ht]
  \centering
  \includegraphics[width=15cm]{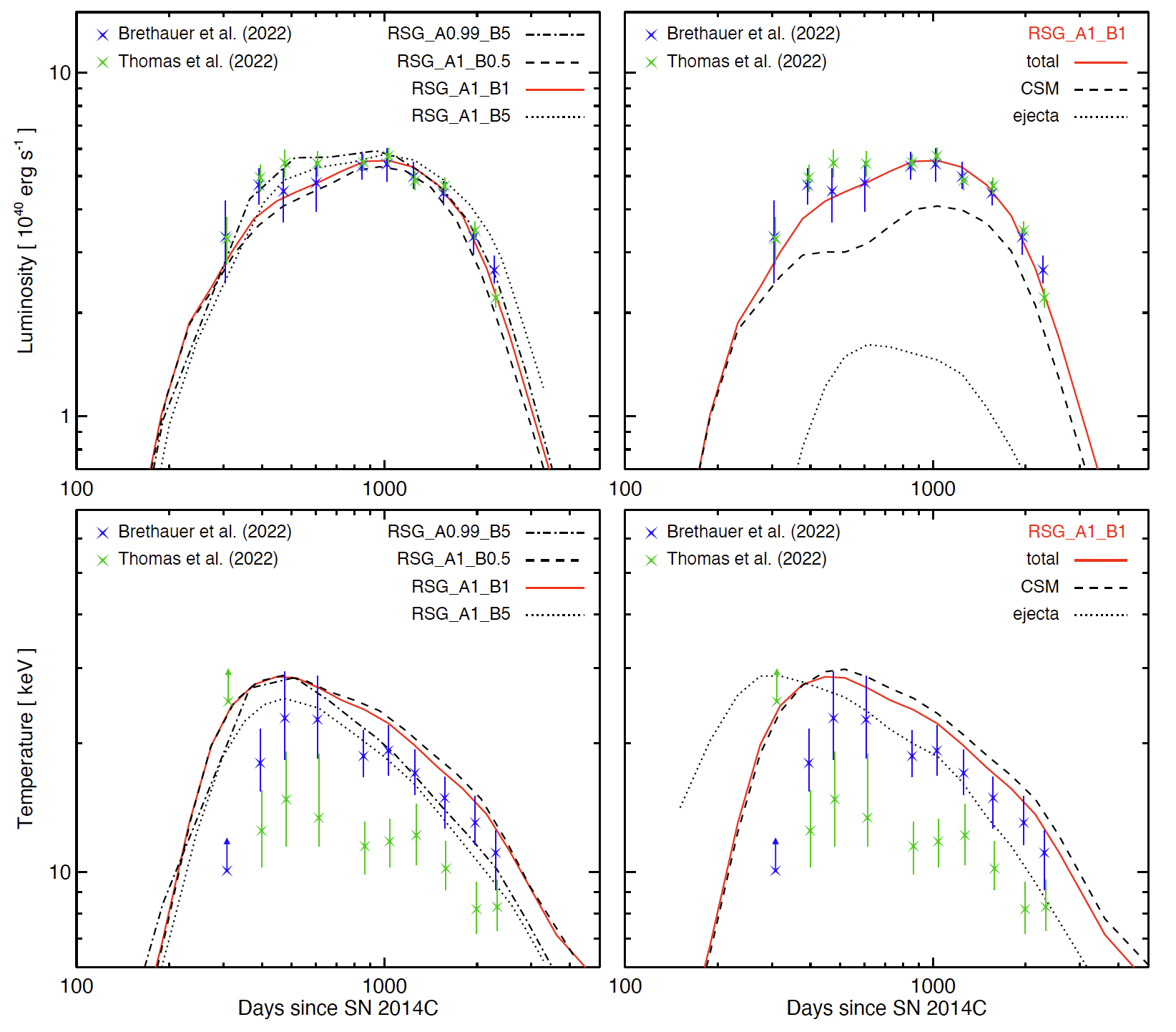}
  \caption{Upper panels: X-ray lightcurve in the $[0.5, 100]$~keV band synthesized from the RSG models compared with the observed lightcurve of \sn14c. The lightcurves derived from the different RSG models are compared on the left. The reference model RSG\_A1\_B1 is also shown on the right with the contributions to emission from the shocked plasma from the nebula (dashed line) and the shocked ejecta (dotted line). Lower panels: Corresponding to the upper panels, the plots show the evolution of average X-ray emission-weighted electron temperature.}
  \label{lc_rsg}
\end{figure*}

\begin{figure*}[!ht]
  \centering
  \includegraphics[width=15cm]{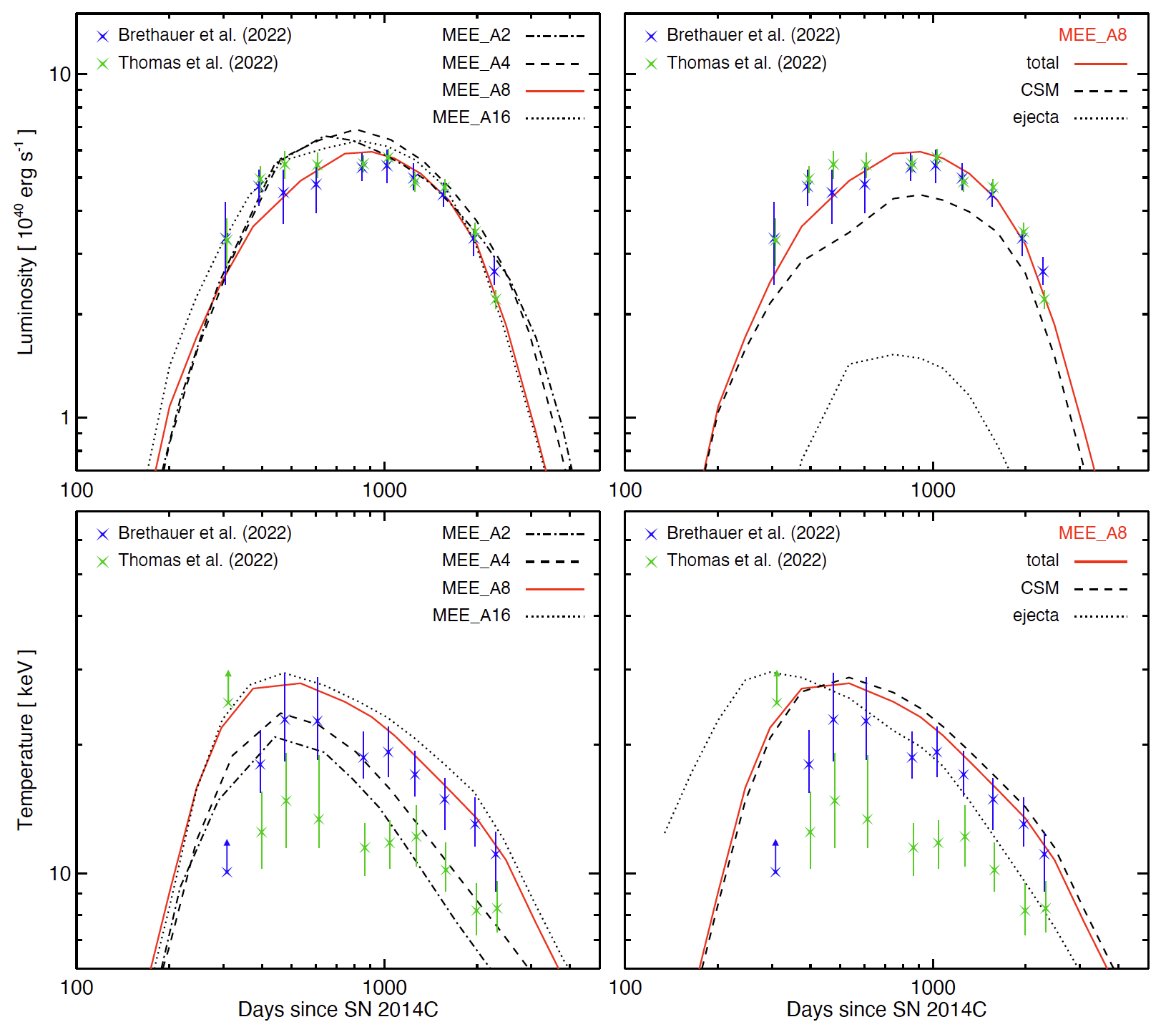}
  \caption{As in Figure~\ref{lc_rsg} but for the MEE models listed in Table~\ref{tab:mod_wind}.}
  \label{lc_mee}
\end{figure*}

\subsection{Evolution of the SNR}
\label{snr_evol}

The evolution of the SNR for each of the RSG and MEE models mirrors that described in Sect.~\ref{sec:mod_par} for the TOY models. Consequently, in these cases, the SN blast exhibits a bipolar morphology following interaction with the dense CSM nebula. Figure~\ref{snr_wind} displays the remnant morphology when the blast has extended to approximately $5\times 10^{17}$~cm from the center of explosion polewards\footnote{Note that the epoch varies for each model due to the distinct velocity propagation of the blast wave through a CSM with different densities.}. The differences in bipolar morphology among different models stem from variations in the structure of the nebula and the inner cavity (see Figure~\ref{csm_map}). The blast collimation is most effective in RSG models with low values of $B$ (such as runs RSG\_A1\_B1 and RSG\_A1\_B0.5), resulting in jet-like protrusions rapidly expanding polewards at velocities of some $10000$~\kms. In the other cases, the SN blast exhibits a morphology characterized by extended lobes again propagating polewards. In these latter models, we note that the wider the lobes, the slower the forward shock, with the lowest expanding velocity observed in run RSG\_A0.99\_B5 due to relatively high values of outflow density at the poles (see upper panel in Figure~\ref{csm_map}). Across all cases, the blast wave propagates at velocities around $4000-5000$~{\kms} across the nebula. These velocity values are consistent with those inferred from the analysis of radio and X-ray observations (\citealt{2021MNRAS.502.1694B, bmm22}).

Figures~\ref{lc_rsg} and \ref{lc_mee} illustrate the X-ray lightcurve (upper panels) and the evolution of X-ray emission-weighted average electron temperature (lower panels) synthesized from the RSG and MEE models, respectively. Despite the variations in model characteristics, all exhibit similar X-ray lightcurves, capable of approximately reproducing the observed X-ray lightcurve in {\sn14c} (blue and green symbols in the figures). However, notable discrepancies emerge in the evolution of temperature, reflecting differences in the average density of the nebula across the various models. This contrast is particularly pronounced within MEE models, where temperatures can diverge by approximately 50\% between different runs (as observed when comparing run MEE\_A2 and MEE\_A16 in the lower left panel of Figure~\ref{lc_mee}), whereas in RSG models, such differences are more constrained. This difference arises due to more significant variations in nebula density among different MEE models compared to RSG models (as illustrated in the right panels of Figure~\ref{csm_map}). It is worth noting that the majority of the models considered align with the temperature evolution results derived by \cite{bmm22}. The exceptions are runs MEE\_A2 and MEE\_A4, which better match the findings of \cite{twd22} after day 1000, although they tend to overestimate the observed peak temperature. 

It is interesting to note that all models replicating the X-ray observations of {\sn14c} necessitate a mass-loss history similar to that shown in Figure~\ref{ml_histo}, implying that our models successfully constrain the mass-loss history of the progenitor star of {\sn14c} in the millennia leading up to the SN event. Among the RSG models, those which more accurately emulate the {\sn14c} observations in terms of temperature exhibit parameter values $B > 1$. As highlighted by \cite{1996ApJ...472..257B}, large values of $A$ and $B$ in Eq.~\ref{wind:blondin} yield a CSM density distribution resembling a dense circumstellar disk, whereas values below 1 result in more gradual density variations typical of rotating variable red giants (see also \citealt{1995ApJ...455..286A}).

The presence of a circumstellar disk suggests a binary nature for the progenitor stellar system, in agreement with previous studies (e.g., \citealt{2020MNRAS.497.5118S}). Lastly, the outflow scenario that appears to best align with the observations requires heavy mass-loss concentrated primarily in the equatorial plane, as seen in MEE models and RSG models with high outflow asymmetry.


\section{Interpretation of X-ray data with the hydrodynamic models}
\label{sec:model_obs}

In this section, we revisit the analysis of the X-ray data of {\sn14c}, using our models for their interpretation. To this end, we consider the MEE models described in Sect.~\ref{sec:phys_wind}, which best replicates the X-ray lightcurve and temperature evolution of {\sn14c} derived by \cite{twd22} and \cite{bmm22}.

\subsection{Reduction of Chandra and NuSTAR data}
\label{sec:reduction}

We analyzed all available X-ray data of \sn14c, comprising a total of 13 Chandra observations spanning 11 distinct epochs\footnote{The complete Chandra dataset, obtained by the Chandra X-ray Observatory, is contained in \dataset[DOI: https://doi.org/10.25574/cdc.301]{https://doi.org/10.25574/cdc.301}.}, as well as 9 NuSTAR observations, each corresponding to a different epoch.. The Chandra and NuSTAR observations were conducted almost simultaneously, with each epoch's observations occurring within two weeks. Due to its superior spatial resolution, Chandra can effectively isolate the emission from {\sn14c} from other nearby X-ray sources. Conversely, NuSTAR's Point Spread Function (PSF) unavoidably results in significant contamination from other sources within the host galaxy of \sn14c. This disparity is illustrated in Figure~\ref{fig:psf_chvsnu} in Appendix \ref{app:obs}, which presents a comparison of the two fields of view using the same extraction regions. Additional information regarding the observations can be found in Table \ref{tab:log_obs} in Appendix \ref{app:obs}.

We reduced the Chandra data by using the task \texttt{chandra\_repro} available within the standard Chandra X-ray Center (CXC) software \texttt{CIAO}, V4.12, with CALDB4-4.1.1. We extracted Chandra spectra from {\sn14c} through the command \texttt{specextract} selecting three different circles, all centered on {\sn14c} but with different radii: $3^{\prime\prime}$, as in \citet{twd22};  $1.5^{\prime\prime}$, as in \citet{bmm22}; $2.5^{\prime\prime}$, an intermediate value tested by us. The background regions were chosen, analogously, following the approaches by \citet{twd22} and \citet{bmm22} selecting a circle centered in the source-free region of the Field of View (FOV) having radii of $8^{\prime\prime}$ and $20^{\prime\prime}$, respectively. We also extracted a background region from an annulus having radii of $4^{\prime\prime}$ and $8^{\prime\prime}$, centered on \sn14c. By comparing the spectra extracted from each of the source extraction regions with each of the background regions we noticed that: i) the choice of the background has minimal to negligible effects on the resulting spectra; ii) employing a smaller extraction region results in a slight underestimation of flux (a few percent) at higher energies compared to the other two cases, due to the effects of the PSF; iii) spectra extracted from regions with radii of $2.5^{\prime\prime}$ and $3^{\prime\prime}$ exhibited perfect overlap. Considering these findings, we opted for source and background regions defined by circles with radii of $2.5^{\prime\prime}$ and $20^{\prime\prime}$, respectively.   

NuSTAR data were reprocessed through the task \texttt{nupipeline} available within the software \texttt{NuSTARDAS}, version v2.1.2, with CALDB4-4.1.1. We set the keyword \texttt{saacalc=3}, \texttt{saamode=OPTIMIZED} and \texttt{tentacle=yes} to optimize the filtering of background events that occurred during each passage of NuSTAR across the South Atlantic Anomaly (SAA). NuSTAR spectral extraction was based on an approach similar to the Chandra one, i.e. selecting regions with the same characteristics as those listed by \citet{twd22} and \citet{bmm22}. The authors of both papers chose a source region as a circle with a radius of $60^{\prime\prime}$ centered on {\sn14c}. However, they considered different background regions: \cite{twd22} used a $60^{\prime\prime}$-radius circle from a source-free region, while \cite{bmm22} used an annulus ranging from $66^{\prime\prime}-180^{\prime\prime}$ centered on {\sn14c}. As we have done for the Chandra spectra, we also explored other regions, e.g. a radius of $40^{\prime\prime}$ for the source region, more than twice the Full Width at Half Maximum ($18^{\prime\prime}$, FWHM) of the NuSTAR PSF and an annulus having radii of $45^{\prime\prime}$ and $85^{\prime\prime}$, centered of \sn14c. We noticed that the selection of the $40^{\prime\prime}$ circle leads to source spectra significantly fainter than in the $60^{\prime\prime}$, due to the tail of the PSF diluting the emission on an area bigger than the selected region. The same effect caused the background annulus region with radii of $45^{\prime\prime}$ and $85^{\prime\prime}$ to significantly lower the counts. Both these two regions were then discarded. We then compared the background-subtracted spectra with the $60^{\prime\prime}$ circle and $66^{\prime\prime}-180^{\prime\prime}$ annulus background regions and found differences noticeable only at energies higher than 10~keV and well within the 1$\sigma$ error bars. The results we present in the rest of the paper are obtained through the adoption of the $60^{\prime\prime}$ source-free circle, but these are unchanged if assuming the other background region.

All the Chandra and NuSTAR spectra were optimally binned through the task \texttt{ftgrouppha}, following the recipe by \citet{kb16}. \texttt{Cstat} (\citealt{cas79}) is used to evaluate the goodness of fits.

\subsection{Spectral analysis revisited: Fit with one \\ thermal component}
\label{sec:fit_1t}

We simultaneously analyzed Chandra and NuSTAR spectra for each epoch, replicating the analysis performed by \cite{twd22} and \cite{bmm22}. Specifically, we employed two distinct models for spectral fitting. The first model (in the following the VAPEC model) used a collisionally ionized diffuse gas in ionization equilibrium (\textsc{vapec} model in XSPEC; \citealt{1996ASPC..101...17A}) along with an absorption component (\textsc{tbabs} in XSPEC), as described by \cite{twd22}. The second model (in the following the BREMS+GAUSS model) incorporated an absorbed thermal bremsstrahlung (\textsc{tbabs}$\times$\textsc{ztbabs}$\times$\textsc{bremss}) and a Gaussian component in the energy range $5-9$~keV to describe the presumed Fe K$\alpha$ emission line, following the approach of \cite{bmm22}. We note that even if the BREMSS+GAUSS model divides the total absorption between the foreground (integrated along the line of sight) and the intrinsic (due to the dense CSM surrounding the explosion site) ones, the column density of the foreground material $N_{\rm H}=6.14\times 10^{20}$~cm$^{-2}$ is a small fraction ($<~10$\%) of the intrinsic one. 

The BREMS+GAUSS model also includes a \textsc{power-law} component, applied to NuSTAR spectra only, to take into account the emission from other sources enclosed in the NuSTAR extraction region (see again Figure~\ref{fig:psf_chvsnu} in Appendix \ref{app:obs}). The latter correction was not performed by \cite{twd22} since they considered this spurious emission negligible. In both BREMSS+GAUSS and VAPEC models, a constant between Chandra and NuSTAR spectra is included to account for cross-calibration effects between the two instruments, that could be as high as 20\% (\citealt{mhm15,mfg22}).

The BREMSS+GAUSS model has a total of 6 free parameters (excluding the power-law for addressing contamination from other sources), while the VAPEC model has 4. The main difference between the two approaches consists in the treatment of the Fe K line: in the BREMSS+GAUSS model, this is described by a dedicated Gaussian component with 3 free parameters, while the continuum emission is represented by the \textsc{bremss} component; conversely, in the VAPEC model, all thermal contributions from shock-heated plasma are included, with the assumption that the Fe K line originates from the same plasma responsible for the continuum emission. We note that using the \texttt{bremss} component overlooks any potential line emission arising from the plasma component responsible for the continuum emission.

Our spectral analysis produced results comparable to those reported by \citet{twd22} and \citet{bmm22}\footnote{Note that \citet{bmm22} do not report best-fit values and uncertainties for the power-law applied to the NuSTAR data to account for contamination from other sources; hence a direct comparison for this component is not feasible.}. We found that the BREMSS+GAUSS model overall describes the data significantly better than the VAPEC model. Particularly, the inclusion of the power-law factor, addressing emission from other sources, significantly enhanced the overall goodness-of-fit, as indicated by the \texttt{cstat} value. The contribution from this spurious component is a fraction ($<10$\%) of the total emission from {\sn14c}, and it predominantly manifests at high energies ($> 10$~keV). Hence, it cannot account for the discrepancy in the best-fit temperature values as reported by \citet{bmm22} and \citet{twd22}. Notably, even in the absence of the additional power-law in the NuSTAR spectra, we observed a similar temperature discrepancy between the two phenomenological models.

Hence, the temperature discrepancy between the findings of \citet{bmm22} and \citet{twd22} is clearly tied to the differing modeling approaches used in the two studies, rather than to variations in background treatment. To investigate this issue further, we generated a synthetic NuSTAR spectrum using a BREMSS+GAUSS model with the parameters outlined in \citet{bmm22}. Then we fitted this spectrum with a VAPEC model, following the approach outlined by \citet{twd22}. We found that the VAPEC model described very well the synthetic spectrum, although the best-fit temperature derived was $\sim$ 10 keV at odds with that assumed in the BREMSS+GAUSS model that was $20$~keV. This discrepancy arises from the fact that the \textsc{vapec} component includes all emission processes, including line emission, thus being effectively constrained by the prominent Fe K line observed in the data. Conversely, the \textsc{bremss} component within the BREMSS+GAUSS model solely accounts for continuum emission and remains unaffected by the flux of the Fe K line. This different treatment of the same spectral feature is the main reason for the temperature discrepancy reported by \citet{bmm22} and \citet{twd22}. Specifically, the assumption made by \citet{twd22} that the Fe K line originates from the same plasma responsible for continuum emission may not hold true, as we will show later in this section.

Another factor that might contribute to the discrepancy between the temperature values reported by \citet{bmm22} and \citet{twd22} is related to the ionization stage of the plasma. The \textsc{vapec} model assumes the plasma is in CIE, which is valid if the ionization age is $\gtrsim 10^{12}$~s~cm$^{-3}$. If the plasma in \sn14c is actually out of ionization equilibrium (NEI), then fitting it with a CIE model would significantly underestimate the actual electron temperature. This is because the spectral features of a NEI plasma can only be partially recovered by a CIE model, and only by lowering the electron temperature. This diagnostic issue cannot be addressed with the BREMSS+GAUSS model, as it describes the sole visible emission line in the spectra with a phenomenological Gaussian. This line is the only emission process related to ionization age in the model. In the following part of this section, we will address this issue by fitting the observed spectra with a model that properly describes the X-ray emission of an optically thin plasma in NEI, specifically the \textsc{vnei} model in the XSPEC package.

In light of the above discussion, we repeated the analysis performed by \citet{twd22}, as described in the first part of this section, but replaced the \textsc{vapec} component with a \textsc{vnei} component (in the following the VNEI model). The detailed results of this fitting are shown in Table \ref{tab:fit_values} in Appendix \ref{app:obs}. We found that the best-fit temperatures inferred are significantly higher than those reported by \citet{twd22} and align with those found by \citet{bmm22}. Since the Fe K emission line is the only spectral feature that constrains the ionization state of the plasma, it is evident that the plasma emitting the line must be underionized. This confirms our previous hypothesis: fitting an NEI plasma with a CIE model leads to a significant underestimate of the actual electron temperature. The best-fit values of the ionization age, $\tau$, which remain constant at approximately $4 \times 10^{11}$~s~cm$^{-3}$ across all epochs, further indicate that NEI effects significantly influence the X-ray emission during the whole observed evolution. We also note that the Fe abundance is always super-solar and peaks at epoch 3 (day 606), reaching a super-solar value of $\approx 5$, before decreasing in subsequent epochs (Table \ref{tab:fit_values} in Appendix \ref{app:obs}). This suggests that the plasma component responsible for the Fe line has its greatest relative contribution around day 600.

Although using only one spectral component provides a satisfactory description of the data, it is worth to emphasize that it offers only a partial view of the complex state of the plasma. For example, if the Fe emission originates from both the CSM and the ejecta, the VNEI model would capture only a weighted average of the two plasma components.

\begin{figure*}[!ht]
  \centering
  \includegraphics[width=1\textwidth]{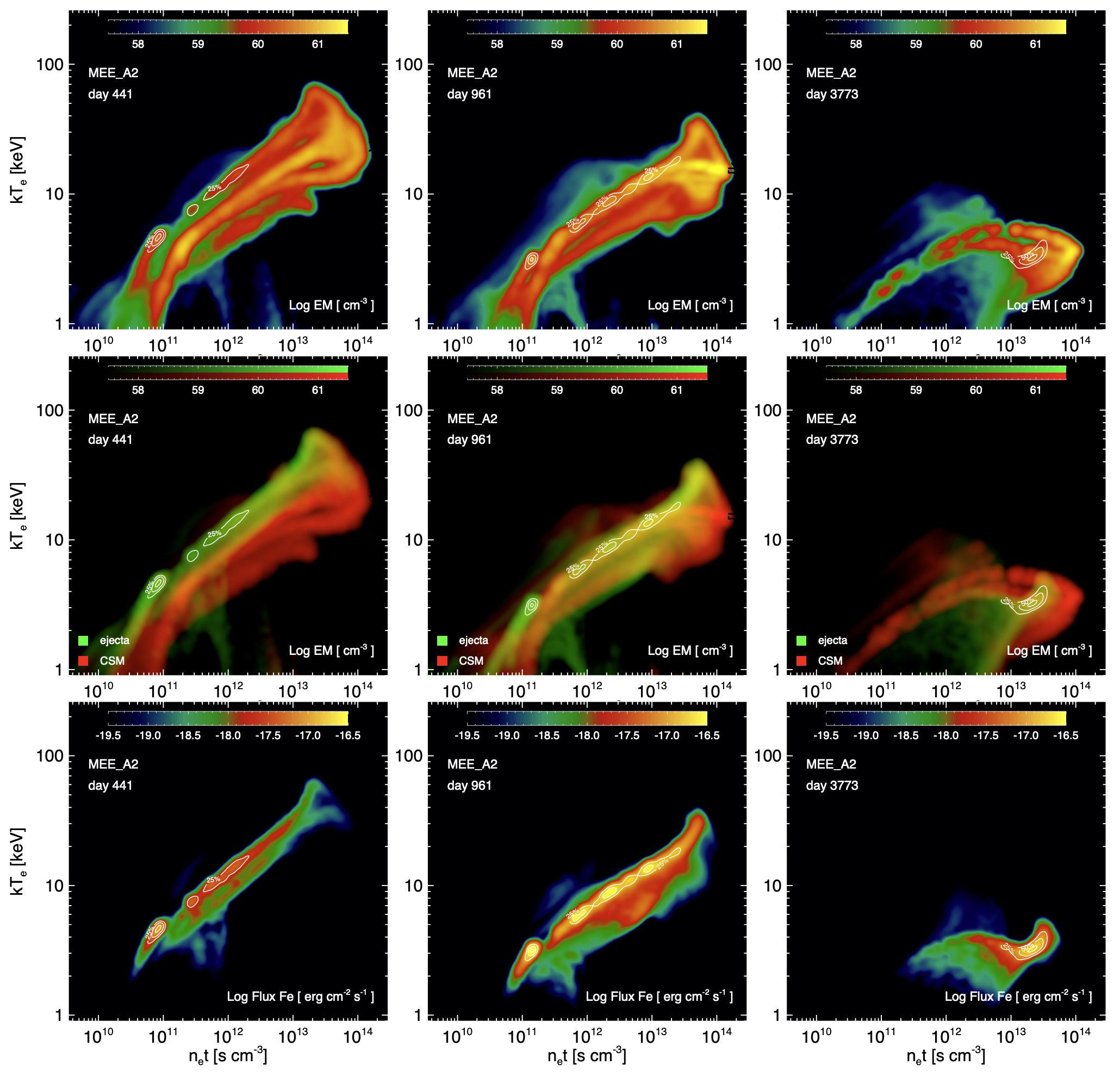}
  \caption{Upper panels: Distributions of emission measure vs. electron temperature $kT{\rm e}$ and ionization timescale $\tau = n_{\rm e} t$ at the labeled times for model MEE\_A2. Middle panels: Corresponding three-color composite images of the emission measure distributions. The colors show the contribution to emission measure from the different shocked plasma components, namely the ejecta (green) and the nebula (red). Lower panels: Maps of the continuum-subtracted Fe K$\alpha$ line flux at the same times. The white contours identify the regions where 25\%, 50\%, and 90\% of Fe K$\alpha$ line flux originates.}
  \label{dem_mee_a2}
\end{figure*}

\begin{figure*}[!ht]
  \centering
  \includegraphics[width=1\textwidth]{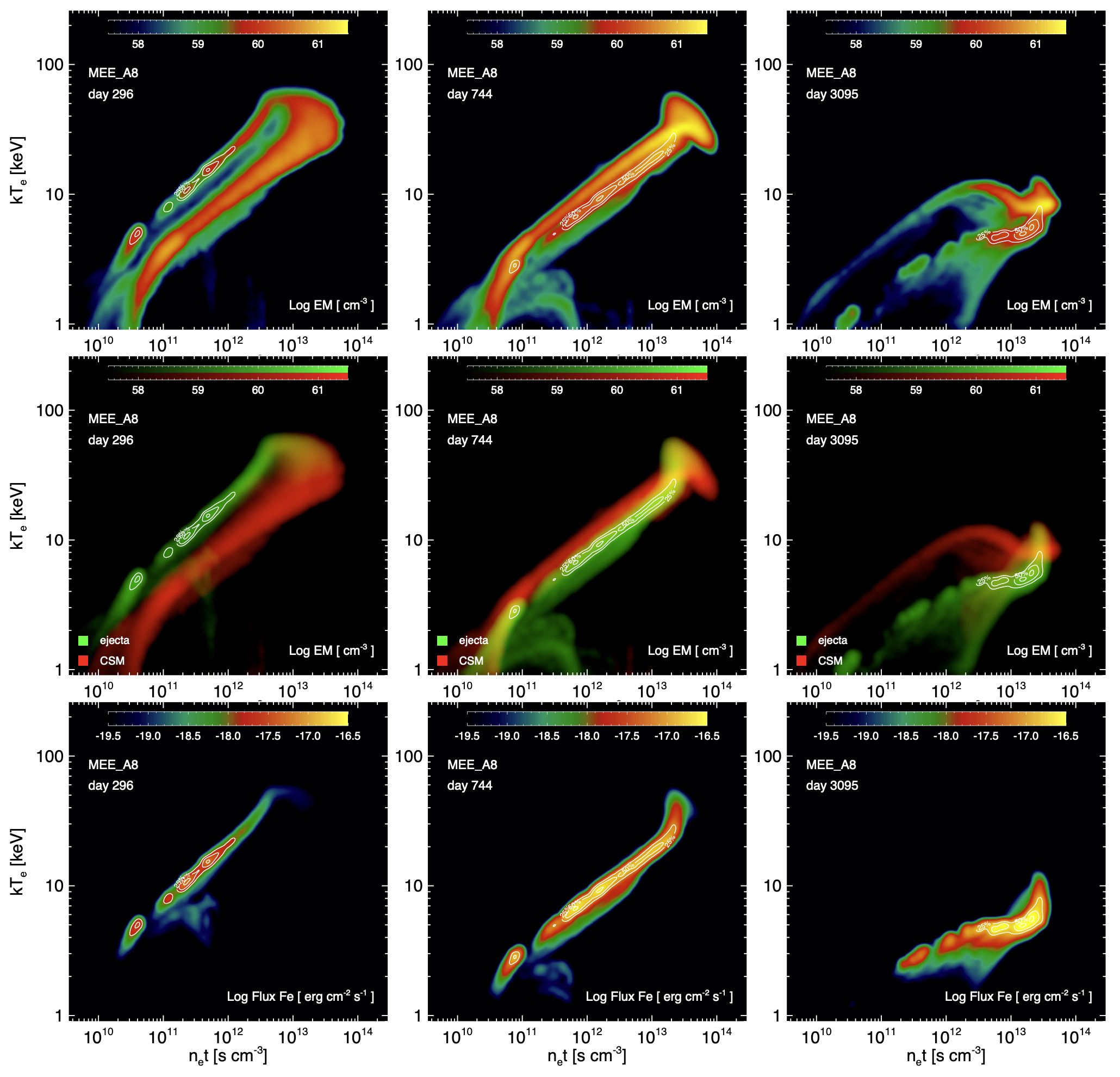}
  \caption{As in Figure~\ref{dem_mee_a2} but for model MEE\_A8.}
  \label{dem_mee_a8}
\end{figure*}

\subsection{Emission measure distribution versus temperature and ionization time from MEE models}
\label{sec:EM_maps}
The hydrodynamic models detailed in Sect.~\ref{sec:phys_wind} allow us to clarify even further the properties of the X-ray emitting plasma, providing insights into its emission measure (EM) distribution as a function of electron temperature ($kT_{\rm e}$) and ionization timescale ($\tau$). Figures~\ref{dem_mee_a2} and \ref{dem_mee_a8} illustrate the evolution of the EM distribution for models MEE\_A2 and MEE\_A8, respectively. These figures reveal complex EM distributions in both hydrodynamic models, spanning temperatures from 1 to 70 keV. They also demonstrate that the emitting plasma may deviate significantly from ionization equilibrium throughout various phases of evolution, thus supporting the need to use a \textsc{vnei} component in the spectral fitting (see Sect.~\ref{sec:fit_1t}). Both shocked material from the nebula and shocked ejecta contribute to these distributions (as observed in the middle row of the figures). However, the two components generally do not overlap, except near the peak of luminosity around day 1000. The spectral models employed by \cite{twd22} and \cite{bmm22} produce temperatures that represent averaged quantities from these distributions. 

According to our hydrodynamic models, at the onset of the interaction between the remnant and the nebula, the EM distribution exhibits two primary regions with peak EM (as seen in the left column of Figures~\ref{dem_mee_a2} and \ref{dem_mee_a8}): one at $kT_{\rm e} \approx 2-4$~keV and $\tau$ ranging from $5\times 10^{10}$ to $5\times 10^{11}$~s~cm$^{-3}$ (indicating plasma out of ionization equilibrium), and the other at $kT_{\rm e} > 10$~keV and $\tau > 10^{13}$~s~cm$^{-3}$ (suggesting plasma in equilibrium). These two regions are connected by an almost continuous distribution of shocked plasma. During this phase, the majority of the emission arises from the shocked CSM, particularly in model MEE\_A2. On average, during the early phase of evolution, the shocked ejecta display higher temperatures compared to the shocked CSM (highlighted by the green and red colors respectively in the middle row of the figures). This primarily arises because, at this stage, the shocked ejecta are those from the outer layers with lower densities, while the initially shocked portion of the nebula corresponds to regions with higher densities (refer to Figure~\ref{csm_map}).

During the rise of the X-ray lightcurve, the EM gradually increases at higher temperatures and ionization times, which explains the observed rise in average temperature in the X-ray data. As the luminosity peaks (around days $700-1000$ from the SN; see center column in Figures~\ref{dem_mee_a2} and \ref{dem_mee_a8}), the EM distributions of shocked ejecta and shocked CSM show more similar temperatures. This is particularly evident in model MEE\_A8, where the two EM distributions almost superimpose, resulting in similar average temperatures for both plasma components. This convergence is also evident in the lower right panel of Figure~\ref{lc_mee}, where the average temperatures of ejecta and CSM in model MEE\_A8 become increasingly closer over time and even coincide around day 400. 

In the following phase, the temperatures of the EM distributions gradually decrease, with the temperature of the shocked ejecta declining more rapidly. This trend occurs due to the propagation of the reverse shock through the innermost ejecta, which has higher densities, leading to lower temperatures. Again this trend is also evident in Figure~\ref{lc_mee}, where the ejecta temperature is even lower than the CSM temperature in model MEE\_A8 after day 400. During this phase, the ionization time increases, with most of the emitting plasma exhibiting values greater than $10^{13}$~s~cm$^{-3}$ after day 2000 (as shown in the right column of Figures~\ref{dem_mee_a2} and \ref{dem_mee_a8}).

\subsection{Fit with two thermal components}
\label{sec:fit_2t}

In Sect. \ref{sec:fit_1t} we demonstrated that a model consisting of a single \textsc{vnei} component satisfactorily describes the multi-epoch X-ray spectra. However, the complexity observed in the modeled EM distributions of the X-ray emitting plasma suggests that a spectral model assuming an isothermal component might not adequately represent the average properties of the emitting plasma, making data interpretation difficult. 

Motivated by the evidence of plasma components with different temperatures, we further analyzed the spectra with two (instead of one) absorbed isothermal components. We tested different spectral models, consisting of a combination of a \textsc{vapec} and \textsc{vnei} components. The latter component was also considered in light of the evidence of plasma out of equilibrium of ionization in Figures~\ref{dem_mee_a2} and \ref{dem_mee_a8} and from the results of the single-component \textsc{vnei} fitting. More specifically, we explored the following models in XSPEC format: i) \textsc{tbabs} $\times$ (\textsc{tbabs} $\times$ (\textsc{vapec} + \textsc{vapec}) + \textsc{power-law}) $\times$ \textsc{constant}; ii) \textsc{tbabs} $\times$ (\textsc{tbabs} $\times$ (\textsc{vapec} + \textsc{vnei}) + \textsc{power-law}) $\times$ \textsc{constant}. Hereafter, we will refer to these phenomenological models as 2VAPEC and VAPEC+VNEI, respectively.

\begin{figure*}[!ht]
  \centering
  \includegraphics[width=15cm]{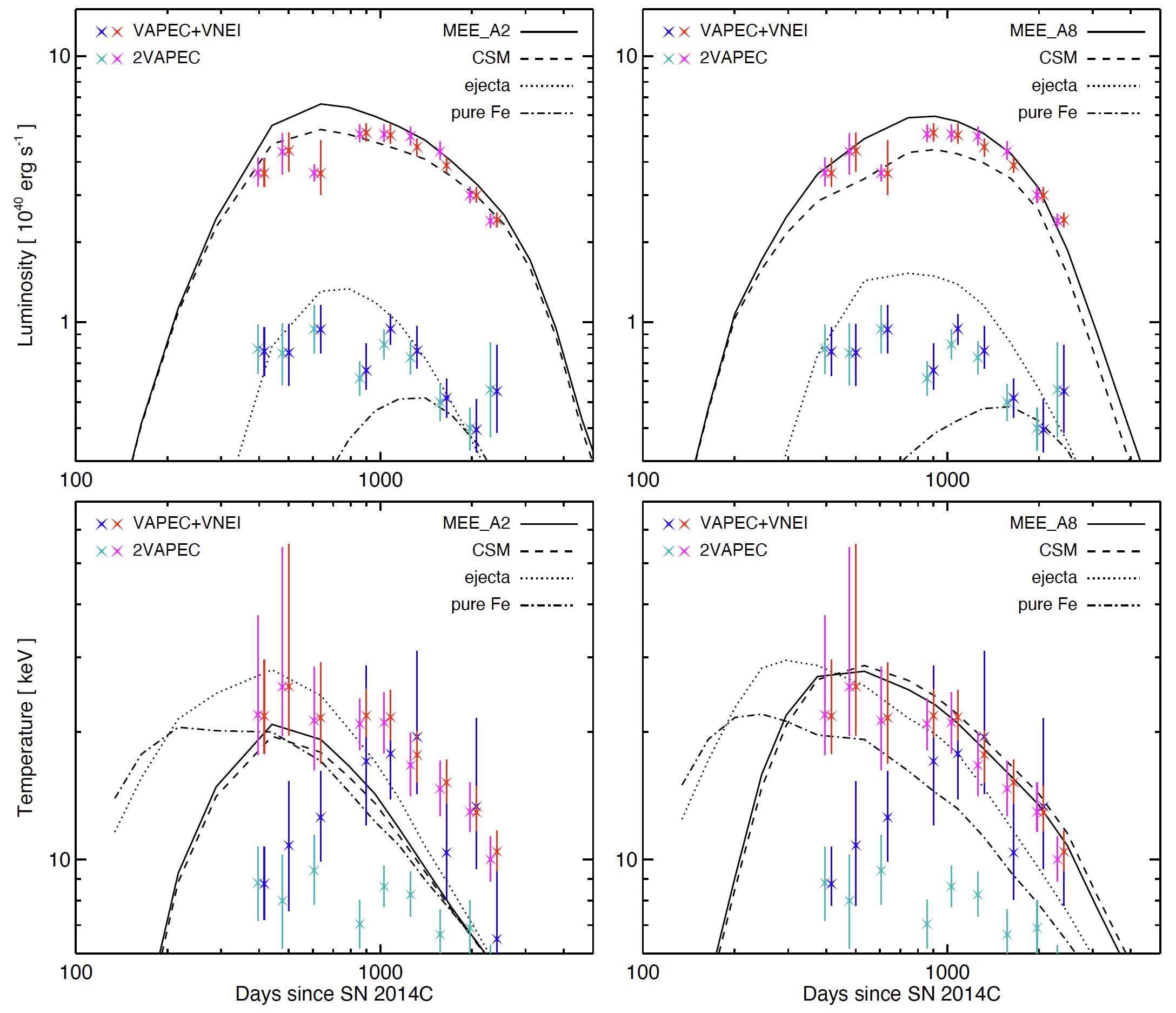}
  \caption{X-ray lightcurve in the $[0.5, 100]$ keV band (upper panels) and evolution of the average X-ray emission-weighted electron temperature (lower panels) synthesized from models MEE\_A2 (left panels) and MEE\_A8 (right panels). These are compared with the luminosity and temperature derived at various epochs from the fitting of joint Chandra and NuSTAR spectra of {\sn14c} using the spectral model 2VAPEC (cyan and magenta symbols) or VAPEC+VNEI (blue and red symbols). Dashed, dotted, and dot-dashed lines show the contribution from shocked CSM, shocked ejecta, and shocked Fe, respectively, synthesized from the MEE models.}
  \label{lc_mee_greco}
\end{figure*}

We performed an extensive analysis using the two-thermal models discussed above for all epochs where Chandra and NuSTAR observations were available (i.e., from day 396 to 2301). Our fitting procedure is detailed in Appendix~\ref{app:obs}. The best-fit values obtained using the 2VAPEC and VAPEC+VNEI models are listed in Table~\ref{tab:fit_values} in Appendix~\ref{app:obs}.

Figure~\ref{lc_mee_greco} displays the X-ray luminosities and temperatures obtained from fitting the data with the 2VAPEC or the VAPEC+VNEI model, together with the X-ray lightcurve and temperature evolution simulated from models MEE\_A2 and MEE\_A8. We note that while the 2VAPEC and VAPEC+VNEI models yield comparable results regarding X-ray luminosity and the thermal properties of the hot plasma component, they diverge significantly on the properties of the cold plasma component, described by the \textsc{vnei} component in the VAPEC+VNEI model. Specifically, the 2VAPEC model predicts temperatures of the cold component consistently below 10~keV, exhibiting a slight declining trend after day 1000. In contrast, the VAPEC+VNEI model indicates more pronounced temperature variations: initially, the temperature rises, peaking around 20~keV at approximately day 1500, followed by a subsequent decline to approximately 6~keV in the next days.

Based on the X-ray lightcurves (upper panels in Figure~\ref{lc_mee_greco}), we observe that the hot \textsc{vapec} component in both phenomenological models (magenta and red symbols in the figure) corresponds closely to the shocked CSM as described by the MEE models (dashed line). Conversely, the cold component (represented by either \textsc{vapec} in 2VAPEC or \textsc{vnei} in VAPEC+VNEI; shown in cyan and blue symbols in the figure, respectively) roughly aligns with the shocked ejecta as derived by the MEE models (dotted line in the figure). Given that the thermal properties of the cold component are expected to be constrained by the Fe K line evident in the spectra, we also report in Figure~\ref{lc_mee_greco} the luminosity and evolution of the average temperature of the shocked pure-Fe ejecta as derived by the MEE models (indicated by the dot-dashed line). 

We note that the thermal properties of the hot component closely match those predicted by model MEE\_A8 (lower right panel in Figure~\ref{lc_mee_greco}), while they are inconsistent with model MEE\_A2 (lower left panel). Specifically, MEE\_A2 notably underestimates the temperature of the hot component, leading to consistently lower temperature values. Regarding the cold plasma component, in general, its thermal properties derived from either the 2VAPEC or the VAPEC+VNEI model do not align well with the MEE models. However, it is worth noting that, after day 1000, the thermal properties of the cold component derived from the VAPEC+VNEI model closely match those of the shocked ejecta predicted by model MEE\_A8.

We note that the VAPEC+VNEI model yields results in which the hot component aligns closely with the findings of \cite{bmm22}. This agreement is not surprising because the temperature in the BREMSS+GAUSS model is determined by the bremsstrahlung component, which effectively represents the contribution to emission from the shocked CSM. In contrast, the temperature obtained using the VAPEC model (\citealt{twd22}) assumes a plasma in CIE and is influenced by the presence of the Fe K line. In fact, when we employed two \textsc{vapec} components instead of one (the 2VAPEC model), we found that the hot component once again aligns with the results of \cite{bmm22}, while the cold component (constrained by the Fe line) exhibits temperatures around $7-10$~keV (see lower panels in Figure~\ref{lc_mee_greco}). As already mentioned, the obvious interpretation is that the first component describes the shocked CSM, while the cold one the shocked ejecta. The temperature reported by \cite{twd22} therefore reflects some weighted average between the temperatures of these two \textsc{vapec} components. The same argument applies to the analysis performed with the VNEI model in Sect.~\ref{sec:fit_1t}. Since this model can account for NEI effects in the X-ray spectra, its best-fit electron temperatures are more accurate than those inferred using the VAPEC model.

Moreover, we note that the temperature of the cold component in the 2VAPEC model is significantly lower than that found in the VAPEC+VNEI model adopted (see lower panels in Figure~\ref{lc_mee_greco}). This discrepancy, as already mentioned, can be attributed to the underionization of shocked Fe caused by NEI effects, resulting in the cold \textsc{vapec} component underestimating the plasma temperature. To further explore this point, we generated maps of continuum-subtracted line flux for the Fe K$\alpha$ line (in the $[6.2,7.0]$~keV energy range) from our MEE models (lower row of Figures~\ref{dem_mee_a2} and \ref{dem_mee_a8}). The white contours indicate the regions mainly contributing to the Fe K emission.

These figures show that the line emission predominantly originates from the ejecta (compare the white contours with green areas in the middle row of the figures), with the bulk of emission occurring at temperatures ranging from $3$ to $20$~keV and ionization timescales spanning over two orders of magnitude, from $10^{11}$~s~cm$^{-3}$ (in the early phase of the interaction of the remnant with the nebula) to more than $10^{13}$~s~cm$^{-3}$ (at later phases, thousands of days after the SN). This evidence, combined with the results from the VNEI model analyzed in Sect.~\ref{sec:fit_1t}, which suggest an ionization timescale of $3-5 \times 10^{11}$~s~cm$^{-3}$ (see Table~\ref{tab:fit_values}, in Appendix~\ref{app:obs}), justifies using the \textsc{vnei} model to describe the cold component in our VAPEC+VNEI model.

According to our analysis of phenomenological models, we suggest that the CSM configuration described by model MEE\_A8 better represents the structure of the nebula around {\sn14c} than model MEE\_A2. Based on Figure~\ref{lc_mee}, we expect that a model intermediate between MEE\_A8 and MEE\_A4 would offer a closer match with the observations. Moreover, in light of the complexity of the EM distribution of the X-ray emitting plasma evidenced by the hydrodynamic models, we suggest caution when interpreting the physical meaning of the thermal fitting components.

\subsection{Comparison of actual and synthetic spectra and origin of Fe K emission line}
\label{sec:compare_spec}

The analysis of the X-ray spectra of {\sn14c} with phenomenological spectral models has shown the limits of using this approach, particularly when dealing with sources exhibiting intrinsic multi-temperature characteristics and potential deviations from equilibrium, as indicated in Figures~\ref{dem_mee_a2} and \ref{dem_mee_a8}. In such scenarios, a more effective method for extracting valuable insights into the emitting plasma involves synthesizing spectra using multi-dimensional hydrodynamic models that accurately capture the evolution and structure of the emitting source (e.g., \citealt{2019NatAs...3..236M, 2022ApJ...931..132G, 2024ApJ...961L...9S}). 

In light of this, we used model MEE\_A8 to generate synthetic Chandra and NuSTAR spectra, aiming to closely match the epochs of the actual observations. As described in Sect.~\ref{sec:synth}, the synthesis includes the photoelectric absorption, adopting the values of column density derived by \cite{bmm22} for each epoch (and reproduced by our spectral analysis), and Doppler broadening effects of emission lines due to the bulk motion of emitting material along the line of sight. The ejecta abundances are calculated self-consistently with the mass fractions of the species derived from the SN model (see Sect.~\ref{sec:sn}). Since we do not know the true abundances of the CSM, we assumed solar values consistently with the findings of \cite{2015ApJ...815..120M}. The exposure times assumed for the synthetic spectra are 30~ks for Chandra and 50~ks for NuSTAR, similar to the actual observations.

\begin{figure}[!t]
  \centering
  \includegraphics[width=8cm]{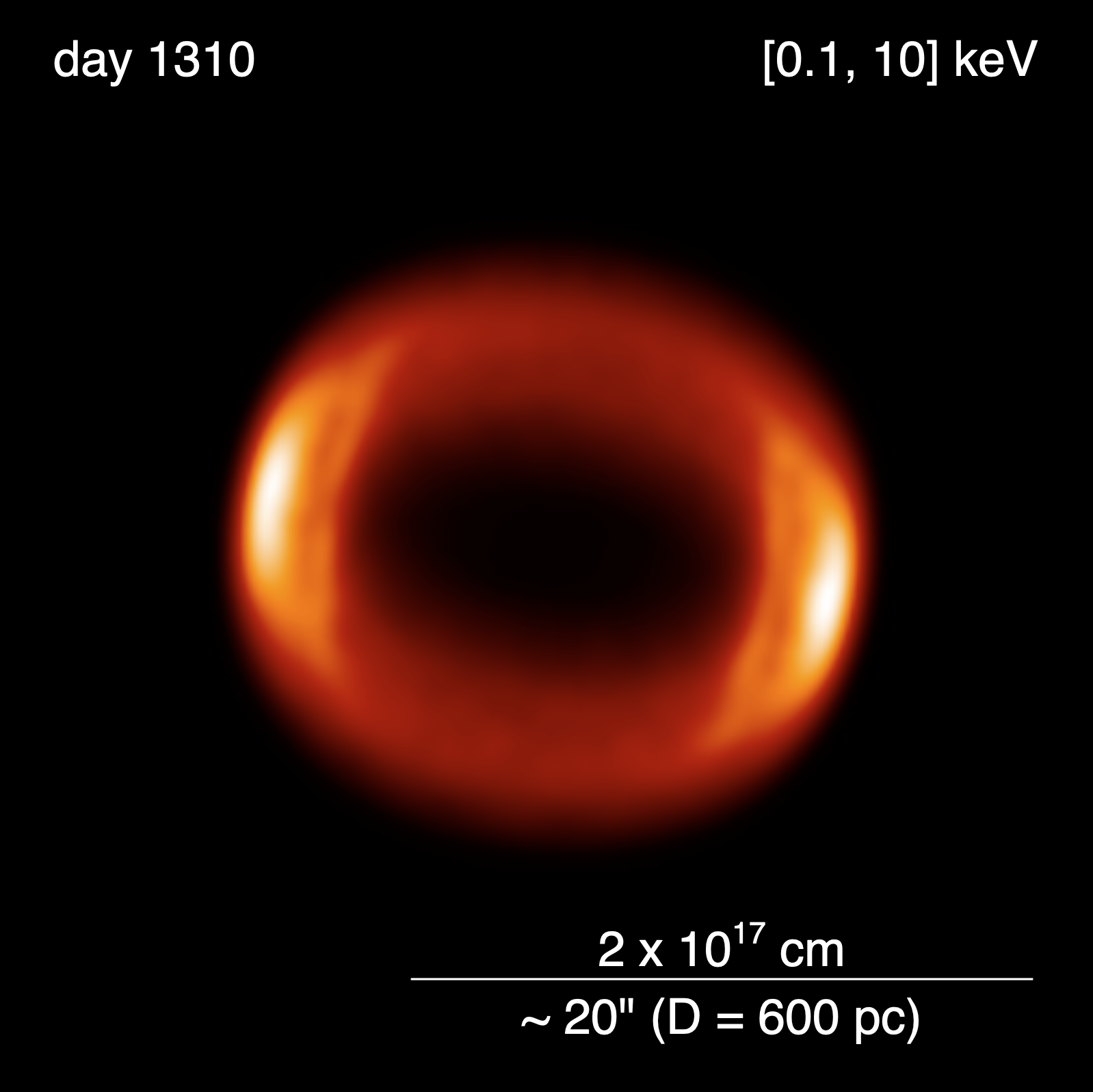}
  \caption{Synthetic X-ray emission map from model MEE\_A8 in the $[0.1-10]$ keV band at day 1310 as it would be captured by Chandra if {\sn14c} were located at a distance of 600~pc.}
  \label{xray_image}
\end{figure}

\begin{figure*}[!ht]
  \centering
  \includegraphics[width=1\textwidth]{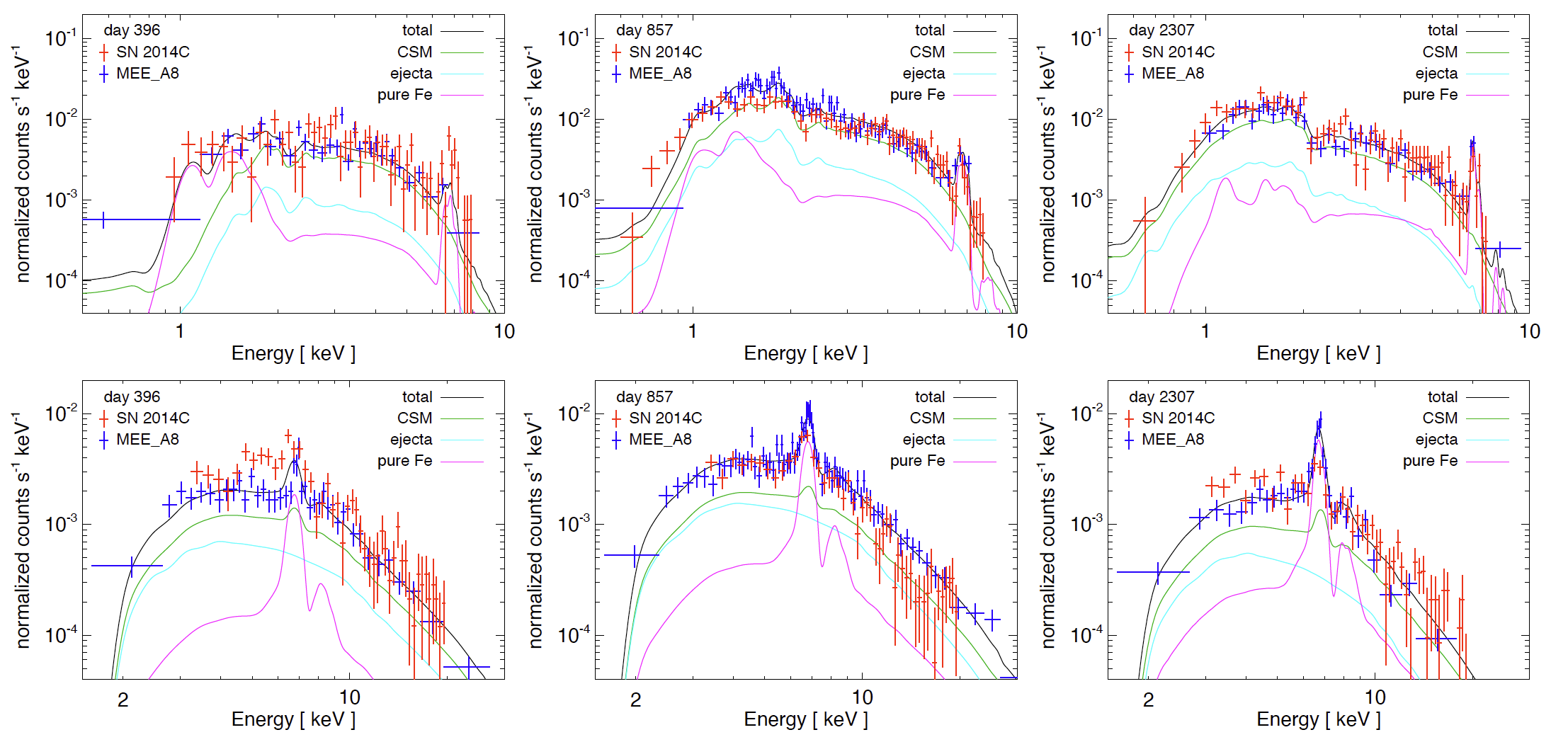}
  \caption{Comparison of actual (red symbols) and synthetic (blue symbols) spectra of {\sn14c} collected with Chandra (upper panels) and NuSTAR (lower panels) at the labeled times. The synthetic spectra have been derived from model MEE\_A8 at epochs close to those of observations, assuming an exposure time of 30~ks for Chandra and 50~ks for NuSTAR. The synthetic ideal spectra (black lines) and the contributions from different plasma components are also reported: shocked CSM (green) shocked ejecta (without the contribution of iron; lightblue) and shocked pure-Fe ejecta (magenta).}
  \label{spectra}
\end{figure*}

\begin{figure*}[!ht]
  \centering
  \includegraphics[width=1\textwidth]{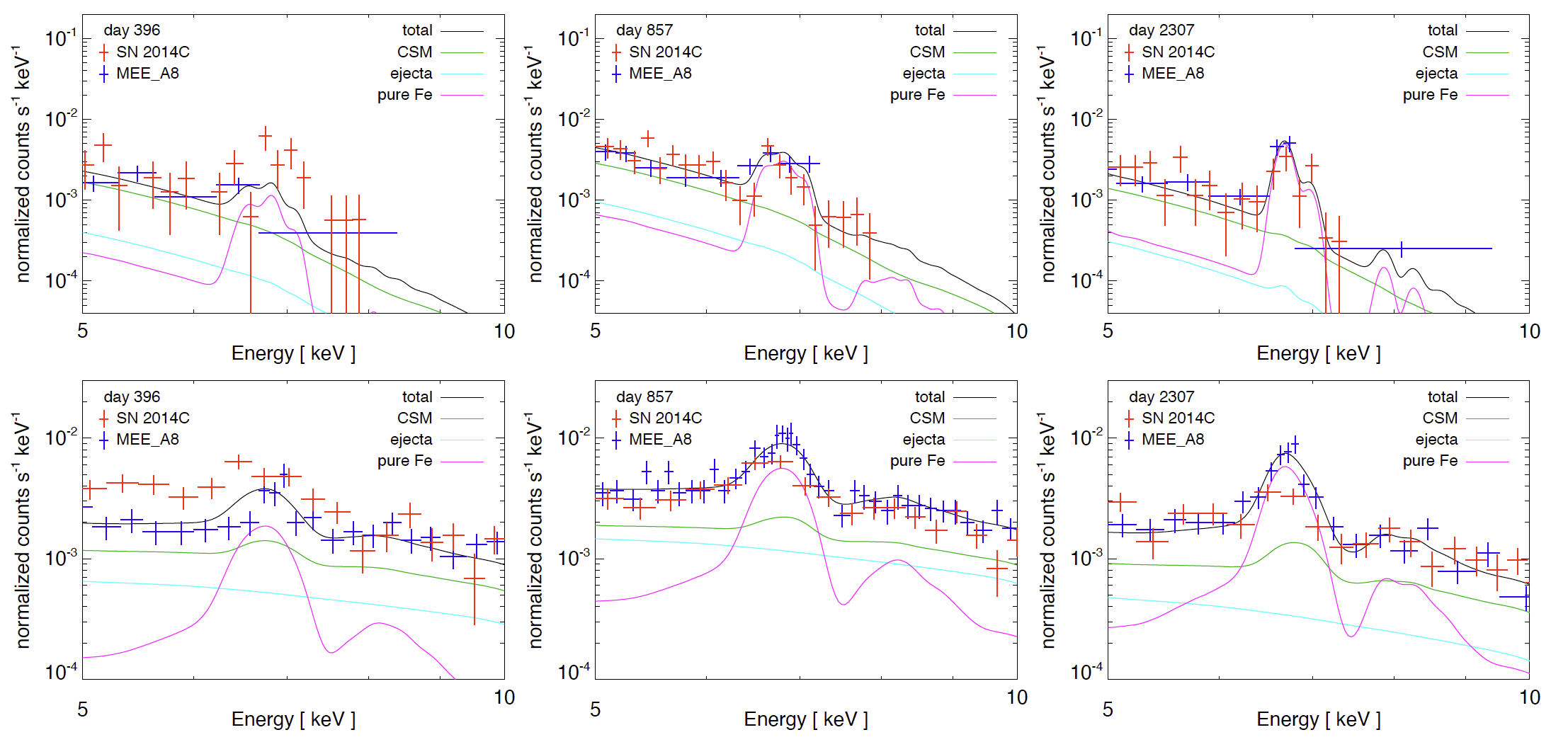}
  \caption{Spectra as in Figure~\ref{spectra} but restricted to the $[5,10]$~keV band to show the details of the Fe K$\alpha$ line.}
  \label{fe_line}
\end{figure*}

Figure~\ref{xray_image} shows an example of X-ray emission map in the $[0.1,10]$ keV band as it would be captured by Chandra if {\sn14c} were located at a distance of 600~pc (instead of 14.7~Mpc). We arbitrarily rotated the system\footnote{The true orientation of the system with respect to the observer is unknown.} about the $x$-axis by $45^\circ$ and about the $z$-axis by $10^\circ$ to clearly reveal the remnant morphology in the X-ray band. As anticipated, the predominant X-ray emission originates from the shocked material of the nebula. The two bright limbs evident in the image are the result of projection effects, as the amount of X-ray emitting material along the line of sight is the highest in these regions. Throughout the entire period of remnant-nebula interaction, the remnant morphology exhibits a ring-like structure in the X-ray band, resembling the X-ray morphology of SN\,1987A (e.g., \citealt{2016ApJ...829...40F, 2024ApJ...966..147R}).

\begin{figure*}[!ht]
  \centering
  \includegraphics[width=1\textwidth]{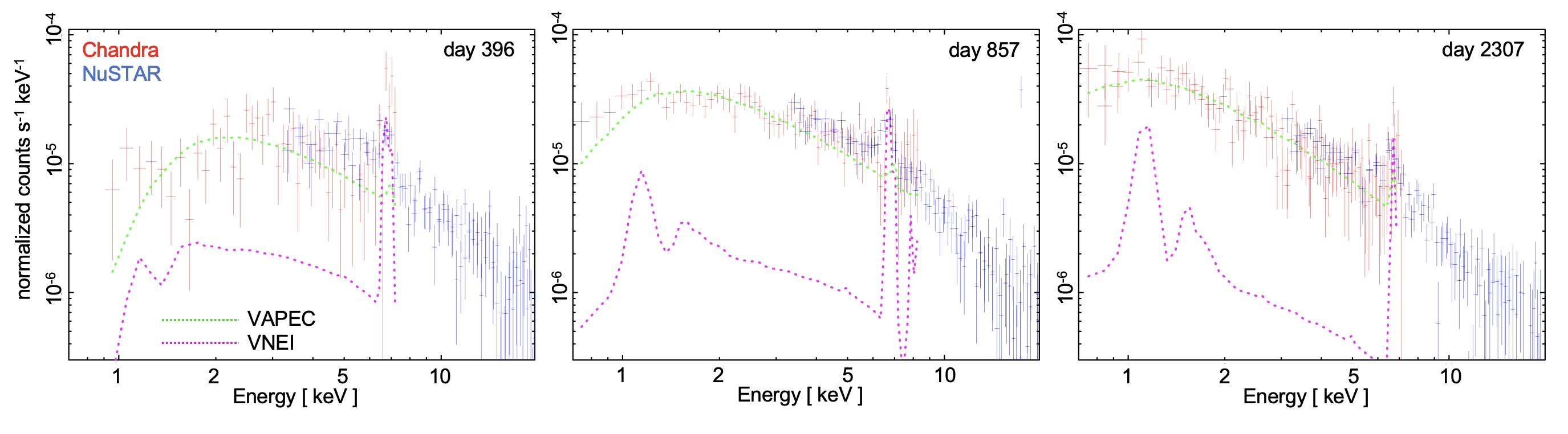}
  \caption{Chandra (red data points) and NuSTAR (blue data points) spectra collected at the labeled times together with the VAPEC+VNEI models (green and magenta curves) resulting from the spectral fitting (see Sect.~\ref{sec:fit_2t} and Appendix~\ref{app:obs}).}
  \label{fit_vapec-vnei}
\end{figure*}

Figures~\ref{spectra} and \ref{fe_line} present a comparison between the actual and synthetic observations in three selected epochs (see Appendix \ref{app:spectra} for all epochs). Additionally, the model allowed us to discern the contribution of different shocked plasma components to the spectra. The figure illustrates the contributions of shocked CSM, shocked ejecta (excluding the contribution of pure-Fe ejecta), and shocked pure-Fe ejecta. We emphasize that no spectral fitting or renormalization of synthetic spectra was performed to match the observed ones; rather, the synthetic spectra are directly overlaid onto the actual observations. It is noteworthy that the synthetic spectra well reproduce both the Chandra and NuSTAR observations at all epochs. In other words, a single physical model (MEE\_A8) successfully reproduces the observed spectra across all available epochs, effectively capturing the entire interaction of the SNR with the dense circumstellar nebula.

As expected, the synthetic spectra show that the majority of the X-ray emission originates from the shocked CSM (see green line in Figure~\ref{spectra}). However, during the initial phase of the remnant-nebula interaction (day 396 in the figure), the emission below 2~keV in Chandra spectra is predominantly attributed to the contribution from shocked pure-Fe ejecta (magenta line in the upper left panel of the figure). This soft X-ray excess likely arises from a ``forest'' of Fe lines originating from ionization states XIX to XXIV. This excess reduces as the evolution progresses, becoming lower than the contribution from shocked CSM after day 606 (see Appendix~\ref{app:spectra}). This is most likely due to the increasing temperature and ionization time of the emitting plasma (see Figure~\ref{dem_mee_a8}). It is also worth mentioning that the model slightly overestimates the emission at energies between 1 and 2~keV in the period between days 477 and 1257 (see Appendix~\ref{app:spectra} and Figure~\ref{all_chandra}). This might indicate that we overestimated the abundances adopted for the CSM.

A striking feature in all observed spectra is the distinct emission line in the energy interval between 6.2 and 7.0~keV (see Figures~\ref{spectra} and \ref{fe_line}). Our model shows that, starting from day 477, this feature can be attributed almost entirely to the shocked pure-Fe ejecta (magenta line in the figures), thus confirming it to be the Fe K$\alpha$ line (6.7~keV), as proposed by \cite{2017ApJ...835..140M} and \cite{bmm22}. The zoom in the energy interval between 6.2 and 7.0~keV (see also Figures~\ref{all_chandra_line} and \ref{all_nustar_line}) shows that the MEE\_A8 model slightly overestimates the line flux in synthetic NuSTAR spectra but not in synthetic Chandra spectra. This may be due to cross-calibration effects between the two instruments. 

Nevertheless, the figures demonstrate that the MEE\_A8 model successfully reproduces both the flux and the broadening of the Fe K$\alpha$ line throughout the entire remnant-nebula interaction, without relying on any ad-hoc assumptions. In fact the Fe line flux in all the synthetic spectra generated by our model originates naturally from the shocked ejecta with the abundance of species calculated in the SN explosion. Notably, we did not impose any specific iron abundance in the ejecta; instead, the Fe content was derived self-consistently from the stellar evolution and SN models (see Sect.\ref{sec:sn}). In the MEE\_A8 model, as well as in the other models presented, a portion of the pure Fe-rich ejecta becomes shocked due to the reflected shock generated by the remnant's interaction with the dense nebula.

It is worth to mention that the Fe K line has also been observed in other interacting SN events, including SNe\,1996cr (\citealt{2010MNRAS.407..812D}), 2010jl (\citealt{2012ApJ...750L...2C}), 2009ip (\citealt{2014ApJ...780...21M}), and more recently, SN\,2023ixf (\citealt{2023ApJ...952L...3G}). The conventional interpretation is that the Fe K emission arises either from a metallicity of approximately 5~Z$_{\rm e}$ or from a clumpy/multiphase medium (e.g., \citealt{2015ApJ...815..120M, 2017ApJ...835..140M, bmm22}). However, our study proposes an alternative or complementary scenario where the Fe K line may originate from the shocked ejecta itself.

To further explore this possibility, we compared the results of the comparison between actual and synthetic spectra (Figures~\ref{spectra} and \ref{fe_line}) with the spectral fitting results obtained using the VAPEC+VNEI model, as discussed in Sect.~\ref{sec:fit_2t} (see also Appendix~\ref{app:obs}). As an example, Figure~\ref{fit_vapec-vnei} presents the fitting results for the three epochs shown in Figure~\ref{spectra}. Note that the Chandra and NuSTAR spectra in the figure have been renormalized for the simultaneous spectral fitting to account for differences in their effective area using the XSPEC task \textsc{setplot area}; this is why the fluxes in this figure differ from those shown in Figures~\ref{spectra} and \ref{fe_line}. 

Through spectral fitting, we determined that the Chandra and NuSTAR spectra can be described by two isothermal components: a \textsc{vapec} model representing a hot plasma in CIE with solar abundances, and a \textsc{vnei} model accounting for a cooler component in NEI with an ionization timescale of $\tau \leq 10^{12}$~s~cm$^{-3}$ (see Table~\ref{tab:fit_values}). While constraining the Fe abundance in the cooler component is challenging (due to the presence of only one line and the signal-to-noise ratio of the data), the fitting suggests super-solar values (A$_{\rm Fe} > 1$), according to the findings of the fitting with the VNEI model (see Table~\ref{tab:fit_values}). Moreover, a good fitting of the data can be obtained also when assuming A$_{\rm Fe} = 100$, consistent with pure Fe ejecta (for further details, see Appendix~\ref{app:obs}), as shown in Figure~\ref{fit_vapec-vnei}. Interestingly, this figure clearly shows that the colder \textsc{vnei} component is responsible for the Fe line emission in the $[6,8]$~keV band, aligning perfectly with the results from the MEE\_A8 model. These findings substantiate the interpretation that the hot \textsc{vapec} component reflects the average properties of the shocked CSM, while the cold \textsc{vnei} component represents the shocked ejecta. Furthermore, these results corroborate the scenario proposed by the MEE\_A8 model: the Fe K line emission originates predominantly from the shocked ejecta.

It is important to note that the core-collapse SN model used in this study is 1D, thus assuming a spherically symmetric explosion (see Sect.~\ref{sec:sn}). Consequently, this model neglects the effects of instabilities, such as the standing accretion shock instability or the effects of convective overturning from neutrino heating, occurring in the initial seconds after core bounce (e.g., \citealt{2013A&A...552A.126W, 2017hsn..book.1095J}). These instabilities lead to the development of large-scale asymmetries in the distribution of ejecta (for instance the development of extended Fe-rich plumes as seen, e.g., in \citealt{2017ApJ...842...13W}), and are responsible for efficient mixing of different ejecta layers before shock breakout (e.g., \citealt{2020ApJ...888..111O}). Moreover, the comparison of the synthetic bolometric lightcurve with the actual one shows some discrepancies up to a few tens of percent (see Figure~\ref{bolometric_lc}) and this may lead to some effects on the forward shock velocity and mass of Fe-rich ejecta. 

\begin{figure}[!t]
  \centering
  \includegraphics[width=8.5cm]{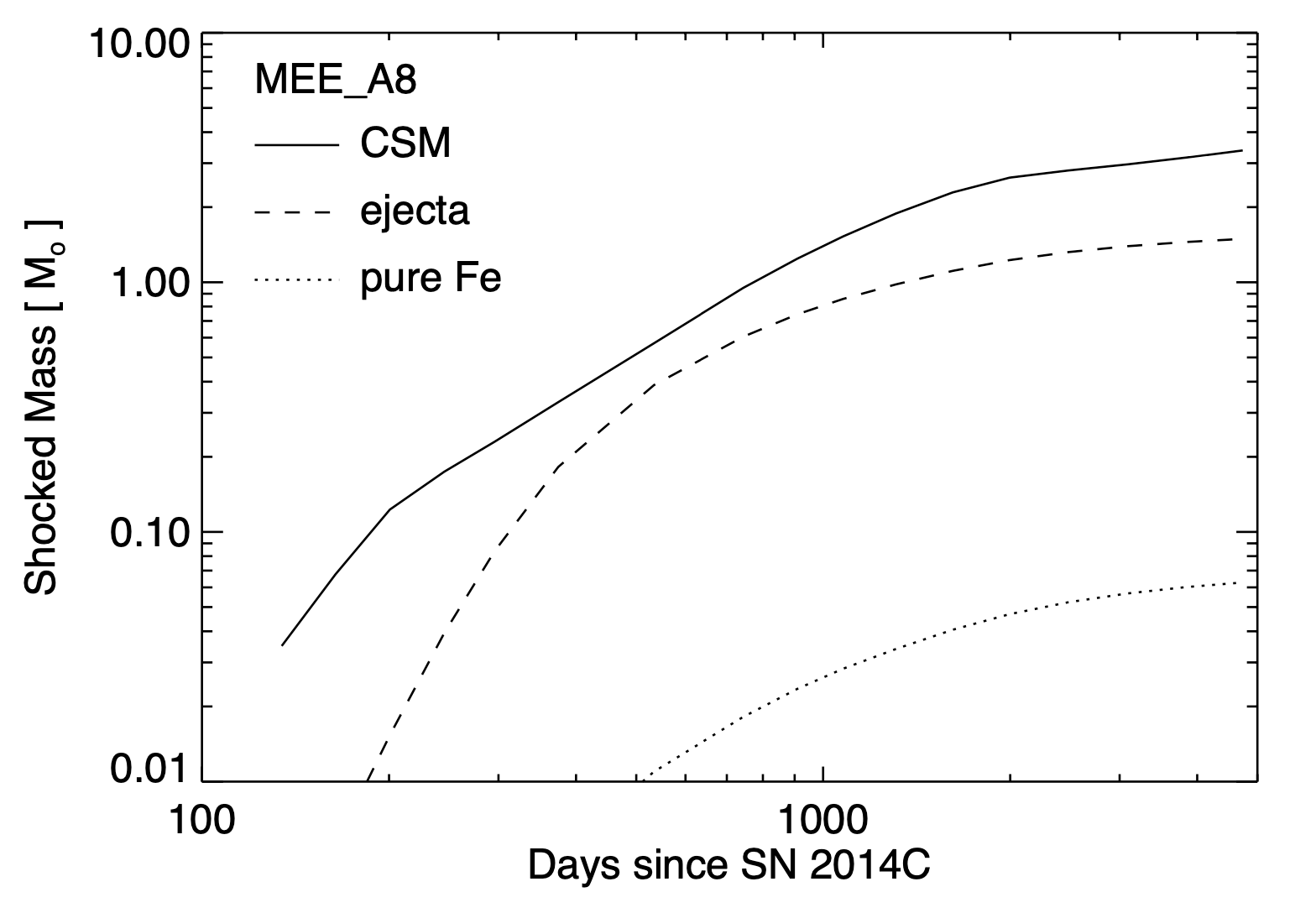}
  \caption{Masses of the shocked plasma components contributing to the observed X-ray emission from \sn14c, as derived from model MEE\_A8: shocked CSM (solid line), shocked ejecta (dashed line), and shocked pure-Fe ejecta (dotted line).}
  \label{shocked_mass}
\end{figure}

Nevertheless, our analysis reveals that the MEE\_A8 model consistently reproduces the flux and broadening of the Fe K line throughout the entire evolutionary process, suggesting that it captures the average properties of Fe-rich ejecta distribution, despite the above limitations. Consequently, the model can provide a reliable and quantitative estimate of the mass of Fe-rich ejecta shocked during the remnant-nebula interaction. Figure~\ref{shocked_mass} presents the masses of the various plasma components contributing to the observed X-ray emission, as derived from the model. Our findings indicate that more than $1\,M_{\odot}$ of CSM, along with a smaller amount of ejecta material, was shocked after day 1000, following the peak of X-ray luminosity. Additionally, during this period, a mass ranging between 0.03 and $0.06\,M_{\odot}$ of pure-Fe ejecta was also shocked.

\section{Summary and conclusion}
\label{sec:summary}
We presented 3D hydrodynamic simulations which describe the evolution of {\sn14c} from the SN event to the interaction of the remnant with a dense circumstellar nebula. The simulations are the result of coupling a stellar evolution model, which accurately replicates the main properties of the progenitor star at collapse based on observational data (\citealt{2015ApJ...815..120M, 2020MNRAS.497.5118S}), with a core-collapse SN model that generates kinetic energy of ejecta ($1.8\times 10^{51}$~erg), ejecta mass ($1.7\,M_{\odot}$), and bolometric lightcurve (with peak emission of $\approx 4\times 10^{42}$~erg~s$^{-1}$) consistent with those inferred from observations (\citealt{2017ApJ...835..140M}), and with a SNR model similar to that used to model SN\,1987A (\citealt{2020A&A...636A..22O}). The modeling includes the evolution of the inhomogeneous CSM, sculpted by the progenitor star's outflows in the millennia leading up to the SN explosion. The aim was to reconstruct the 3D structure and density distribution of the pre-SN CSM surrounding {\sn14c} through a quantitative comparison of the model results with X-ray observations and to constrain the mass-loss history of the progenitor star in the final phases of its life cycle. 

The 3D simulations of the SNR start shortly after the shock breakout at the stellar surface, which occurs approximately $\approx 100$~s post-core-collapse, and span about 15 years, predicting the evolution of {\sn14c} until 2030. During this time, the remnant travels through the inhomogeneous CSM derived from dedicated 2D and 3D hydrodynamic simulations, which capture the mass-loss history of the progenitor star in the millennia leading up to the collapse. 

We found that one of the models closely reproduces the observed X-ray lightcurve, spectra, and evolution of average temperature, as derived from the analysis of Chandra and NuSTAR observations (\citealt{twd22, bmm22}). From this model, we were able to constrain the 3D structure of the circumstellar nebula and the mass-loss history of the progenitor star. Our main findings are summarized as follows.

(1) \underline{\textit{3D structure of the pre-SN CSM.}} At the time of core-collapse, the CSM around {\sn14c} was characterized by a cavity carved out by the fast and tenuous outflow emanating from the naked helium star, extending in the equatorial plane up to a distance of approximately $4.3\times 10^{16}$~cm, and by a dense toroidal nebula at larger distances. According to our favourite model, the nebula extended from $4.3 \times 10^{16}$~cm to $15.0 \times 10^{16}$~cm in the equatorial plane, with a thickness of approximately $11.6 \times 10^{16}$~cm at a distance of about $6 \times 10^{16}$~cm from the center of the explosion. The nebula had a peak density of $\approx 3 \times 10^6$~cm$^{-3}$ at the inner boundary, gradually decreasing as $\approx r^{-2}$ at greater distances. The geometry and density distribution of the modeled nebula are consistent with the properties of the nebula inferred from the analysis of observations collected in different wavelength bands (e.g., \citealt{2017ApJ...835..140M, 2019ApJ...887...75T}).

(2) \underline{\textit{Mass-loss history of the progenitor star.}} The nebula surrounding {\sn14c} formed as a result of intense mass-loss from the progenitor star in the millennia leading up to the SN explosion. X-ray observations of {\sn14c} are well reproduced if the mass-loss eruption began not later than 5000 years before the SN, with the mass-loss rate increasing from $\sim 10^{-4}\,M_{\odot}$~yr$^{-1}$ (or lower values) to $\sim 8\times 10^{-4}\,M_{\odot}$~yr$^{-1}$ over a few centuries\footnote{The X-ray lightcurve does not allow us to constrain well the onset of the eruption.}. It is interesting that this finding closely matches the mass-loss rate inferred by \cite{2019ApJ...887...75T} based on their analysis of infrared data. This eruption phase ended roughly 1000 years before the SN, with the mass-loss rate rapidly decreasing to below $10^{-5}\,M_{\odot}$~yr$^{-1}$ within a few centuries. In the final phase of evolution, the fast and tenuous outflow emitted by the progenitor helium star gradually displaced the dense outflow from the preceding mass eruption, creating the nebula and a cavity in the surrounding environment of the star constrained by our favourite model at the time of core-collapse. During the period of mass-loss from the progenitor star a total of $\approx 2.5\,M_{\odot}$ of stellar material were released into the CSM, forming the nebula.

(3) \underline{\textit{Origin of the Fe K line.}} Our study reveals that the X-ray emitting plasma in {\sn14c} exhibits intrinsic multi-temperature characteristics and potential deviations from equilibrium. Analysis of the available spectra indicates that they can be described by two distinct isothermal components: the hottest component in CIE with solar abundances, and the colder component in NEI with super-solar abundances, which primarily accounts for the Fe K line emission in the $[6,8]$~keV band. We interpret the hot component as originating from the shocked CSM and the cold component from the shocked ejecta. This interpretation is corroborated by our preferred hydrodynamic model, which accurately reproduces the Chandra and NuSTAR spectra, including the Fe K line feature, throughout the remnant's evolution. Specifically, the model shows that the shocked CSM contributes predominantly to the emission detected by Chandra and NuSTAR through thermal bremsstrahlung. In contrast, the shocked ejecta are responsible for the Fe K line emission in the $[6,8]$~keV band and a soft component below $1.5$~keV, which arises from a forest of Fe lines in ionization states XIX to XXIV, especially during the early phases of evolution. According to our model, approximately $0.05\,M_{\odot}$ of pure Fe ejecta were shocked during the interaction between the remnant and the nebula. Since the Fe K line could also originate from either super-solar metallicities in the CSM or a clumpy/multiphase medium (e.g., \citealt{2015ApJ...815..120M, 2017ApJ...835..140M, bmm22}), it would be interesting to explore the alternative scenarios with observational data in future studies. Such investigations could help determine whether the Fe K line is primarily a result of the shocked ejecta or CSM.

(4) \underline{\em Origin of temperature discrepancy in literature.} Our analysis identifies NEI effects as the primary cause of the discrepancy in the average temperature values of the X-ray emitting plasma derived by \cite{twd22} and \cite{bmm22}. In fact, a one-component model of plasma out of ionization equilibrium with enhanced Fe abundance provides an equally good description of the data as a model comprising thermal bremsstrahlung plus a Gaussian component for the Fe K line. According to the hydrodynamic model, the Fe K line is expected to arise from shocked ejecta, while the thermal bremsstrahlung dominating the spectra at other energies is expected to arise from the shocked CSM. This suggests that at least two isothermal plasma components should be employed in the spectral fitting to adequately describe the data. This approach separates the contribution from the shocked ejecta, responsible for the Fe K line emission, from that of the shocked CSM, which accounts for most of the emission in the $[0.5,100]$~keV band. We found that the analysis by \citet{bmm22}, who used a model comprising thermal bremsstrahlung plus a Gaussian component for the Fe line, yielded results similar to those of the hot component in our fitting with two isothermal plasma components. This suggests that their analysis effectively described the properties of the shocked CSM, which is dominated by thermal bremsstrahlung. Conversely, the analysis by \cite{twd22}, based on a collisionally ionized diffuse gas in ionization equilibrium, produced results that were intermediate between those of the cold and hot components in our fitting.

\bigskip
We note that the mass-loss history constrained by our preferred MEE\_A8 model is somewhat consistent with that found for binary systems that experience Case-C mass transfer via common envelope (CE) ejection (as, for instance, models bpass1 and bpass2 in \citealt{2020MNRAS.497.5118S}). The CE evolution is a plausible scenario and could indeed be quite common. In fact, approximately 70\% of massive stars are found in binary systems, and of these, 30\% are likely to undergo a merger event (\citealt{2012Sci...337..444S}). Therefore, it is possible that the progenitor of {\sn14c} followed this evolutionary path. However, we also note that it is unlikely that a merger event involving a star with a moderately high mass could lead to an eruptive phase with a mass-loss rate of the order of $10^{-3}\,M_{\odot}$~yr$^{-1}$ that lasts for approximately 4000 years, as suggested by our study.

In the models analyzed by \cite{2020MNRAS.497.5118S}, mass transfer via CE ejection begins near the end of core carbon burning, approximately 20,000 years before the SN, which is considerably earlier than the 5,000 years predicted by our model. On the other hand, it is important to note that the timescale of CE evolution in {\em bpass} models may be overestimated, as these models are 1D and rely on a numerical approach designed to prevent instabilities (see \citealt{2017PASA...34...58E}). The CE phase is inherently a 3D process without spherical symmetry. The 1D approach of {\em bpass} models may overlook critical aspects of the inspiral process, such as dynamical drag, tidal torques, and asymmetric mass loss, leading to an inaccurate CE timescale (however, see \citealt{2022ApJ...937L..42H} for a discussion on CE timescales for massive star systems).

The mass-loss rate required in the CE scenario of \cite{2020MNRAS.497.5118S} reaches values similar to those found in model MEE\_A8 ($\approx 10^{-3}\,M_{\odot}$~yr$^{-1}$). The main difference between our model and this CE scenario is that these authors considered an additional phase (not required by our model) in which a violent mass ejection, raising the mass-loss rate to values of the order of $3–5 \times 10^{-2}\,M_{\odot}$~yr$^{-1}$ for $2-3$ decades, occurred approximately 200 years before the SN event. However, we note that such high mass-loss rates over a short time interval would require an additional mechanism not accounted for in {\em bpass} models. The violent eruption proposed by \cite{2020MNRAS.497.5118S} was needed to explain the presence of a dense H-rich shell, with an estimated mass of $1–1.5\,M_{\odot}$, as suggested by observational studies (\citealt{2015ApJ...815..120M, 2017ApJ...835..140M}). We note that the models of \cite{2020MNRAS.497.5118S}, being 1D, assume spherically symmetric outflows from the star. Interestingly, by introducing asymmetry in the stellar outflows, as done in our models, the solution space expands. In these asymmetric models, the formation of dense inner regions does not necessarily require a new eruption from the progenitor.

Nevertheless, the high mass-loss rates proposed by \cite{2020MNRAS.497.5118S} over just a few decades resemble those expected for a stellar merger event, such as the one responsible for AT\,2018bwo. In that case, the primary star (estimated to have a mass around $15\,M_{\odot}$, similar to the progenitor of {\sn14c}) underwent a dramatic increase in mass-loss, ranging between $3\times 10^{-3}\,M_{\odot}$~yr$^{-1}$ and $5\times 10^{-2}\,M_{\odot}$~yr$^{-1}$ in the decade preceding the merger (\citealt{2021A&A...653A.134B}). Such extreme rates, driven by the gravitational interaction and eventual coalescence of the two stars, are far higher than the mass-loss rates we derived in our study. On the other hand, our findings suggest that the observed X-ray emissions from {\sn14c} can be accurately reproduced by our model without invoking the additional mass-loss mechanisms that would be expected if a merger had occurred. This is a critical point because it implies that, while a CE evolution scenario (where the companion star orbits within the envelope of the primary) remains plausible, the actual merger of the two stars appears unlikely in the case of {\sn14c}. Thus, our results support a scenario where the progenitor system experienced significant interaction, possibly through CE evolution, but did not culminate in a full merger event, thereby leading to more moderate mass-loss rates consistent with observations.

\cite{2020MNRAS.497.5118S} also discussed a second scenario that might explain the observations of {\sn14c}: a binary system with a larger initial secondary-to-primary mass ratio and a shorter initial orbital period compared to the CE scenario, experiencing Roche lobe overflow (RLOF; model bpass3 in \citealt{2020MNRAS.497.5118S}). According to the RLOF scenario, two episodes of mass transfer occurred: the first when the primary star was undergoing H-shell burning, approximately 2 million years before the SN, and the second when the star had finished core helium burning, between 600 and 10,000 years before the SN. However, the mass-loss rate in the RLOF scenario is much lower than that in the CE scenario and that found in our hydrodynamic models.

Additionally, numerical simulations describing mass transfer in binary systems indicate that a structured, high-density circumstellar environment with geometry similar to that found around {\sn14c} can result from the tidal ejection of a portion of the star's H-rich envelope, predominantly along the equator, during the CE phase preceding the merger of a binary system (e.g., \citealt{2012ApJ...744...52P, 2012ApJ...746...74R, 2015MNRAS.446.1716C, 2019MNRAS.484..631R, 2020A&A...644A..60S, 2023LRCA....9....2R}). This similarity, combined with evidence that our findings align more closely with the CE scenario than the RLOF scenario discussed in \cite{2020MNRAS.497.5118S}, suggests that the 3D geometry and density distribution of the CSM around {\sn14c}, as well as the mass-loss history of its progenitor star, are consistent with a scenario where the star was stripped by binary interaction through CE evolution. This scenario offers a coherent explanation for the observed characteristics of {\sn14c} in the X-ray band, aligning with the complex mass-loss processes expected during the final stages of massive star evolution within a binary system.

Observations of {\sn14c} demonstrate that the interactions between SNRs and the CSM provide valuable insights into eruptive mass-loss events occurring centuries to millennia before the SN explosion. The analysis of these interactions through the development of 3D hydrodynamic models are able to follow the evolution from the progenitor star to the SN and SNR and their comparison with observations can enhance our understanding of the final stages of massive star evolution and the complex mechanisms governing their mass-loss. By studying these interactions, we gain a deeper insight into the processes leading up to core-collapse SNe, ultimately advancing our knowledge of the life cycles of massive stars.



\bigskip
\bigskip
\begin{acknowledgments}
We thank the anonymous referee for useful suggestions that have allowed us to improve the paper. We are grateful to Andrea Pastorello (INAF), Miguel-Angel Aloy (University of Valencia), Yuki Takei (Kyoto University), and Toshikazu Shigeyama (University of Tokyo) for the insightful discussions on the physics of interacting supernovae and their progenitor stellar systems. The \pluto\ code is developed at the Turin Astronomical Observatory (Italy) in collaboration with the Department of General Physics of Turin University (Italy) and the SCAI Department of CINECA (Italy). We acknowledge the INAF \textit{Pleiadi} high performance computing (HPC) infrastructure and the SCAN (Sistema di Calcolo per l'Astrofisica Numerica) HPC facility at INAF-Osservatorio Astronomico di Palermo for the availability of HPC resources and support. S.O., M.M., and F.B. acknowledge financial contribution from the PRIN 2022 (20224MNC5A) - ``Life, death and after-death of massive stars'' funded by European Union – Next Generation EU, and the INAF Theory Grant ``Supernova remnants as probes for the structure and mass-loss history of the progenitor systems''. S.N. and M.O. thank supports from JSPS KAKENHI Grant Numbers JP19H00693 and JP21K03545, respectively. S.N. and M.O. also thank supports from ``Pioneering Program of RIKEN for Evolution of Matter in the Universe (r-EMU)'' and ``Interdisciplinary Theoretical and Mathematical Sciences Program of RIKEN''. S.N. is supported by the ASPIRE project for top scientists, JST 'RIKEN-Berkeley Mathematical Quantum Science Initiative. T.M., M.O., and K.C. are supported by the National Science and Technology Council, Taiwan, under grant No. MOST 110-2112-M-001-068-MY3, NSTC 113-2112-M-001-028-, and Academia Sinica, Taiwan, under a career development award under grant No. AS-CDA-111-M04.
\end{acknowledgments}

%

\vspace{5mm}
\facilities{
Chandra\footnote{https://cxc.harvard.edu/index.html}, 
NuSTAR \citep{2013ApJ...770..103H}.
    }


\software{
\mesa\ \citep{2011ApJS..192....3P,2013ApJS..208....4P,2015ApJS..220...15P,2018ApJS..234...34P,2019ApJS..243...10P,2023ApJS..265...15J}, 
\snec\ \citep{2015ApJ...814...63M}, 
\pluto\ \citep{2012ApJS..198....7M}, 
CIAO \citep{2006SPIE.6270E..1VF},
HEASOFT\footnote{https://heasarc.gsfc.nasa.gov/docs/software/heasoft/},
XSPEC \citep{1996ASPC..101...17A}.
}



\newpage
\appendix

\section{Convergence test}
\label{app:convergence}

\begin{figure*}[!t]
  \centering
  \includegraphics[width=1\textwidth]{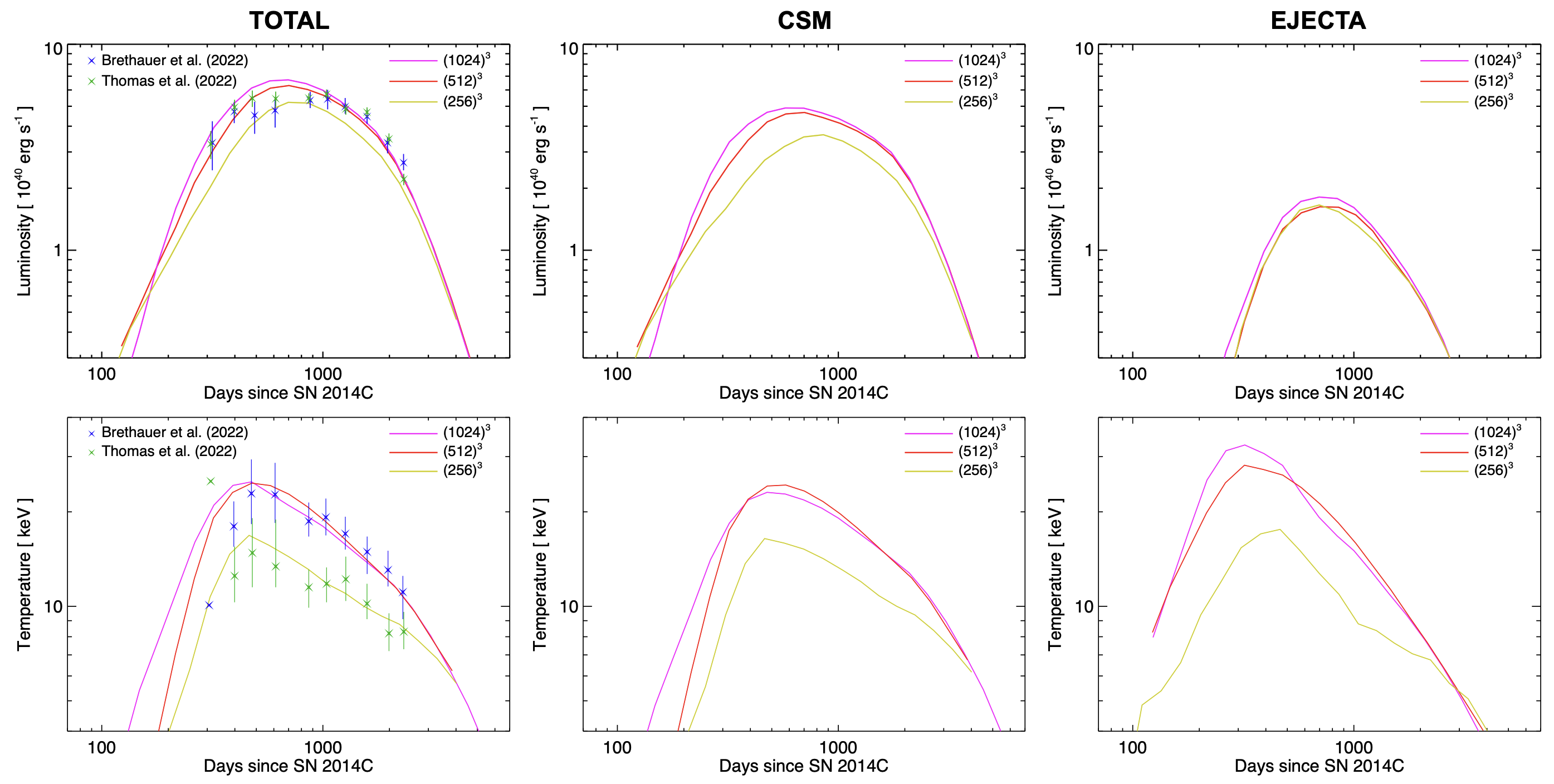}
  \caption{Upper panels: X-ray lightcurves in the $[0.5, 100]$~keV band, synthesized from the model TOY\_T50\_D6.8, are shown for the entire emitting plasma (left panel), the shocked CSM (center panel), and the shocked ejecta (right panel). The figure compares simulation results at three different resolutions: low resolution ($256^3$ grid points; yellow line), medium resolution ($512^3$ grid points; red line), and high resolution ($1024^3$ grid points; magenta line). In the left panel, these results are compared with the observed lightcurve of {\sn14c}, as obtained with Chandra and NuSTAR (blue and green symbols; \citealt{bmm22,twd22}). Lower panels: Corresponding to the upper panels, these plots show the evolution of the average X-ray emission-weighted electron temperature.}
  \label{test_conv}
\end{figure*}

We evaluated how well the simulations used in this paper capture the basic evolution of the remnant as it interacts with the inhomogeneous CSM. To this end, we conducted three simulations for model TOY\_T50\_D6.8, each with a different spatial resolution: $256^3$ (low resolution), $512^3$ (medium resolution - that adopted for most of the simulations in the paper), and $1024^3$ (high resolution) grid points. Figure~\ref{test_conv} presents the results for the synthetic lightcurve and the evolution of the average X-ray emission-weighted electron temperature, derived for the entire X-ray emitting plasma (left panels) and for the main emission components: the shocked CSM (center panels) and the shocked ejecta (right panels).

Our analysis reveals that all three simulations adequately reproduce the observed lightcurve (represented by symbols in the figure), with differences from each other within 40\%. Notably, higher spatial resolution simulations generally predict larger luminosities. The low-resolution simulation exhibits the most significant differences compared to the other two, with lower luminosity values ranging from 12\% to 35\% around the emission peak between days 300 and 3000 (coinciding with available Chandra and NuSTAR observations). These discrepancies are most pronounced in the synthesis of emission from the shocked CSM. In contrast, the medium- and high-resolution simulations show closer agreement. They differ by less than 25\% during the rise phase of luminosity (between days 300 and 400) and by less than 10\% at later times.

The differences among the three simulations become more pronounced when examining the average X-ray emission-weighted electron temperature (shown in the lower panels of Figure~\ref{test_conv}). This is particularly evident when comparing the synthetic curves with those derived from observational analyses (\citealt{twd22,bmm22}). Notably, while the medium- and high-resolution simulations align closely with the results of \cite{bmm22}, the low-resolution simulation corresponds more closely to the findings of \cite{twd22}. Consistent with the lightcurve analysis, higher resolution simulations yield higher average electron temperatures. The low-resolution simulation exhibits the most significant deviations from the medium- and high-resolution simulations, with temperatures up to 45\% lower around the emission peak between days 300 and 3000. These disparities are most prominent in the temperature of the shocked ejecta, which can be up to 50\% lower than those obtained from the medium-resolution simulation. In contrast, the medium- and high-resolution simulations demonstrate remarkable consistency, with temperature differences not exceeding 7\% throughout the entire simulated period.

Based on these findings, we conclude that the medium-resolution simulation effectively captures the fundamental evolution of the remnant as it interacts with the dense nebula. In contrast, the results obtained from the low spatial resolution simulation are unreliable. Our analysis suggests that simulations employing higher spatial resolutions than that used in the paper are likely to produce luminosity and average electron temperature values within a 10\% margin of those discussed in this paper.

\section{Analysis of X-ray observations}
\label{app:obs}

\begin{figure}
    \centering
    \includegraphics[width=0.99\textwidth]{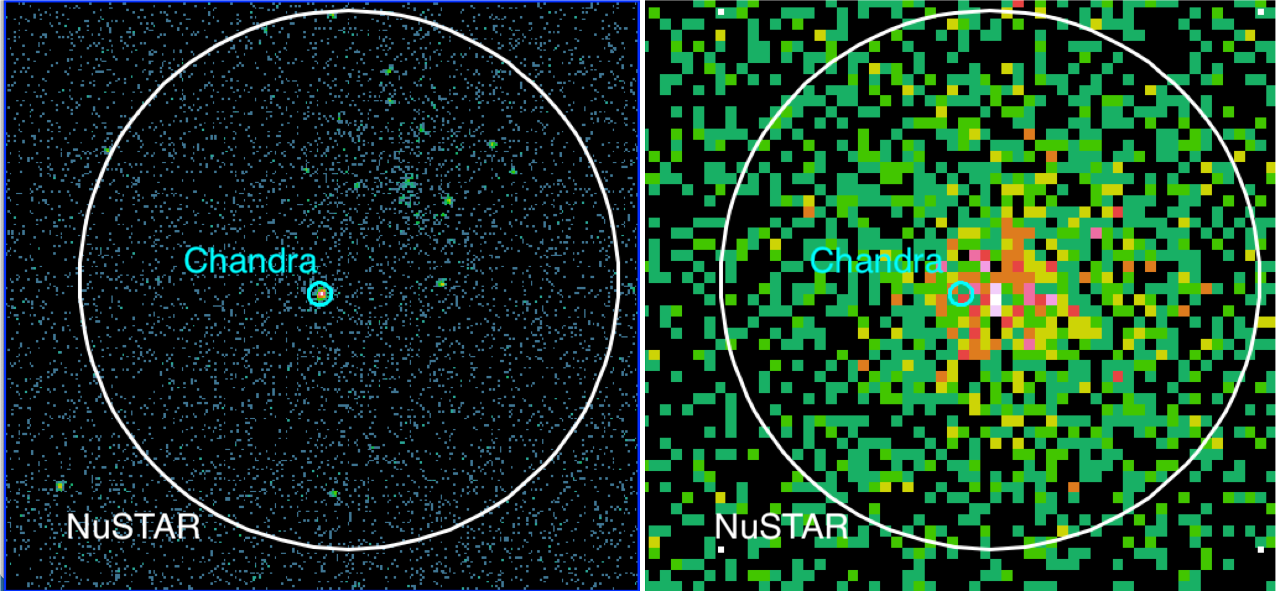}
    \caption{Comparison between the Chandra and NuSTAR images and corresponding source regions in epoch 5. Chandra 2.5'' radius extraction circle is marked in cyan, whereas the NuSTAR one having a radius of 60'' in white. \textit{Left panel.} Chandra 0.5-8 keV image. \textit{Right panel.} NuSTAR 3-20 keV image. Note that the NuSTAR circle encloses the emission from other sources in the host galaxy.}
    \label{fig:psf_chvsnu}
\end{figure}

We conducted a comprehensive analysis of all available X-ray data for \sn14c, encompassing a total of 13 Chandra observations across 11 distinct epochs, as well as 9 NuSTAR observations, each corresponding to a different epoch. The complete Chandra dataset, obtained by the Chandra X-ray Observatory, is contained in  ~\dataset[DOI: https://doi.org/10.25574/cdc.301]{https://doi.org/10.25574/cdc.301}. Table~\ref{tab:log_obs} provides a log of Chandra and NuSTAR observations. The data reduction process is detailed in Sect.~\ref{sec:reduction}. As illustrated in Figure~\ref{fig:psf_chvsnu}, which shows the extraction regions, Chandra and NuSTAR images are compared alongside their corresponding source regions for epoch 5. The figure demonstrates that Chandra, with its superior spatial resolution, can effectively isolate the emission from {\sn14c} amidst other nearby X-ray sources. Conversely, NuSTAR's Point Spread Function (PSF) unavoidably leads to significant contamination from other sources within the host galaxy of \sn14c.

We analyzed the X-ray data using the two-thermal models discussed in Sect.~\ref{sec:fit_2t} for all epochs where Chandra and NuSTAR observations were available (i.e., from day 396 to 2301). Our fitting procedure was performed as follows. 

\begin{table}
    \centering
    \caption{Chandra and NuSTAR observations log table.}
    \label{tab:log_obs}
    \begin{tabular}{c|c|c|c|c}
    \hline\hline
     Chandra ID & NuSTAR ID & Days from SN & Epoch & PI      \\
     \hline
     16005 & / & 308 & 0 & Soderberg\\
     17569 &80001085002& 396 & 1 & Margutti\\
     17570	& 40102014001& 477 & 2& Margutti\\
     17571& 40102014003& 606 & 3& Margutti\\
     18340&	40202013002& 857& 4& Margutti\\
     18341& 40202013004& 1030& 5& Margutti\\
     18342& 40302002002& 1258& 6& Margutti\\
	 18343 and 21077& 40302002004 & 1571& 7& Margutti\\
     21639& 40502001002& 1971 & 8& Margutti\\
     21640 and 23216&	40502001004& 2301& 9& Margutti\\
     25191    & / &2945&  10&Margutti\\
     \hline\hline
    \end{tabular}
\end{table}

i) Initially, we performed spectral fits without including corrections for cross-calibration or contamination from spurious sources in the NuSTAR data; the intrinsic column density $N_{\rm{H,int}}$, electron temperatures $kT$, and normalization of the components were left free to vary, while abundances were frozen to solar values (in agreement with \citealt{2015ApJ...815..120M} and \citealt{2020MNRAS.497.5118S}).

ii) Noting significant residuals at the Fe K line, we allowed the Fe abundance in one of the thermal components (either \textsc{vapec} or \textsc{vnei}) to vary; this adjustment significantly improved the fit, suggesting super-solar abundances (A$_{\rm{Fe}} > 1$); we also verified that abundances A$_{\rm{Fe}} > 100$ are able to produce best-fit values indicating emissions consistent with pure-Fe plasma.
    
iii) Due to the known degeneracy between chemical abundance and emission measure (as discussed in \citealt{gvm21}), we constrained the Fe abundance to 100 to ensure a consistent estimate of ejecta flux across different epochs.
    
iv) To account for emission from spurious sources in NuSTAR, we introduced a power-law component (\textsc{power-law} model in XSPEC), further optimizing the fit and significantly reducing the C-statistics.
    
v) Subsequently, we fixed the power-law parameters to their best-fit values and adjusted the NuSTAR cross-calibration constants, yielding values consistent with nominal uncertainties \citep{mhm15,mfg22}.

vi) The uncertainties for the parameters were computed at the 1$\sigma$ level, with the cross-calibration constants and power-law component frozen at their best-fit values.

vii) We estimated the unabsorbed flux and its 1$\sigma$ uncertainty for both components in the $[0.3,100]$~keV band using the \textsc{cflux} component.

The only difference in the procedure between the 2VAPEC and VAPEC+VNEI models was the inclusion of the ionization parameter $\tau$ in the latter. Its uncertainty was computed while maintaining the temperature of the ejecta component at its best-fit value, and vice versa. The best-fit values obtained using the 2VAPEC and VAPEC+VNEI models are listed in Table~\ref{tab:fit_values}.

\begin{table*}[!t]
    \centering
    \caption{Best-fit values on the Chandra and NuSTAR spectra with the VNEI, 2VAPEC and VAPEC+VNEI models. The fluxes are unabsorbed, estimated in the 0.3-100 keV range, and expressed in units of 10$^{-12}$ erg s$^{-1}$ cm$^{-2}$.}
    \begin{tabular}{c|c|c|c|c|c|c|c|c|c}
    \hline\hline
    & & \multicolumn{8}{l}{ } \\
    & & \multicolumn{8}{l}{\bf VNEI} \\
    \hline
    Epoch & Days from SN & NH$_{\rm int}$ & kT & Fe & \multicolumn{2}{c|}{$\tau$} & \multicolumn{2}{c|}{{Flux$_{\rm{total}}$}} & cstat/dof\\
     &  &  (10$^{22}$ cm$^{-2}$) &  (keV) &  & \multicolumn{2}{c|}{(10$^{11}$ s/cm$^3$)}  & \multicolumn{2}{c|}{ } & \\
    \hline
    1 & 396& 2.2$\pm 0.3 $ & 23$^{+6}_{-5}$ & 3.7$^{+1.0}_{-0.7}$ & \multicolumn{2}{c|}{5.3$^{+1.5}_{-1.1}$}& \multicolumn{2}{c|}{$1.85_{-0.12}^{+0.12}$} & 191/146\\
    2 & 477& 1.17$\pm0.19 $ & 31$^{+12}_{-8}$ & 3.82$^{+1.31}_{-0.94}$ & \multicolumn{2}{c|}{4.1$^{+1.35}_{-1.04}$}& \multicolumn{2}{c|}{$1.82_{-0.10}^{+0.15}$} & 152/139\\
    3 & 606& 1.04$\pm0.17 $ & 22$^{+6}_{-5}$ & 4.6$^{+1.4}_{-0.9}$ & \multicolumn{2}{c|}{5.5$^{+1.7}_{-1.2}$}& \multicolumn{2}{c|}{$2.17_{-0.09}^{+0.18}$} & 143/143\\
    4 & 857& 0.82$\pm0.07 $ & 24$^{+4}_{-3}$ & 2.4$^{+0.3}_{-0.4}$ & \multicolumn{2}{c|}{3.5$^{+0.6}_{-0.4}$}& \multicolumn{2}{c|}{$1.90_{-0.05}^{+0.10}$} & 152/164\\
    5 & 1030& 0.60$\pm 0.06 $ & 22$^{+3}_{-2}$ & 3.2$^{+0.5}_{-0.4}$ & \multicolumn{2}{c|}{4.4$^{+0.7}_{-0.6}$}& \multicolumn{2}{c|}{$2.14_{-0.08}^{+0.08}$} & 201/161\\
    6 & 1258& 0.49$^{+0.05}_{-0.06}$ & 17.2$^{+2.1}_{-1.2}$ & 2.8$^{+0.4}_{-0.4}$ & \multicolumn{2}{c|}{4.4$^{+1.0}_{-0.7}$}& \multicolumn{2}{c|}{$1.84_{-0.07}^{+0.07}$} & 161/166\\
    7 & 1571& 0.36$^{+0.06}_{-0.05}$ & 17.1$^{+2.0}_{-1.6}$ & 2.1$^{+0.3}_{-0.3}$ & \multicolumn{2}{c|}{3.1$^{+0.6}_{-0.4}$}& \multicolumn{2}{c|}{$1.77_{-0.05}^{+0.09}$} & 232/207\\
    8 & 1971& 0.15$\pm 0.05 $ & 16.8$^{+2.2}_{-1.8}$ & 2.1$^{+0.3}_{-0.3}$ & \multicolumn{2}{c|}{3.2$^{+0.6}_{-0.4}$}& \multicolumn{2}{c|}{$1.33_{-0.06}^{+0.06}$} & 154/158\\
    9 & 2301& 0.25$\pm 0.07$ & 11.74$^{+1.4}_{-1.2}$ & 1.75$^{+0.3}_{-0.3}$ & \multicolumn{2}{c|}{2.9$^{+0.8}_{-0.4}$}& \multicolumn{2}{c|}{$1.12_{-0.05}^{+0.05}$} & 222/202\\
    
    \hline\hline
    \multicolumn{2}{c|}{} & \multicolumn{8}{l}{ } \\
    \multicolumn{2}{c|}{} & \multicolumn{8}{l}{\bf 2VAPEC} \\
    \hline
     Epoch & Days from SN & NH$_{\rm int}$ & kT$_{\rm{CSM}}$ &  Flux$_{\rm{CSM}}$ & kT$_{\rm{ej}}$ & $\tau_{\rm{ej}}$ & Flux$_{\rm{ej}}$ &  Flux$_{\rm{total}}$ & cstat/dof\\
     & &(10$^{22}$ cm$^{-2}$)& (keV) & & (keV) & (10$^{11}$ s/cm$^3$) & &  \\
\hline    
1&396&$2.28_{-0.28}^{+0.25}$  &$22_{-4}^{+16}$ & $1.40_{-0.15}^{+0.20}$&$8.8_{-1.6}^{+1.9}$&/&$0.31_{-0.06}^{+0.07}$&$1.71_{-0.15}^{+0.25}$& 189/146\\
2&477& $1.33_{-0.24}^{+0.20}$ &$26_{-6}^{+29}$ & $1.69_{-0.30}^{+0.29}$&$8.0_{-1.8}^{+2}$&/&$0.30_{-0.07}^{+0.09}$&$2.0_{-0.3}^{+0.3}$& 147/139\\
3&606& $1.12_{-0.15}^{+0.16}$ &$21_{-5}^{+7}$ & $1.41_{-0.10}^{+0.11}$&$9.4_{-1.6}^{+2}$&/&$0.36_{-0.07}^{+0.08}$&$1.7_{-0.2}^{+0.3}$& 139/143\\
4&857& $0.90_{-0.07}^{+0.08}$ &$21_{-3}^{+3}$ & $1.96_{-0.12}^{+0.17}$&$7.0_{-0.9}^{+1.0}$&/&$0.24_{-0.03}^{+0.04}$&$2.19_{-0.11}^{+0.17}$& 156/164\\
5&1030&$0.64_{-0.06}^{+0.06}$ &$21_{-3}^{+4}$ & $1.96_{-0.12}^{+0.17}$&$8.6_{-0.9}^{+1.0}$&/&$0.32_{-0.04}^{+0.04}$&$2.24_{-0.14}^{+0.16}$& 204/161\\
6&1258&$0.52_{-0.06}^{+0.06}$ &$17_{-3}^{+3}$ & $1.93_{-0.14}^{+0.17}$&$8.3_{-0.9}^{+1.1}$&/&$0.28_{-0.04}^{+0.04}$&$1.99_{-0.12}^{+0.14}$& 168/166\\
7&1571&$0.42_{-0.06}^{+0.07}$ &$15_{-2}^{+2}$ & $1.70_{-0.12}^{+0.15}$&$6.7_{-0.9}^{+1.0}$&/&$0.19_{-0.03}^{+1.03}$&$1.68_{-0.09}^{+0.10}$& 236/207\\
8&1971&$0.21_{-0.05}^{+0.06}$ &$13.0_{-1.3}^{+2.3}$ &$1.16_{-0.08}^{+0.09}$&$6.9_{-1.0}^{+1.1}$&/&$0.15_{-0.03}^{+0.03}$&$1.3_{-0.06}^{+0.09}$& 159/158\\
9&2301&$0.33_{-0.08}^{+0.09}$ &$10.0_{-1.1}^{+1.3}$ &$0.93_{-0.06}^{+0.06}$&$4.9_{-1.0}^{+1.3}$&/&$0.21_{-0.07}^{+0.01}$&$1.03_{-0.05}^{+0.06}$& 224/202\\
    \hline\hline

    \multicolumn{2}{c|}{} & \multicolumn{8}{l}{ } \\
    \multicolumn{2}{c|}{} & \multicolumn{8}{l}{\bf VAPEC+VNEI} \\
    \hline
    Epoch & Days from SN & NH$_{\rm int}$ & kT$_{\rm{CSM}}$ &  Flux$_{\rm{CSM}}$ & kT$_{\rm{ej}}$ & $\tau_{\rm{ej}}$ & Flux$_{\rm{ej}}$ &  Flux$_{\rm{total}}$ & cstat/dof\\
      & &(10$^{22}$ cm$^{-2}$)& (keV) &  & (keV) & (10$^{11}$ s/cm$^3$) &  & \\
     \hline
1&396&$2.28_{-0.27}^{+0.29}$ &$21_{-4}^{+8}$ & $1.40_{-0.15}^{+0.20}$&$9_{-2}^{+2}$&$34_{-24}^{+50}$&$0.30_{-0.06}^{+0.07}$&$1.7_{-0.2}^{+0.2}$& 189/145\\
2&477& $1.24_{-0.10}^{+0.21}$&$26_{-6}^{+30}$ & $1.7_{-0.28}^{+0.29}$&$11_{-3}^{+4}$&$8.4_{-3.0}^{+12}$&$0.30_{-0.07}^{+0.08}$&$1.60_{-0.14}^{+0.12}$& 147/138\\
3&606& $1.04_{-0.18}^{+0.15}$&$22_{-5}^{+8}$ & $1.40_{-0.24}^{+0.46}$&$13_{-3}^{+4}$&$9.0_{-3.0}^{+7.0}$&$0.36_{-0.07}^{+0.08}$&$1.7_{-0.2}^{+0.3}$& 140/142\\
4&857&$0.86_{-0.07}^{+0.08}$ &$22_{-2}^{+3}$ & $1.99_{-0.14}^{+0.17}$&$17_{-5}^{+12}$&$3.4_{-0.6}^{+0.8}$&$0.25_{-0.04}^{+0.07}$&$2.2_{-0.1}^{+0.2}$& 152/163\\
5&1030&$0.61_{-0.06}^{+0.06}$ &$22_{-3}^{+3}$ & $1.95_{-0.14}^{+0.20}$&$18_{-4}^{+7}$&$4.5_{-0.7}^{+0.9}$&$0.36_{-0.05}^{+0.05}$&$2.28_{-0.14}^{+0.18}$& 201/160\\
6&1258&$0.48_{-0.06}^{+0.06}$ &$18_{-2}^{+2}$ & $1.76_{-0.14}^{+0.13}$&$19_{-5}^{+12}$&$4.1_{-0.6}^{+0.8}$&$0.30_{-0.04}^{+0.07}$&$2.06_{-0.13}^{+0.13}$& 164/165\\
7&1571&$0.39_{-0.06}^{+0.07}$ &$15_{-2}^{+2}$ & $1.50_{-0.09}^{+0.11}$&$10_{-2}^{+4}$&$4.2_{-1.0}^{+1.9}$&$0.20_{-0.03}^{+0.04}$&$1.70_{-0.09}^{+0.10}$& 235/206\\
8&1971&$0.19_{-0.05}^{+0.06}$ &$12.9_{-1.3}^{+2.0}$ &$1.16_{-0.07}^{+0.08}$&$13_{-4}^{+8}$&$3.6_{-0.7}^{+1.2}$&$0.15_{-0.03}^{+0.05}$&$1.3_{-0.6}^{+0.8}$& 157/157\\
9&2301&$0.31_{-0.08}^{+0.09}$ &$10.4_{-1.1}^{+1.2}$ &$0.93_{-0.06}^{+0.06}$&$6_{-2}^{+4}$&$3.8_{-1.1}^{+9.4}$&$0.21_{-0.06}^{+0.01}$&$1.03_{-0.05}^{+0.06}$& 224/201\\
\hline\hline
    \end{tabular}
    \label{tab:fit_values}
\end{table*}

\section{Synthetic spectra of SN~2014c}
\label{app:spectra}
 
\begin{figure*}[!ht]
  \centering
  \includegraphics[width=1\textwidth]{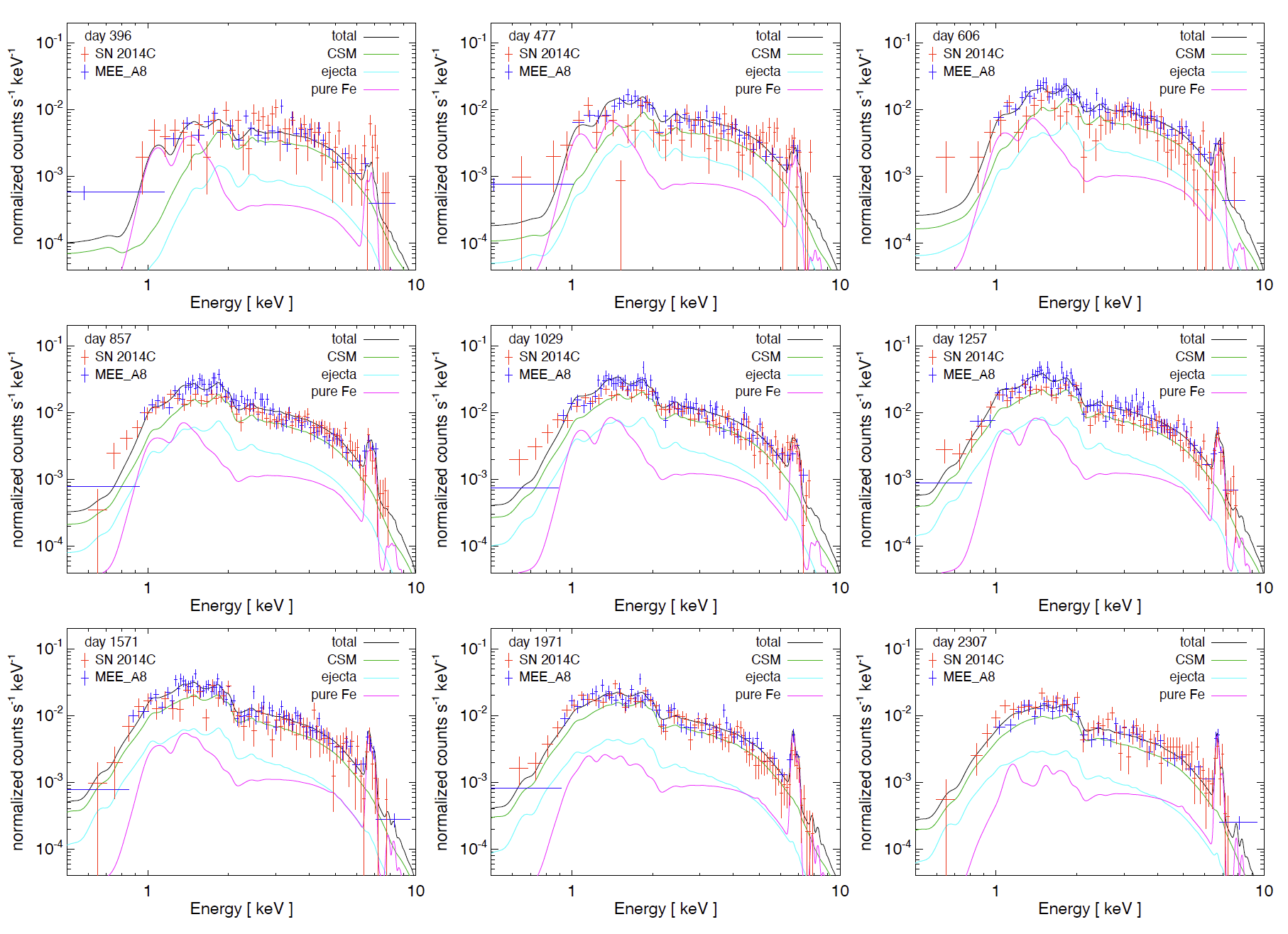}
  \caption{As in Figure~\ref{spectra} for all Chandra observations.}
  \label{all_chandra}
\end{figure*}

\begin{figure*}[!ht]
  \centering
  \includegraphics[width=1\textwidth]{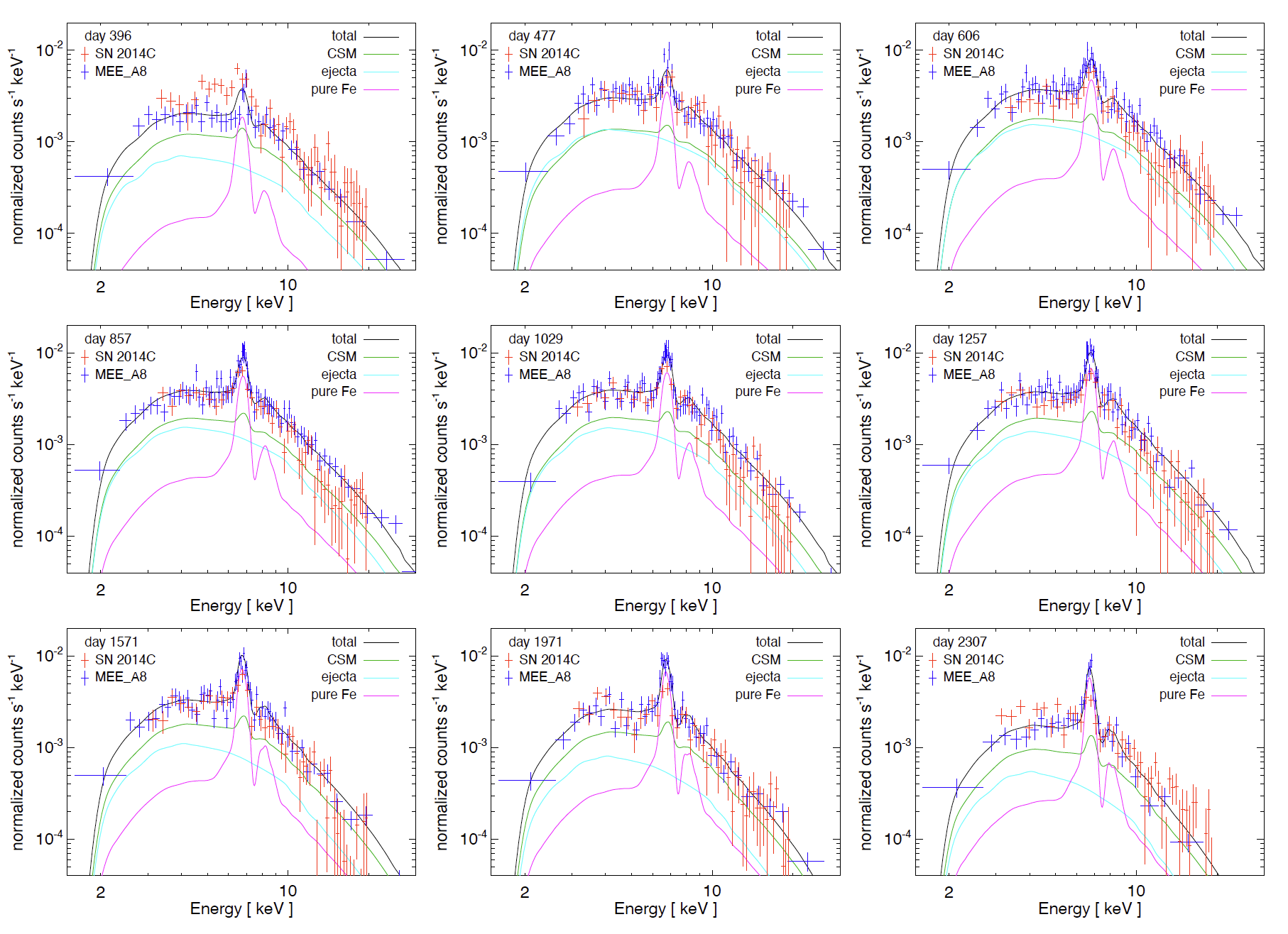}
  \caption{As in Figure~\ref{spectra} for all NuSTAR observations.}
  \label{all_nustar}
\end{figure*}

\begin{figure*}[!ht]
  \centering
  \includegraphics[width=1\textwidth]{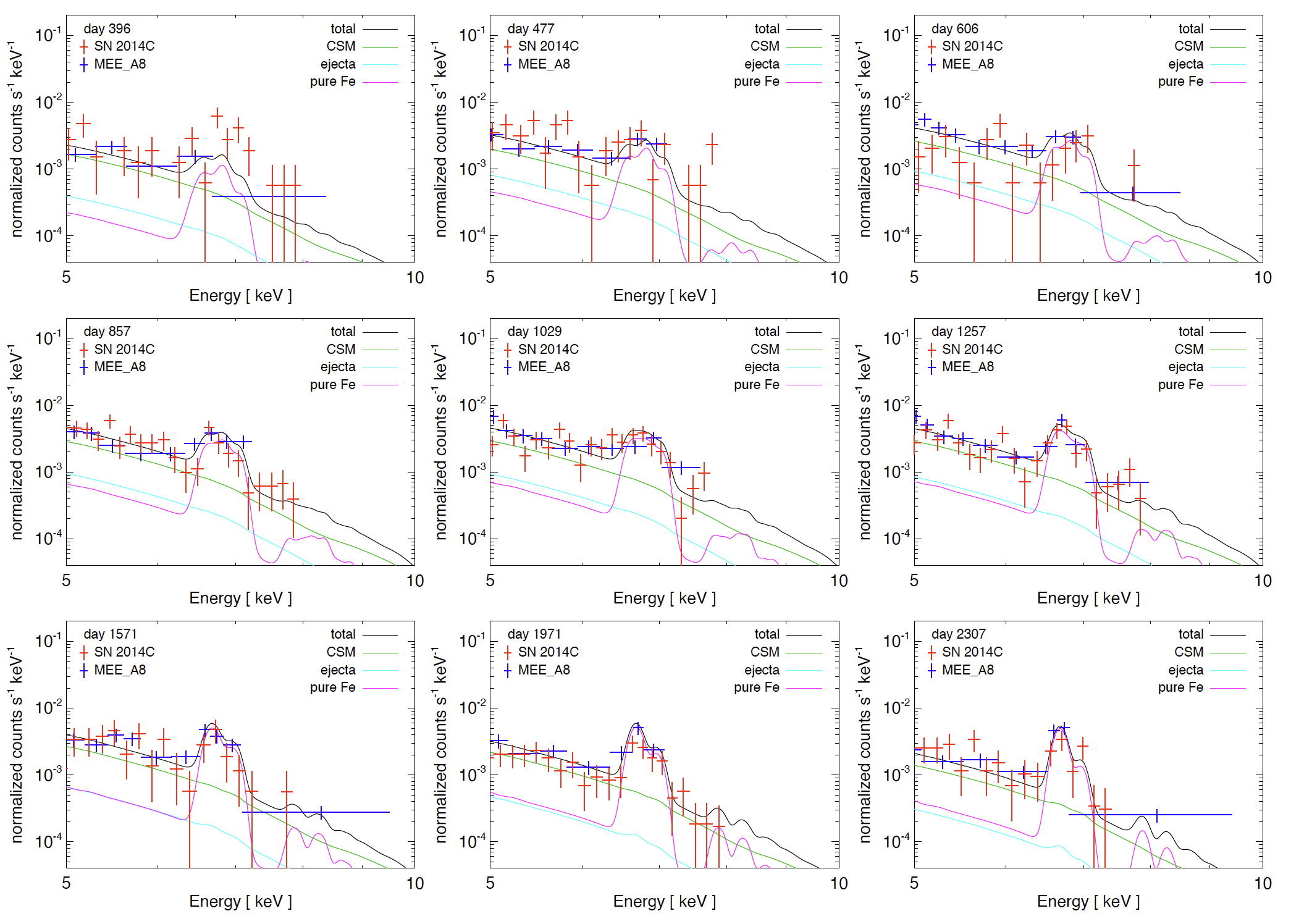}
  \caption{As in Figure~\ref{fe_line} for all Chandra observations.}
  \label{all_chandra_line}
\end{figure*}

\begin{figure*}[!ht]
  \centering
  \includegraphics[width=1\textwidth]{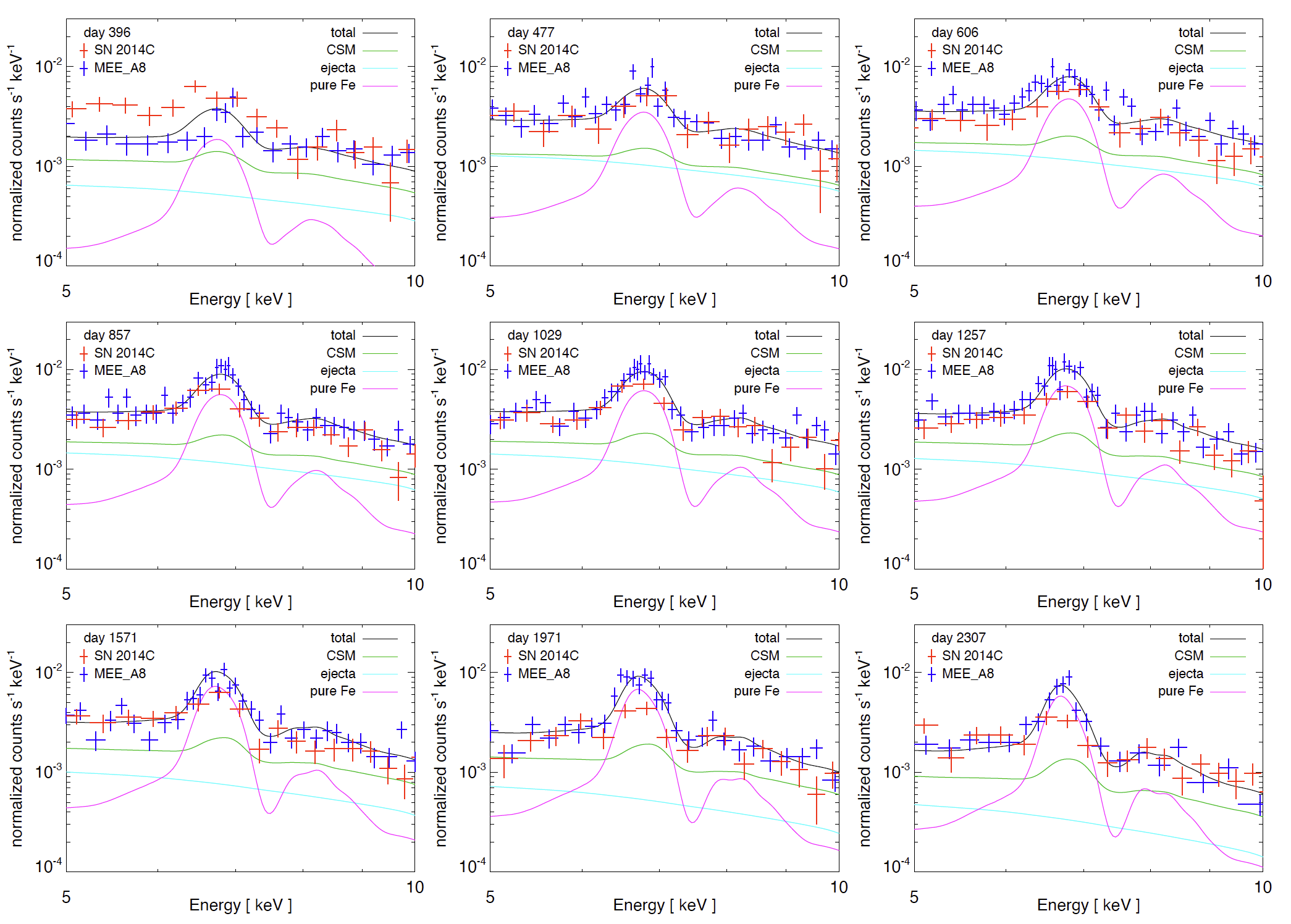}
  \caption{As in Figure~\ref{fe_line} for all NuSTAR observations.}
  \label{all_nustar_line}
\end{figure*}

In this section, we present a comprehensive comparison of all the data for which almost simultaneous Chandra and NuSTAR observations are available, with synthetic spectra derived from model MEE\_A8. Specifically, Figures~\ref{all_chandra} and \ref{all_nustar} illustrate both observed and synthetic spectra spanning from day 396 to day 2301, as detailed in Table~\ref{tab:log_obs}. Remarkably, the synthetic spectra faithfully capture the observed data without requiring any ad-hoc renormalization. To provide a closer examination of the comparison, Figures~\ref{all_chandra_line} and \ref{all_nustar_line} offer a focused view within the energy band of $[5,10]$~keV, emphasizing the details around the prominent emission line. Notably, the synthetic spectra consistently replicate the line flux across various observed epochs, effectively reproducing even the line profile, as evident from the comparison of the theoretical spectra (black lines in the figures) and the observed spectra (represented by red crosses) in Figures~\ref{all_chandra_line} and \ref{all_nustar_line}.


\bibliography{references}{}

\begin{thebibliography}{}
\expandafter\ifx\csname natexlab\endcsname\relax\def\natexlab#1{#1}\fi
\providecommand{\url}[1]{\href{#1}{#1}}
\providecommand{\dodoi}[1]{doi:~\href{http://doi.org/#1}{\nolinkurl{#1}}}
\providecommand{\doeprint}[1]{\href{http://ascl.net/#1}{\nolinkurl{http://ascl.net/#1}}}
\providecommand{\doarXiv}[1]{\href{https://arxiv.org/abs/#1}{\nolinkurl{https://arxiv.org/abs/#1}}}

\bibitem[{{Anderson} {et~al.}(2017){Anderson}, {Horesh}, {Mooley}, {Rushton}, {Fender}, {Staley}, {Argo}, {Beswick}, {Hancock}, {P{\'e}rez-Torres}, {Perrott}, {Plotkin}, {Pretorius}, {Rumsey}, \& {Titterington}}]{2017MNRAS.466.3648A}
{Anderson}, G.~E., {Horesh}, A., {Mooley}, K.~P., {et~al.} 2017, \mnras, 466, 3648, \dodoi{10.1093/mnras/stw3310}

\bibitem[{{Arnaud}(1996)}]{1996ASPC..101...17A}
{Arnaud}, K.~A. 1996, in Astronomical Society of the Pacific Conference Series, Vol. 101, Astronomical Data Analysis Software and Systems V, ed. G.~H. {Jacoby} \& J.~{Barnes}, 17

\bibitem[{{Asida} \& {Tuchman}(1995)}]{1995ApJ...455..286A}
{Asida}, S.~M., \& {Tuchman}, Y. 1995, \apj, 455, 286, \dodoi{10.1086/176576}

\bibitem[{{Bietenholz} {et~al.}(2021){Bietenholz}, {Bartel}, {Kamble}, {Margutti}, {Matthews}, \& {Milisavljevic}}]{2021MNRAS.502.1694B}
{Bietenholz}, M.~F., {Bartel}, N., {Kamble}, A., {et~al.} 2021, \mnras, 502, 1694, \dodoi{10.1093/mnras/staa4003}

\bibitem[{{Blagorodnova} {et~al.}(2021){Blagorodnova}, {Klencki}, {Pejcha}, {Vreeswijk}, {Bond}, {Burdge}, {De}, {Fremling}, {Gehrz}, {Jencson}, {Kasliwal}, {Kupfer}, {Lau}, {Masci}, \& {Rich}}]{2021A&A...653A.134B}
{Blagorodnova}, N., {Klencki}, J., {Pejcha}, O., {et~al.} 2021, \aap, 653, A134, \dodoi{10.1051/0004-6361/202140525}

\bibitem[{{Blondin} {et~al.}(1996){Blondin}, {Lundqvist}, \& {Chevalier}}]{1996ApJ...472..257B}
{Blondin}, J.~M., {Lundqvist}, P., \& {Chevalier}, R.~A. 1996, \apj, 472, 257, \dodoi{10.1086/178060}

\bibitem[{{Blondin} {et~al.}(1998){Blondin}, {Wright}, {Borkowski}, \& {Reynolds}}]{1998ApJ...500..342B}
{Blondin}, J.~M., {Wright}, E.~B., {Borkowski}, K.~J., \& {Reynolds}, S.~P. 1998, \apj, 500, 342, \dodoi{10.1086/305708}

\bibitem[{{Brethauer} {et~al.}(2022){Brethauer}, {Margutti}, {Milisavljevic}, {Bietenholz}, {Chornock}, {Coppejans}, {De Colle}, {Hajela}, {Terreran}, {Vargas}, {DeMarchi}, {Harris}, {Jacobson-Gal{\'a}n}, {Kamble}, {Patnaude}, \& {Stroh}}]{bmm22}
{Brethauer}, D., {Margutti}, R., {Milisavljevic}, D., {et~al.} 2022, \apj, 939, 105, \dodoi{10.3847/1538-4357/ac8b14}

\bibitem[{{Cash}(1979)}]{cas79}
{Cash}, W. 1979, \apj, 228, 939.
\newblock \url{http://cdsads.u-strasbg.fr/cgi-bin/nph-bib_query?bibcode=1979ApJ...228..939C&db_key=AST}

\bibitem[{{Chandra} {et~al.}(2012){Chandra}, {Chevalier}, {Irwin}, {Chugai}, {Fransson}, \& {Soderberg}}]{2012ApJ...750L...2C}
{Chandra}, P., {Chevalier}, R.~A., {Irwin}, C.~M., {et~al.} 2012, \apjl, 750, L2, \dodoi{10.1088/2041-8205/750/1/L2}

\bibitem[{Chaplin \& Colella(2015)}]{chaplin15}
Chaplin, C., \& Colella, P. 2015, Communications in Applied Mathematics and Computational Science, 12, \dodoi{10.2140/camcos.2017.12.1}

\bibitem[{{Chen} {et~al.}(2018){Chen}, {Inserra}, {Fraser}, {Moriya}, {Schady}, {Schweyer}, {Filippenko}, {Perley}, {Ruiter}, {Seitenzahl}, {Sollerman}, {Taddia}, {Anderson}, {Foley}, {Jerkstrand}, {Ngeow}, {Pan}, {Pastorello}, {Points}, {Smartt}, {Smith}, {Taubenberger}, {Wiseman}, {Young}, {Benetti}, {Berton}, {Bufano}, {Clark}, {Della Valle}, {Galbany}, {Gal-Yam}, {Gromadzki}, {Guti{\'e}rrez}, {Heinze}, {Kankare}, {Kilpatrick}, {Kuncarayakti}, {Leloudas}, {Lin}, {Maguire}, {Mazzali}, {McBrien}, {Prentice}, {Rau}, {Rest}, {Siebert}, {Stalder}, {Tonry}, \& {Yu}}]{2018ApJ...867L..31C}
{Chen}, T.~W., {Inserra}, C., {Fraser}, M., {et~al.} 2018, \apjl, 867, L31, \dodoi{10.3847/2041-8213/aaeb2e}

\bibitem[{{Colella}(1990)}]{1990JCoPh..87..171C}
{Colella}, P. 1990, Journal of Computational Physics, 87, 171, \dodoi{10.1016/0021-9991(90)90233-Q}

\bibitem[{{Cuesta-Mart{\'\i}nez} {et~al.}(2015){Cuesta-Mart{\'\i}nez}, {Aloy}, \& {Mimica}}]{2015MNRAS.446.1716C}
{Cuesta-Mart{\'\i}nez}, C., {Aloy}, M.~A., \& {Mimica}, P. 2015, \mnras, 446, 1716, \dodoi{10.1093/mnras/stu2186}

\bibitem[{{Drake} \& {Orlando}(2010)}]{2010ApJ...720L.195D}
{Drake}, J.~J., \& {Orlando}, S. 2010, \apjl, 720, L195, \dodoi{10.1088/2041-8205/720/2/L195}

\bibitem[{{Dwarkadas} {et~al.}(2010){Dwarkadas}, {Dewey}, \& {Bauer}}]{2010MNRAS.407..812D}
{Dwarkadas}, V.~V., {Dewey}, D., \& {Bauer}, F. 2010, \mnras, 407, 812, \dodoi{10.1111/j.1365-2966.2010.16966.x}

\bibitem[{{Eldridge} {et~al.}(2017){Eldridge}, {Stanway}, {Xiao}, {McClelland}, {Taylor}, {Ng}, {Greis}, \& {Bray}}]{2017PASA...34...58E}
{Eldridge}, J.~J., {Stanway}, E.~R., {Xiao}, L., {et~al.} 2017, \pasa, 34, e058, \dodoi{10.1017/pasa.2017.51}

\bibitem[{{Elias-Rosa} {et~al.}(2024){Elias-Rosa}, {Brennan}, {Benetti}, {Cappellaro}, {Pastorello}, {Kozyreva}, {Lundqvist}, {Fraser}, {Anderso}, {Cai}, {Chen}, {Dennefeld}, {Gromadzki}, {Guti{\'e}rrez}, {Ihanec}, {Inserra}, {Kankare}, {Kotak}, {Mattila}, {Moran}, {M{\"u}ller-Bravo}, {Pessi}, {Pignata}, {Reguitti}, {Reynolds}, {Smartt}, {Smith}, {Tartaglia}, {Valerin}, {de Boer}, {Chambers}, {Gal-Yam}, {Gao}, {Geier}, {Mazzali}, {Nicholl}, {Ragosta}, {Rest}, {Yaron}, \& {Young}}]{2024arXiv240202924E}
{Elias-Rosa}, N., {Brennan}, S.~J., {Benetti}, S., {et~al.} 2024, arXiv e-prints, arXiv:2402.02924, \dodoi{10.48550/arXiv.2402.02924}

\bibitem[{{Frank} {et~al.}(1995){Frank}, {Balick}, \& {Davidson}}]{1995ApJ...441L..77F}
{Frank}, A., {Balick}, B., \& {Davidson}, K. 1995, \apjl, 441, L77, \dodoi{10.1086/187794}

\bibitem[{{Frank} {et~al.}(2016){Frank}, {Zhekov}, {Park}, {McCray}, {Dwek}, \& {Burrows}}]{2016ApJ...829...40F}
{Frank}, K.~A., {Zhekov}, S.~A., {Park}, S., {et~al.} 2016, \apj, 829, 40, \dodoi{10.3847/0004-637X/829/1/40}

\bibitem[{{Fraser} {et~al.}(2013){Fraser}, {Magee}, {Kotak}, {Smartt}, {Smith}, {Polshaw}, {Drake}, {Boles}, {Lee}, {Burgett}, {Chambers}, {Draper}, {Flewelling}, {Hodapp}, {Kaiser}, {Kudritzki}, {Magnier}, {Price}, {Tonry}, {Wainscoat}, \& {Waters}}]{2013ApJ...779L...8F}
{Fraser}, M., {Magee}, M., {Kotak}, R., {et~al.} 2013, \apjl, 779, L8, \dodoi{10.1088/2041-8205/779/1/L8}

\bibitem[{{Freedman} {et~al.}(2001){Freedman}, {Madore}, {Gibson}, {Ferrarese}, {Kelson}, {Sakai}, {Mould}, {Kennicutt}, {Ford}, {Graham}, {Huchra}, {Hughes}, {Illingworth}, {Macri}, \& {Stetson}}]{2001ApJ...553...47F}
{Freedman}, W.~L., {Madore}, B.~F., {Gibson}, B.~K., {et~al.} 2001, \apj, 553, 47, \dodoi{10.1086/320638}

\bibitem[{{Fruscione} {et~al.}(2006){Fruscione}, {McDowell}, {Allen}, {Brickhouse}, {Burke}, {Davis}, {Durham}, {Elvis}, {Galle}, {Harris}, {Huenemoerder}, {Houck}, {Ishibashi}, {Karovska}, {Nicastro}, {Noble}, {Nowak}, {Primini}, {Siemiginowska}, {Smith}, \& {Wise}}]{2006SPIE.6270E..1VF}
{Fruscione}, A., {McDowell}, J.~C., {Allen}, G.~E., {et~al.} 2006, in Society of Photo-Optical Instrumentation Engineers (SPIE) Conference Series, Vol. 6270, Observatory Operations: Strategies, Processes, and Systems, ed. D.~R. {Silva} \& R.~E. {Doxsey}, 62701V, \dodoi{10.1117/12.671760}

\bibitem[{{Gardiner} \& {Stone}(2005)}]{2005JCoPh.205..509G}
{Gardiner}, T.~A., \& {Stone}, J.~M. 2005, Journal of Computational Physics, 205, 509, \dodoi{10.1016/j.jcp.2004.11.016}

\bibitem[{{Ghavamian} {et~al.}(2007){Ghavamian}, {Laming}, \& {Rakowski}}]{2007ApJ...654L..69G}
{Ghavamian}, P., {Laming}, J.~M., \& {Rakowski}, C.~E. 2007, \apjl, 654, L69, \dodoi{10.1086/510740}

\bibitem[{{Greco} {et~al.}(2020){Greco}, {Vink}, {Miceli}, {Orlando}, {Dom{\v{c}}ek}, {Zhou}, {Bocchino}, \& {Peres}}]{gvm21}
{Greco}, E., {Vink}, J., {Miceli}, M., {et~al.} 2020, \aap, 638, A101, \dodoi{10.1051/0004-6361/202038092}

\bibitem[{{Greco} {et~al.}(2022){Greco}, {Miceli}, {Orlando}, {Olmi}, {Bocchino}, {Nagataki}, {Sun}, {Vink}, {Sapienza}, {Ono}, {Dohi}, \& {Peres}}]{2022ApJ...931..132G}
{Greco}, E., {Miceli}, M., {Orlando}, S., {et~al.} 2022, \apj, 931, 132, \dodoi{10.3847/1538-4357/ac679d}

\bibitem[{{Grefenstette} {et~al.}(2023){Grefenstette}, {Brightman}, {Earnshaw}, {Harrison}, \& {Margutti}}]{2023ApJ...952L...3G}
{Grefenstette}, B.~W., {Brightman}, M., {Earnshaw}, H.~P., {Harrison}, F.~A., \& {Margutti}, R. 2023, \apjl, 952, L3, \dodoi{10.3847/2041-8213/acdf4e}

\bibitem[{{Harrison} {et~al.}(2013){Harrison}, {Craig}, {Christensen}, {Hailey}, {Zhang}, {Boggs}, {Stern}, {Cook}, {Forster}, {Giommi}, {Grefenstette}, {Kim}, {Kitaguchi}, {Koglin}, {Madsen}, {Mao}, {Miyasaka}, {Mori}, {Perri}, {Pivovaroff}, {Puccetti}, {Rana}, {Westergaard}, {Willis}, {Zoglauer}, {An}, {Bachetti}, {Barri{\`e}re}, {Bellm}, {Bhalerao}, {Brejnholt}, {Fuerst}, {Liebe}, {Markwardt}, {Nynka}, {Vogel}, {Walton}, {Wik}, {Alexander}, {Cominsky}, {Hornschemeier}, {Hornstrup}, {Kaspi}, {Madejski}, {Matt}, {Molendi}, {Smith}, {Tomsick}, {Ajello}, {Ballantyne}, {Balokovi{\'c}}, {Barret}, {Bauer}, {Blandford}, {Brandt}, {Brenneman}, {Chiang}, {Chakrabarty}, {Chenevez}, {Comastri}, {Dufour}, {Elvis}, {Fabian}, {Farrah}, {Fryer}, {Gotthelf}, {Grindlay}, {Helfand}, {Krivonos}, {Meier}, {Miller}, {Natalucci}, {Ogle}, {Ofek}, {Ptak}, {Reynolds}, {Rigby}, {Tagliaferri}, {Thorsett}, {Treister}, \& {Urry}}]{2013ApJ...770..103H}
{Harrison}, F.~A., {Craig}, W.~W., {Christensen}, F.~E., {et~al.} 2013, \apj, 770, 103, \dodoi{10.1088/0004-637X/770/2/103}

\bibitem[{{Hirai} \& {Mandel}(2022)}]{2022ApJ...937L..42H}
{Hirai}, R., \& {Mandel}, I. 2022, \apjl, 937, L42, \dodoi{10.3847/2041-8213/ac9519}

\bibitem[{{Hirai} {et~al.}(2021){Hirai}, {Podsiadlowski}, {Owocki}, {Schneider}, \& {Smith}}]{2021MNRAS.503.4276H}
{Hirai}, R., {Podsiadlowski}, P., {Owocki}, S.~P., {Schneider}, F. R.~N., \& {Smith}, N. 2021, \mnras, 503, 4276, \dodoi{10.1093/mnras/stab571}

\bibitem[{{Hiramatsu} {et~al.}(2023){Hiramatsu}, {Tsuna}, {Berger}, {Itagaki}, {Goldberg}, {Gomez}, {Kishalay}, {Hosseinzadeh}, {Bostroem}, {Brown}, {Arcavi}, {Bieryla}, {Blanchard}, {Esquerdo}, {Farah}, {Howell}, {Matsumoto}, {McCully}, {Newsome}, {Gonzalez}, {Pellegrino}, {Rhee}, {Terreran}, {Vink{\'o}}, \& {Wheeler}}]{2023ApJ...955L...8H}
{Hiramatsu}, D., {Tsuna}, D., {Berger}, E., {et~al.} 2023, \apjl, 955, L8, \dodoi{10.3847/2041-8213/acf299}

\bibitem[{{Hiramatsu} {et~al.}(2024){Hiramatsu}, {Matsumoto}, {Berger}, {Ransome}, {Villar}, {Gomez}, {Cendes}, {De}, {Bostroem}, {Farah}, {Howell}, {McCully}, {Newsome}, {Padilla Gonzalez}, {Pellegrino}, {Suzuki}, \& {Terreran}}]{2024ApJ...964..181H}
{Hiramatsu}, D., {Matsumoto}, T., {Berger}, E., {et~al.} 2024, \apj, 964, 181, \dodoi{10.3847/1538-4357/ad2854}

\bibitem[{{Janka}(2017)}]{2017hsn..book.1095J}
{Janka}, H.-T. 2017, {``Neutrino-Driven Explosions'' chapter in {\it Handbook of Supernovae}} (edited by Athem W. Alsabti and Paul Murdin, ISBN 978-3-319-21845-8. Springer International Publishing, Switzerland), p.~1095, \dodoi{10.1007/978-3-319-21846-5_109}

\bibitem[{{Jermyn} {et~al.}(2023){Jermyn}, {Bauer}, {Schwab}, {Farmer}, {Ball}, {Bellinger}, {Dotter}, {Joyce}, {Marchant}, {Mombarg}, {Wolf}, {Sunny Wong}, {Cinquegrana}, {Farrell}, {Smolec}, {Thoul}, {Cantiello}, {Herwig}, {Toloza}, {Bildsten}, {Townsend}, \& {Timmes}}]{2023ApJS..265...15J}
{Jermyn}, A.~S., {Bauer}, E.~B., {Schwab}, J., {et~al.} 2023, \apjs, 265, 15, \dodoi{10.3847/1538-4365/acae8d}

\bibitem[{{Kaastra} \& {Bleeker}(2016)}]{kb16}
{Kaastra}, J.~S., \& {Bleeker}, J.~A.~M. 2016, \aap, 587, A151, \dodoi{10.1051/0004-6361/201527395}

\bibitem[{{Kaastra} \& {Mewe}(2000)}]{2000adnx.conf..161K}
{Kaastra}, J.~S., \& {Mewe}, R. 2000, in Atomic Data Needs for X-ray Astronomy, p. 161

\bibitem[{{Kilpatrick} {et~al.}(2021){Kilpatrick}, {Drout}, {Auchettl}, {Dimitriadis}, {Foley}, {Jones}, {DeMarchi}, {French}, {Gall}, {Hjorth}, {Jacobson-Gal{\'a}n}, {Margutti}, {Piro}, {Ramirez-Ruiz}, {Rest}, \& {Rojas-Bravo}}]{2021MNRAS.504.2073K}
{Kilpatrick}, C.~D., {Drout}, M.~R., {Auchettl}, K., {et~al.} 2021, \mnras, 504, 2073, \dodoi{10.1093/mnras/stab838}

\bibitem[{{Kuncarayakti} {et~al.}(2018){Kuncarayakti}, {Maeda}, {Ashall}, {Prentice}, {Mattila}, {Kankare}, {Fransson}, {Lundqvist}, {Pastorello}, {Leloudas}, {Anderson}, {Benetti}, {Bersten}, {Cappellaro}, {Cartier}, {Denneau}, {Della Valle}, {Elias-Rosa}, {Folatelli}, {Fraser}, {Galbany}, {Gall}, {Gal-Yam}, {Guti{\'e}rrez}, {Hamanowicz}, {Heinze}, {Inserra}, {Kangas}, {Mazzali}, {Melandri}, {Pignata}, {Rest}, {Reynolds}, {Roy}, {Smartt}, {Smith}, {Sollerman}, {Somero}, {Stalder}, {Stritzinger}, {Taddia}, {Tomasella}, {Tonry}, {Weiland}, \& {Young}}]{2018ApJ...854L..14K}
{Kuncarayakti}, H., {Maeda}, K., {Ashall}, C.~J., {et~al.} 2018, \apjl, 854, L14, \dodoi{10.3847/2041-8213/aaaa1a}

\bibitem[{{Langer} {et~al.}(1999){Langer}, {Garc{\'\i}a-Segura}, \& {Mac Low}}]{1999ApJ...520L..49L}
{Langer}, N., {Garc{\'\i}a-Segura}, G., \& {Mac Low}, M.-M. 1999, \apjl, 520, L49, \dodoi{10.1086/312131}

\bibitem[{{Madsen} {et~al.}(2022){Madsen}, {Forster}, {Grefenstette}, {Harrison}, \& {Miyasaka}}]{mfg22}
{Madsen}, K.~K., {Forster}, K., {Grefenstette}, B., {Harrison}, F.~A., \& {Miyasaka}, H. 2022, Journal of Astronomical Telescopes, Instruments, and Systems, 8, 034003, \dodoi{10.1117/1.JATIS.8.3.034003}

\bibitem[{{Madsen} {et~al.}(2015){Madsen}, {Harrison}, {Markwardt}, {An}, {Grefenstette}, {Bachetti}, {Miyasaka}, {Kitaguchi}, {Bhalerao}, {Boggs}, {Christensen}, {Craig}, {Forster}, {Fuerst}, {Hailey}, {Perri}, {Puccetti}, {Rana}, {Stern}, {Walton}, {J{\o}rgen Westergaard}, \& {Zhang}}]{mhm15}
{Madsen}, K.~K., {Harrison}, F.~A., {Markwardt}, C.~B., {et~al.} 2015, \apjs, 220, 8, \dodoi{10.1088/0067-0049/220/1/8}

\bibitem[{{Margutti} {et~al.}(2014){Margutti}, {Milisavljevic}, {Soderberg}, {Chornock}, {Zauderer}, {Murase}, {Guidorzi}, {Sanders}, {Kuin}, {Fransson}, {Levesque}, {Chandra}, {Berger}, {Bianco}, {Brown}, {Challis}, {Chatzopoulos}, {Cheung}, {Choi}, {Chomiuk}, {Chugai}, {Contreras}, {Drout}, {Fesen}, {Foley}, {Fong}, {Friedman}, {Gall}, {Gehrels}, {Hjorth}, {Hsiao}, {Kirshner}, {Im}, {Leloudas}, {Lunnan}, {Marion}, {Martin}, {Morrell}, {Neugent}, {Omodei}, {Phillips}, {Rest}, {Silverman}, {Strader}, {Stritzinger}, {Szalai}, {Utterback}, {Vinko}, {Wheeler}, {Arnett}, {Campana}, {Chevalier}, {Ginsburg}, {Kamble}, {Roming}, {Pritchard}, \& {Stringfellow}}]{2014ApJ...780...21M}
{Margutti}, R., {Milisavljevic}, D., {Soderberg}, A.~M., {et~al.} 2014, \apj, 780, 21, \dodoi{10.1088/0004-637X/780/1/21}

\bibitem[{{Margutti} {et~al.}(2017){Margutti}, {Kamble}, {Milisavljevic}, {Zapartas}, {de Mink}, {Drout}, {Chornock}, {Risaliti}, {Zauderer}, {Bietenholz}, {Cantiello}, {Chakraborti}, {Chomiuk}, {Fong}, {Grefenstette}, {Guidorzi}, {Kirshner}, {Parrent}, {Patnaude}, {Soderberg}, {Gehrels}, \& {Harrison}}]{2017ApJ...835..140M}
{Margutti}, R., {Kamble}, A., {Milisavljevic}, D., {et~al.} 2017, \apj, 835, 140, \dodoi{10.3847/1538-4357/835/2/140}

\bibitem[{{Mewe} {et~al.}(1985){Mewe}, {Gronenschild}, \& {van den Oord}}]{mgv85}
{Mewe}, R., {Gronenschild}, E.~H.~B.~M., \& {van den Oord}, G.~H.~J. 1985, \aaps, 62, 197

\bibitem[{{Miceli} {et~al.}(2019){Miceli}, {Orlando}, {Burrows}, {Frank}, {Argiroffi}, {Reale}, {Peres}, {Petruk}, \& {Bocchino}}]{2019NatAs...3..236M}
{Miceli}, M., {Orlando}, S., {Burrows}, D.~N., {et~al.} 2019, Nature Astronomy, 3, 236, \dodoi{10.1038/s41550-018-0677-8}

\bibitem[{{Mignone} {et~al.}(2005){Mignone}, {Plewa}, \& {Bodo}}]{2005ApJS..160..199M}
{Mignone}, A., {Plewa}, T., \& {Bodo}, G. 2005, \apjs, 160, 199, \dodoi{10.1086/430905}

\bibitem[{{Mignone} {et~al.}(2012){Mignone}, {Zanni}, {Tzeferacos}, {van Straalen}, {Colella}, \& {Bodo}}]{2012ApJS..198....7M}
{Mignone}, A., {Zanni}, C., {Tzeferacos}, P., {et~al.} 2012, \apjs, 198, 7, \dodoi{10.1088/0067-0049/198/1/7}

\bibitem[{{Milisavljevic} {et~al.}(2015){Milisavljevic}, {Margutti}, {Kamble}, {Patnaude}, {Raymond}, {Eldridge}, {Fong}, {Bietenholz}, {Challis}, {Chornock}, {Drout}, {Fransson}, {Fesen}, {Grindlay}, {Kirshner}, {Lunnan}, {Mackey}, {Miller}, {Parrent}, {Sanders}, {Soderberg}, \& {Zauderer}}]{2015ApJ...815..120M}
{Milisavljevic}, D., {Margutti}, R., {Kamble}, A., {et~al.} 2015, \apj, 815, 120, \dodoi{10.1088/0004-637X/815/2/120}

\bibitem[{{Morozova} {et~al.}(2015){Morozova}, {Piro}, {Renzo}, {Ott}, {Clausen}, {Couch}, {Ellis}, \& {Roberts}}]{2015ApJ...814...63M}
{Morozova}, V., {Piro}, A.~L., {Renzo}, M., {et~al.} 2015, \apj, 814, 63, \dodoi{10.1088/0004-637X/814/1/63}

\bibitem[{{Neustadt} {et~al.}(2024){Neustadt}, {Kochanek}, \& {Smith}}]{2024MNRAS.527.5366N}
{Neustadt}, J.~M.~M., {Kochanek}, C.~S., \& {Smith}, M.~R. 2024, \mnras, 527, 5366, \dodoi{10.1093/mnras/stad3073}

\bibitem[{{Ono} {et~al.}(2020){Ono}, {Nagataki}, {Ferrand}, {Takahashi}, {Umeda}, {Yoshida}, {Orland o}, \& {Miceli}}]{2020ApJ...888..111O}
{Ono}, M., {Nagataki}, S., {Ferrand}, G., {et~al.} 2020, \apj, 888, 111, \dodoi{10.3847/1538-4357/ab5dba}

\bibitem[{{Orlando}(2023)}]{2023arXiv231105612O}
{Orlando}, S. 2023, arXiv e-prints, arXiv:2311.05612, \dodoi{10.48550/arXiv.2311.05612}

\bibitem[{{Orlando} \& {Drake}(2012)}]{2012MNRAS.419.2329O}
{Orlando}, S., \& {Drake}, J.~J. 2012, \mnras, 419, 2329, \dodoi{10.1111/j.1365-2966.2011.19880.x}

\bibitem[{{Orlando} {et~al.}(2009){Orlando}, {Drake}, \& {Laming}}]{2009A&A...493.1049O}
{Orlando}, S., {Drake}, J.~J., \& {Laming}, J.~M. 2009, \aap, 493, 1049, \dodoi{10.1051/0004-6361:200810109}

\bibitem[{{Orlando} {et~al.}(2015){Orlando}, {Miceli}, {Pumo}, \& {Bocchino}}]{2015ApJ...810..168O}
{Orlando}, S., {Miceli}, M., {Pumo}, M.~L., \& {Bocchino}, F. 2015, \apj, 810, 168, \dodoi{10.1088/0004-637X/810/2/168}

\bibitem[{{Orlando} {et~al.}(2016){Orlando}, {Miceli}, {Pumo}, \& {Bocchino}}]{2016ApJ...822...22O}
---. 2016, \apj, 822, 22, \dodoi{10.3847/0004-637X/822/1/22}

\bibitem[{{Orlando} {et~al.}(2021){Orlando}, {Wongwathanarat}, {Janka}, {Miceli}, {Ono}, {Nagataki}, {Bocchino}, \& {Peres}}]{2021A&A...645A..66O}
{Orlando}, S., {Wongwathanarat}, A., {Janka}, H.~T., {et~al.} 2021, \aap, 645, A66, \dodoi{10.1051/0004-6361/202039335}

\bibitem[{{Orlando} {et~al.}(2019){Orlando}, {Miceli}, {Petruk}, {Ono}, {Nagataki}, {Aloy}, {Mimica}, {Lee}, {Bocchino}, {Peres}, \& {Guarrasi}}]{2019A&A...622A..73O}
{Orlando}, S., {Miceli}, M., {Petruk}, O., {et~al.} 2019, \aap, 622, A73, \dodoi{10.1051/0004-6361/201834487}

\bibitem[{{Orlando} {et~al.}(2020){Orlando}, {Ono}, {Nagataki}, {Miceli}, {Umeda}, {Ferrand}, {Bocchino}, {Petruk}, {Peres}, {Takahashi}, \& {Yoshida}}]{2020A&A...636A..22O}
{Orlando}, S., {Ono}, M., {Nagataki}, S., {et~al.} 2020, \aap, 636, A22, \dodoi{10.1051/0004-6361/201936718}

\bibitem[{{Orlando} {et~al.}(2022){Orlando}, {Wongwathanarat}, {Janka}, {Miceli}, {Nagataki}, {Ono}, {Bocchino}, {Vink}, {Milisavljevic}, {Patnaude}, \& {Peres}}]{2022A&A...666A...2O}
{Orlando}, S., {Wongwathanarat}, A., {Janka}, H.~T., {et~al.} 2022, \aap, 666, A2, \dodoi{10.1051/0004-6361/202243258}

\bibitem[{{Passy} {et~al.}(2012){Passy}, {De Marco}, {Fryer}, {Herwig}, {Diehl}, {Oishi}, {Mac Low}, {Bryan}, \& {Rockefeller}}]{2012ApJ...744...52P}
{Passy}, J.-C., {De Marco}, O., {Fryer}, C.~L., {et~al.} 2012, \apj, 744, 52, \dodoi{10.1088/0004-637X/744/1/52}

\bibitem[{{Pastorello} {et~al.}(2007){Pastorello}, {Smartt}, {Mattila}, {Eldridge}, {Young}, {Itagaki}, {Yamaoka}, {Navasardyan}, {Valenti}, {Patat}, {Agnoletto}, {Augusteijn}, {Benetti}, {Cappellaro}, {Boles}, {Bonnet-Bidaud}, {Botticella}, {Bufano}, {Cao}, {Deng}, {Dennefeld}, {Elias-Rosa}, {Harutyunyan}, {Keenan}, {Iijima}, {Lorenzi}, {Mazzali}, {Meng}, {Nakano}, {Nielsen}, {Smoker}, {Stanishev}, {Turatto}, {Xu}, \& {Zampieri}}]{2007Natur.447..829P}
{Pastorello}, A., {Smartt}, S.~J., {Mattila}, S., {et~al.} 2007, \nat, 447, 829, \dodoi{10.1038/nature05825}

\bibitem[{{Pastorello} {et~al.}(2013){Pastorello}, {Cappellaro}, {Inserra}, {Smartt}, {Pignata}, {Benetti}, {Valenti}, {Fraser}, {Tak{\'a}ts}, {Benitez}, {Botticella}, {Brimacombe}, {Bufano}, {Cellier-Holzem}, {Costado}, {Cupani}, {Curtis}, {Elias-Rosa}, {Ergon}, {Fynbo}, {Hambsch}, {Hamuy}, {Harutyunyan}, {Ivarson}, {Kankare}, {Martin}, {Kotak}, {LaCluyze}, {Maguire}, {Mattila}, {Maza}, {McCrum}, {Miluzio}, {Norgaard-Nielsen}, {Nysewander}, {Ochner}, {Pan}, {Pumo}, {Reichart}, {Tan}, {Taubenberger}, {Tomasella}, {Turatto}, \& {Wright}}]{2013ApJ...767....1P}
{Pastorello}, A., {Cappellaro}, E., {Inserra}, C., {et~al.} 2013, \apj, 767, 1, \dodoi{10.1088/0004-637X/767/1/1}

\bibitem[{{Paxton} {et~al.}(2011){Paxton}, {Bildsten}, {Dotter}, {Herwig}, {Lesaffre}, \& {Timmes}}]{2011ApJS..192....3P}
{Paxton}, B., {Bildsten}, L., {Dotter}, A., {et~al.} 2011, \apjs, 192, 3, \dodoi{10.1088/0067-0049/192/1/3}

\bibitem[{{Paxton} {et~al.}(2013){Paxton}, {Cantiello}, {Arras}, {Bildsten}, {Brown}, {Dotter}, {Mankovich}, {Montgomery}, {Stello}, {Timmes}, \& {Townsend}}]{2013ApJS..208....4P}
{Paxton}, B., {Cantiello}, M., {Arras}, P., {et~al.} 2013, \apjs, 208, 4, \dodoi{10.1088/0067-0049/208/1/4}

\bibitem[{{Paxton} {et~al.}(2015){Paxton}, {Marchant}, {Schwab}, {Bauer}, {Bildsten}, {Cantiello}, {Dessart}, {Farmer}, {Hu}, {Langer}, {Townsend}, {Townsley}, \& {Timmes}}]{2015ApJS..220...15P}
{Paxton}, B., {Marchant}, P., {Schwab}, J., {et~al.} 2015, \apjs, 220, 15, \dodoi{10.1088/0067-0049/220/1/15}

\bibitem[{{Paxton} {et~al.}(2018){Paxton}, {Schwab}, {Bauer}, {Bildsten}, {Blinnikov}, {Duffell}, {Farmer}, {Goldberg}, {Marchant}, {Sorokina}, {Thoul}, {Townsend}, \& {Timmes}}]{2018ApJS..234...34P}
{Paxton}, B., {Schwab}, J., {Bauer}, E.~B., {et~al.} 2018, \apjs, 234, 34, \dodoi{10.3847/1538-4365/aaa5a8}

\bibitem[{{Paxton} {et~al.}(2019){Paxton}, {Smolec}, {Schwab}, {Gautschy}, {Bildsten}, {Cantiello}, {Dotter}, {Farmer}, {Goldberg}, {Jermyn}, {Kanbur}, {Marchant}, {Thoul}, {Townsend}, {Wolf}, {Zhang}, \& {Timmes}}]{2019ApJS..243...10P}
{Paxton}, B., {Smolec}, R., {Schwab}, J., {et~al.} 2019, \apjs, 243, 10, \dodoi{10.3847/1538-4365/ab2241}

\bibitem[{{Ravi} {et~al.}(2024){Ravi}, {Park}, {Zhekov}, {Orlando}, {Miceli}, {Frank}, {Broos}, \& {Burrows}}]{2024ApJ...966..147R}
{Ravi}, A.~P., {Park}, S., {Zhekov}, S.~A., {et~al.} 2024, \apj, 966, 147, \dodoi{10.3847/1538-4357/ad3800}

\bibitem[{{Raymond} \& {Smith}(1977)}]{rs77}
{Raymond}, J.~C., \& {Smith}, B.~W. 1977, \apjs, 35, 419

\bibitem[{{Reichardt} {et~al.}(2019){Reichardt}, {De Marco}, {Iaconi}, {Tout}, \& {Price}}]{2019MNRAS.484..631R}
{Reichardt}, T.~A., {De Marco}, O., {Iaconi}, R., {Tout}, C.~A., \& {Price}, D.~J. 2019, \mnras, 484, 631, \dodoi{10.1093/mnras/sty3485}

\bibitem[{{Ricker} \& {Taam}(2012)}]{2012ApJ...746...74R}
{Ricker}, P.~M., \& {Taam}, R.~E. 2012, \apj, 746, 74, \dodoi{10.1088/0004-637X/746/1/74}

\bibitem[{{R{\"o}pke} \& {De Marco}(2023)}]{2023LRCA....9....2R}
{R{\"o}pke}, F.~K., \& {De Marco}, O. 2023, Living Reviews in Computational Astrophysics, 9, 2, \dodoi{10.1007/s41115-023-00017-x}

\bibitem[{{Sana} {et~al.}(2012){Sana}, {de Mink}, {de Koter}, {Langer}, {Evans}, {Gieles}, {Gosset}, {Izzard}, {Le Bouquin}, \& {Schneider}}]{2012Sci...337..444S}
{Sana}, H., {de Mink}, S.~E., {de Koter}, A., {et~al.} 2012, Science, 337, 444, \dodoi{10.1126/science.1223344}

\bibitem[{{Sand} {et~al.}(2020){Sand}, {Ohlmann}, {Schneider}, {Pakmor}, \& {R{\"o}pke}}]{2020A&A...644A..60S}
{Sand}, C., {Ohlmann}, S.~T., {Schneider}, F. R.~N., {Pakmor}, R., \& {R{\"o}pke}, F.~K. 2020, \aap, 644, A60, \dodoi{10.1051/0004-6361/202038992}

\bibitem[{{Sapienza} {et~al.}(2024){Sapienza}, {Miceli}, {Bamba}, {Orlando}, {Lee}, {Nagataki}, {Ono}, {Katsuda}, {Mori}, {Sawada}, {Terada}, {Giuffrida}, \& {Bocchino}}]{2024ApJ...961L...9S}
{Sapienza}, V., {Miceli}, M., {Bamba}, A., {et~al.} 2024, \apjl, 961, L9, \dodoi{10.3847/2041-8213/ad16e3}

\bibitem[{{Smith}(2017)}]{2017hsn..book..403S}
{Smith}, N. 2017, {``Interacting Supernovae: Types IIn and Ibn'' chapter in {\it Handbook of Supernovae}} (edited by Athem W. Alsabti and Paul Murdin, ISBN 978-3-319-21845-8. Springer International Publishing, Switzerland), 403, \dodoi{10.1007/978-3-319-21846-5_38}

\bibitem[{{Smith} {et~al.}(2015){Smith}, {Mauerhan}, {Cenko}, {Kasliwal}, {Silverman}, {Filippenko}, {Gal-Yam}, {Clubb}, {Graham}, {Leonard}, {Horst}, {Williams}, {Andrews}, {Kulkarni}, {Nugent}, {Sullivan}, {Maguire}, {Xu}, \& {Ben-Ami}}]{2015MNRAS.449.1876S}
{Smith}, N., {Mauerhan}, J.~C., {Cenko}, S.~B., {et~al.} 2015, \mnras, 449, 1876, \dodoi{10.1093/mnras/stv354}

\bibitem[{{Smith} {et~al.}(2001){Smith}, {Brickhouse}, {Liedahl}, \& {Raymond}}]{2001ApJ...556L..91S}
{Smith}, R.~K., {Brickhouse}, N.~S., {Liedahl}, D.~A., \& {Raymond}, J.~C. 2001, \apjl, 556, L91, \dodoi{10.1086/322992}

\bibitem[{{Sollerman} {et~al.}(2020){Sollerman}, {Fransson}, {Barbarino}, {Fremling}, {Horesh}, {Kool}, {Schulze}, {Sfaradi}, {Yang}, {Bellm}, {Burruss}, {Cunningham}, {De}, {Drake}, {Golkhou}, {Green}, {Kasliwal}, {Kulkarni}, {Kupfer}, {Laher}, {Masci}, {Rodriguez}, {Rusholme}, {Williams}, {Yan}, \& {Zolkower}}]{2020A&A...643A..79S}
{Sollerman}, J., {Fransson}, C., {Barbarino}, C., {et~al.} 2020, \aap, 643, A79, \dodoi{10.1051/0004-6361/202038960}

\bibitem[{{Sugerman} {et~al.}(2005){Sugerman}, {Crotts}, {Kunkel}, {Heathcote}, \& {Lawrence}}]{2005ApJS..159...60S}
{Sugerman}, B.~E.~K., {Crotts}, A.~P.~S., {Kunkel}, W.~E., {Heathcote}, S.~R., \& {Lawrence}, S.~S. 2005, Astrophys. J. Suppl. Ser., 159, 60, \dodoi{10.1086/430408}

\bibitem[{{Sukhbold} {et~al.}(2018){Sukhbold}, {Woosley}, \& {Heger}}]{2018ApJ...860...93S}
{Sukhbold}, T., {Woosley}, S.~E., \& {Heger}, A. 2018, \apj, 860, 93, \dodoi{10.3847/1538-4357/aac2da}

\bibitem[{{Sun} {et~al.}(2020){Sun}, {Maund}, \& {Crowther}}]{2020MNRAS.497.5118S}
{Sun}, N.-C., {Maund}, J.~R., \& {Crowther}, P.~A. 2020, \mnras, 497, 5118, \dodoi{10.1093/mnras/staa2277}

\bibitem[{{Thomas} {et~al.}(2022){Thomas}, {Wheeler}, {Dwarkadas}, {Stockdale}, {Vink{\'o}}, {Pooley}, {Xu}, {Zeimann}, \& {MacQueen}}]{twd22}
{Thomas}, B.~P., {Wheeler}, J.~C., {Dwarkadas}, V.~V., {et~al.} 2022, \apj, 930, 57, \dodoi{10.3847/1538-4357/ac5fa6}

\bibitem[{{Tinyanont} {et~al.}(2019){Tinyanont}, {Lau}, {Kasliwal}, {Maeda}, {Smith}, {Fox}, {Gehrz}, {De}, {Jencson}, {Bally}, \& {Masci}}]{2019ApJ...887...75T}
{Tinyanont}, S., {Lau}, R.~M., {Kasliwal}, M.~M., {et~al.} 2019, \apj, 887, 75, \dodoi{10.3847/1538-4357/ab521b}

\bibitem[{{Vargas} {et~al.}(2022){Vargas}, {De Colle}, {Brethauer}, {Margutti}, \& {Bernal}}]{2022ApJ...930..150V}
{Vargas}, F., {De Colle}, F., {Brethauer}, D., {Margutti}, R., \& {Bernal}, C.~G. 2022, \apj, 930, 150, \dodoi{10.3847/1538-4357/ac649d}

\bibitem[{{Vishniac}(1983)}]{1983ApJ...274..152V}
{Vishniac}, E.~T. 1983, \apj, 274, 152, \dodoi{10.1086/161433}

\bibitem[{{Vishniac} \& {Ryu}(1989)}]{1989ApJ...337..917V}
{Vishniac}, E.~T., \& {Ryu}, D. 1989, \apj, 337, 917, \dodoi{10.1086/167161}

\bibitem[{{Wongwathanarat} {et~al.}(2013){Wongwathanarat}, {Janka}, \& {M{\"u}ller}}]{2013A&A...552A.126W}
{Wongwathanarat}, A., {Janka}, H.~T., \& {M{\"u}ller}, E. 2013, \aap, 552, A126, \dodoi{10.1051/0004-6361/201220636}

\bibitem[{{Wongwathanarat} {et~al.}(2017){Wongwathanarat}, {Janka}, {M{\"u}ller}, {Pllumbi}, \& {Wanajo}}]{2017ApJ...842...13W}
{Wongwathanarat}, A., {Janka}, H.-T., {M{\"u}ller}, E., {Pllumbi}, E., \& {Wanajo}, S. 2017, \apj, 842, 13, \dodoi{10.3847/1538-4357/aa72de}

\end{thebibliography}
\bibliographystyle{aasjournal}



\end{document}